\documentclass[showpacs,preprintnumbers,amsmath,amssymb,prb]{revtex4}

\usepackage{graphicx}
\usepackage{dcolumn}
\usepackage{bm}

\providecommand\bnabla{\boldsymbol{\nabla}}

\newcommand\Rey{\mbox{\textit{Re}}}  

\newcommand\etal{\mbox{\textit{et al.}}}

\newcommand\eg{e.g.\ }

\newcommand{\no}[1]{}

\begin{document}


\title{MHD turbulence in a channel with spanwise magnetic field}

\author{Dmitry Krasnov}
\affiliation{Fakult\"at Maschinenbau, Technische Universit\"at Ilmenau,\\
Postfach 100565, 98684 Ilmenau, Germany}
\author{Oleg Zikanov}
\affiliation{ University of Michigan-Dearborn, 48128 MI, USA }
\author{ J\"{o}rg Schumacher and Thomas Boeck}
\affiliation{ Fakult\"at Maschinenbau, Technische Universit\"at Ilmenau,\\
Postfach 100565, 98684 Ilmenau, Germany}

\date{\today}

\begin{abstract}
The effect of a uniform spanwise magnetic field on a turbulent
channel flow is investigated for the case of low magnetic Reynolds
number.  DNS and LES computations are performed for two values of
the hydrodynamic Reynolds number ($10^4$ and $2\times 10^4$) and the
Hartmann number varying in a wide range. It is shown that the main
effect of the magnetic field is the suppression of turbulent
velocity fluctuations and momentum transfer in the wall-normal
direction. This leads to drag reduction and transformation of the
mean flow profile. The centerline velocity grows, the mean velocity
gradients near the wall decrease,  and  the typical horizontal
dimensions of the coherent structures enlarge upon increasing the
Hartmann number.  Comparison between LES and DNS results shows that
the dynamic Smagorinsky model accurately reproduces the flow
transformation.
\end{abstract}

\maketitle

\section{introduction}
\label{sec:intro}

\begin{figure}[t]
\centerline{
\includegraphics[width=0.5\textwidth]{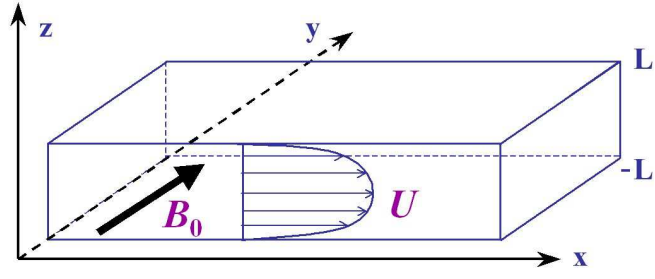}
}
\caption{Channel flow under spanwise magnetic field;
configuration and computational domain.}
\label{fig1:geometry}
\end{figure}

In this paper, we consider a channel flow of an incompressible
electrically conducting fluid (for example, a liquid metal) driven by
a pressure gradient and affected by a uniform steady-state magnetic
field oriented in the spanwise direction (see fig. \ref{fig1:geometry}).
It is assumed that the magnetic Reynolds number is small
\begin{equation}
\Rey_m \equiv \frac{UL}{\eta} \ll 1  \label{rem}.
\end{equation}
Here, $U$ and $L$ are the typical velocity and length scales, chosen
in our case as the centerline velocity of the laminar flow and half
the channel width, and $\eta=(\sigma \mu_0)^{-1}$ is the magnetic
diffusivity, $\sigma$ and $\mu_0$ being the electric conductivity of
the liquid and magnetic permeability of vacuum, respectively.  The
assumption (\ref{rem}) is satisfied by the majority of industrial
and laboratory flows of liquid metals.  The prominent examples of
industrial applications are the continuous casting of
steel\cite{Thomas:Zhang:2001} and growth of large semiconductor
crystals by the Czochralski method,\cite{Ammon:2005} where constant
magnetic fields are imposed with the purpose of controlling the flow
and suppressing unwanted motions. Another example is the liquid
metal (\textsl{Li} or \textsl{Li$_{17}$Pb}) cooling blankets of
breeder type for future fusion reactors, where the inevitable and
very strong magnetic fields play a negative role by suppressing
turbulent heat and mass transfer.\cite{Barleon:2001}

The key feature of the low-$\Rey_m$ magnetohydrodynamics, which
separates it from high-$\Rey_m$ plasma and geo- and astrophysical
applications, is that the perturbations of the magnetic field
induced by fluid motions are much smaller than the imposed magnetic
field and can be neglected.  One can also assume that the magnetic
field perturbations adjust instantaneously to the velocity
fluctuations. As a result of these assumptions, the quasi-static
approximation \cite{Roberts:1967} can be applied, according to which
the Lorentz force is expressed as a linear functional of velocity,
and the governing equations for fluid motion and electric potential
are represented in a closed form as shown in the next section. The
flow is characterized by two non-dimensional parameters.  The first
one is
 the Reynolds number
\begin{equation}\label{reynolds}
\Rey\equiv UL/\nu,
\end{equation}
while the second is either the Hartmann number
\begin{equation}\label{hartmann}
Ha\equiv B L (\sigma/\rho \nu)^{1/2}
\end{equation}
or the magnetic interaction parameter
\begin{equation}
N\equiv Ha^2/Re
\end{equation}
characterizing the ratios of Lorentz to viscous and Lorentz
to inertia forces, respectively. In the formulas above, $B$
is the strength of the applied magnetic field and $\rho$ and
$\nu$ are the density and kinematic viscosity of the fluid.

The transformation of low-$\Rey_m$ turbulence under the impact of
an imposed uniform steady-state magnetic field has been relatively
thoroughly investigated for two simplified flows: homogeneous
turbulence and the Hartmann flow. Let us briefly discuss the aspects
of the results relevant for our study.

Homogeneous turbulence and its computational counterpart, the flow
in a periodic box, are particularly convenient for investigating the
basic features of the MHD transformation of turbulent fluctuations
far from solid walls. This was done analytically, \cite{Moffatt:1967,
Davidson:1997, Alemany:1979} numerically, \cite{Schumann:1976,
Zikanov:1998, Knaepen1:2004, Vorobev:2005} and
experimentally.\cite{Alemany:1979} It has been found that the
magnetic field affects the flow in two different ways. First, the
induced electric currents lead to additional suppression of
turbulence via Joule dissipation. Second, the flow acquires
anisotropy of gradients. A simple explanation of the origins of the
anisotropy is obtained
by considering the rate of Joule dissipation
\begin{equation}\label{mu}
\mu(\bm{k})=\sigma B^2\rho^{-1}
    |\hat{\bm{u}}(\bm{k},t)|^2 \cos^2\theta
\end{equation}
of the Fourier velocity mode $\hat{\bm{u}}(\bm{k})$, where $\theta$
is the angle between the wavenumber vector $\bm{k}$ and the magnetic
field $\bm{B}$. The magnetic field tends to suppress the modes with
small values of $\theta$, that is to eliminate velocity gradients
along the magnetic field lines. The flow becomes anisotropic and may
even approach a two-dimensional state with all fields uniform in the
direction of the magnetic field lines in the limit $N\rightarrow
\infty$.

The picture described above does not take into account the
non-linear interactions that tend to restore the flow isotropy. The
interplay between the two opposing effects was investigated
numerically, most recently in DNS and LES \cite{Vorobev:2005,
Vorobev:2007} of forced homogeneous turbulence. Among other results,
it has been found that the degree of the anisotropy is a relatively
robust function of the magnetic interaction parameter $N$. The
influence of the Reynolds number, the type of the large-scale
forcing, and, remarkably, the length scale, at which the anisotropy
is measured, is weak.

The Hartmann flow, i.e. the flow in a channel with uniform magnetic
field imposed in the direction transverse to the channel walls
\cite{Hartmann:1937} is a classical MHD flow that was the subject of
numerous analytical, experimental, and computational investigations
such as, \eg, Refs. \onlinecite{Sommeria:Moreau:1982,
Reed:Lykoudis:1978, Krasnov:2004, Boeck:2007}.
We would like to stress that, despite the common geometry, there are
fundamental differences between the Hartmann flow and the flow
studied in the present paper. In the Hartmann channel, the magnetic
field interacts directly with the mean flow and changes its profile
into the form with nearly flat core and thin Hartmann boundary
layers. In our case, no Lorentz force is produced by the interaction
between the mean flow and the magnetic field. The MHD effect on the
mean flow can, thus, only be indirect, via the transformation of
turbulent fluctuations.

In essential aspects of field-flow interaction our configuration is
closer to the homogeneous turbulence than to the Hartmann channel
flow.  In particular, similarly to  homogeneous turbulence and in
contrast to the Hartmann flow, the leading role is taken by the
direct interaction between the field and small-scale structures and
the flow is uniform and infinite in the direction of the magnetic
field. As a result, many features of the homogeneous MHD turbulence
are expected to be observed in our flow. For example, the simple
explanation of the origins of the MHD-related gradient anisotropy
based on (\ref{mu}) can be extended straightforwardly to our case if
one limits the wave vector $\bm{k}$ to streamwise and spanwise
components.

In contrast to homogeneous turbulence, the channel flow with
spanwise field provides an opportunity of studying the effect of the
MHD transformation of small-scale turbulent structures on the mean
flow. This phenomenon, undoubtedly playing an important role in
low-$\Rey_m$ MHD flows, cannot be addressed in the framework of
homogeneous turbulence or, in a direct and unequivocal way, of the
Hartmann flow.  Furthermore, we can investigate the effect of the
magnetic field in the presence of a mean shear. This can also be
done using the model of a homogeneous shear flow as in Ref.
\onlinecite{Kassinos:Knaepen:Wray:2006}, where the case of the
moderate magnetic Reynolds number was considered, but the channel
provides a more realistic system since it does not require
artificial forcing.

In the present paper, we also conduct an \emph{a-posteriori}
verification study of the LES (Large-Eddy-Simulation) models for the
case of a channel flow in a spanwise magnetic field.  A brief
discussion of main known facts concerning the LES modeling of
low-$\Rey_m$ MHD turbulence is, therefore, in order. The question is
closely related to the transformation of turbulent fluctuations
discussed above.  An accurate LES model must be capable of adjusting
to the key effects of the transformation: anisotropy of gradients,
suppression of non-linear interactions, steepening of the energy
power spectrum, and reduction of the subgrid-scale dissipation rate.
Accurate reproduction of the last effect seems particularly
important if one takes the view that the primary objective of an LES
model is to produce appropriate amount of subgrid-scale dissipation.
LES modeling of decaying \cite{Knaepen1:2004} and forced
\cite{Vorobev:2005, Vorobev:2007} homogeneous MHD turbulence
provides the basis for an assessment of the performance of Smagorinsky eddy
viscosity models. The main conclusion, achieved through comparison
with DNS data, is that the classical Smagorinsky model with constant
coefficient $C_S$ is overdissipative in the MHD case. On the
contrary, the model based on the dynamic evaluation of $C_S$
\cite{Germano:1991, Lilly:1992} has been found to accurately
reproduce the reduction of $C_S$ and, therefore, of the
subgrid-scale dissipation with growing magnetic field and, in
general, to give much better agreement with DNS.

LES of the Hartmann flow was conducted using conventional and
modified Smagorinsky models,\cite{Shimomura:1991} the dynamic
Smagorinsky model,\cite{Kobayashi:2006, Sarris:Kassinos:Carati:2007}
and the coherent structure model.\cite{Kobayashi:2006}  Albeit
further studies are needed for more extensive verification farther
from the transitional Reynolds numbers, first conclusions can be
made. As in the case of the periodic box, the conventional
Smagorinsky model becomes highly inaccurate as soon as the flow is
significantly affected by the magnetic field. The simple
modification based on modeling SGS Joule
dissipation\cite{Shimomura:1991} also does not provide an adequate
picture of the flow transformation. The models with coefficients
adjustable in accordance to local conditions, such as the dynamic
Smagorinsky or coherent structure models, perform much better
demonstrating good agreement with DNS and experiments. This is not
entirely surprising. The dynamic model has demonstrated its ability
to adjust to variations of the dissipation rate on many occasions,
including flows with rotation, mean shear, or laminar-turbulent
transition.

The only previous work, in which the effect of the spanwise magnetic
field on a turbulent channel flow was studied, was the DNS by Lee
and Choi.\cite{Lee:Choi:2001} Important conclusions were made.  In
particular, it was found that the imposed spanwise magnetic field
suppressed the turbulent fluctuations and reduced the drag.
Interestingly, similar behavior was observed for the streamwise
magnetic field, but not for the case of Hartmann flow, where
suppression of fluctuations was accompanied by increase of drag due
to strong shear in the Hartmann boundary layers.  It did not take a
very strong magnetic field to completely laminarize the flow.  For
example, in the case of the spanwise field, the laminarization
occurred at $N=0.05$.

As will be seen from the following discussion, our results confirm
and extend the findings of Ref. \onlinecite{Lee:Choi:2001}  An
important difference should, however, be noted from the beginning.
Due to low numerical resolution and inherent limitations of DNS, the
calculations of Ref. \onlinecite{Lee:Choi:2001} were conducted at
the low Reynolds number $\Rey\approx 3000$, which is in contrast
with $\Rey=10000$ and $20000$ in our work. Furthermore, we conduct a
much more detailed analysis of the flow transformation and
investigate performance of LES models in the MHD turbulence.

Several experimental studies of turbulent MHD flows in ducts of
large aspect ratio with the magnetic field parallel to the longer
side were conducted in 1970s.  The motivation was essentially the
same as of our work: to investigate the effect of the magnetic field
on turbulence without the dominating influence of the Hartmann
layers.  The experiments most relevant to our
study\cite{Votsish:Kolesnikov:1977} were performed with a flow of
mercury in a rectangular duct of  aspect ratio 10 to 1 and 
values of the Reynolds and Hartmann numbers between 7500 and 15500
and between 0 and 100, respectively.  As discussed in detail below,
the key experimental results are in good qualitative agreement with
our conclusions.  We did not attempt to conduct a quantitative
comparison.  The main reason is that the conditions approaching the
ideal configuration of a uniform magnetic field cannot be realized
in a straight duct experiment.  The magnetic field is inevitably
non-uniform in the streamwise direction.  Apart from making the
quantitative interpretation difficult, this creates the effect of
the so-called `magnetic obstacle'\cite{Votyakov:2007} leading to
further deviation between the experimental and numerical data.

The objectives of our study can be summarized as follows.  First, we
conduct a detailed investigation of the flow transformation caused
by the spanwise magnetic field.  The change of turbulence
statistics, wall friction, and coherent structures is documented and
discussed.  The second aspect of the motivation concerns the
application of LES methods to low-$\Rey_m$ MHD turbulence.  As
discussed above, the dynamic Smagorinsky model has shown reasonably
good accuracy in the latter two cases.  We present a thorough
\emph{a-posteriori} verification of the model for the case of a
channel flow with a spanwise magnetic field.

The paper is organized as follows.  A brief review of the
computational procedure and LES model is provided in the next
section.  After presenting the results of the model verification in
LES-DNS comparison in section \ref{sec:comparison}, we discuss the
flow transformation and its anisotropy characteristics in section
\ref{sec:transform}. A discussion of the effect of drag reduction by
a spanwise magnetic field and its relation to other mechanisms of
drag reduction is given in section \ref{sec:drag}. Finally, 
concluding remarks are provided in section \ref{sec:conclusion}.

\section{Governing Equations, Numerical Method, and Parameters}
\label{sec:equations}

\subsection{Navier-Stokes equations for low-$\Rey_m$ MHD flows}

\indent We consider the flow of an incompressible, electrically
conducting fluid in an infinite plane channel between insulating
walls located at $z=\pm L$, where $x$, $y$ and $z$ denote the
streamwise, spanwise and wall-normal directions, respectively (see
fig. \ref{fig1:geometry}). The flow is driven by a pressure gradient
$\partial P_0/ \partial x$ in the $x$-direction and submitted to a
constant spanwise magnetic field $\bm{B}_0 = {B}_0 \bm{e}$, where
$\bm{e} \equiv (0,1,0)$ is the unit vector in the spanwise
direction.

In the limit of low magnetic Reynolds number $\Rey_m$,
the so-called quasi-static approximation \cite{Roberts:1967} can be applied.
The fluctuating part $\bm{b}$ of the magnetic field due to
fluid motion is much smaller than the external magnetic field:
$\bm{b} \ll \bm{B}_0$. The Lorenz force reduces to
\begin{equation}
\label{Lorenz_force}
\bm{F} = \bm{j}\times \bm{B}_0,
\end{equation}
where the induced electric current density is given by Ohm's law
\begin{equation}
\label{Ohms_law}
\bm{j} =\sigma \left( - \nabla \phi + \bm{v} \times \bm{B}_0\right).
\end{equation}

If displacement currents are neglected and the fluid is assumed
electrically neutral, the current density $\bm{j}$ should satisfy
the divergence-free constraint $\nabla \cdot\bm{j} = 0$. Thus, the
electric potential $\phi$ can be expressed via the following
equation:
\begin{equation}
\nabla^2\phi= \nabla\cdot\left(\bm{v}\times \bm{B}_0\right).
\end{equation}
The problem is solved in a rectangular domain with periodicity
conditions used in the $x$- and $y$-directions following the
assumption of flow homogeneity. The no-slip conditions are
imposed at the walls:
\begin{equation}
\label{MHDbc1} u=v=w=0 \hskip3mm\mbox{at  } z=\pm L.
\end{equation}
The electric potential $\phi$ is also periodic in the $x$- and
$y$-directions. Since no current flows through the electrically
insulating walls and the velocity $\bm{v}$ is zero at these walls,
(\ref{Ohms_law}) leads to the boundary conditions for the electric
potential
\begin{equation}
\label{MHDbc2}
\frac{\partial \phi}{\partial z}=0 \hskip3mm\mbox{at  } z=\pm L.
\end{equation}

As remarked in section \ref{sec:intro}, in our geometry there is no
interaction between the mean flow and the magnetic field. The
non-dimensional laminar basic velocity field retains the classical parabolic
profile
\begin{equation}
\label{BasicFlowNonDim} U(z) = 1 - z^2.
\end{equation}
Hereafter we use the centerline velocity $U$ of the Poiseuille flow
and the half-channel width $L$ as the velocity and length scales
correspondingly. Finally, taking the units of the magnetic field
and the electric potential as ${B}$ and $LU{B}$, the non-dimensional
governing equations and boundary conditions read
\begin{eqnarray}
\label{MHDeqn}
\hskip-15mm& & \frac{\partial \bm{v}}{\partial t} +
(\bm{v}\cdot \nabla)\bm{v} = -\nabla p
+  \frac{1}{Re}\nabla^2 \bm{v} + N \left(
-\nabla \phi \times \bm{e}  + \left(\bm{v}\times \bm{e} \right)\times \bm{e}
\right),\no{MHDeqn}\\
\hskip-15mm& & \nabla \cdot\bm{v}  =  0,\\
 \hskip-15mm& & \nabla^2 \phi =\nabla \cdot\left(\bm{v} \times \bm{e}\right),\\
\label{MHDeqn_bc} \hskip-15mm& & u = v = w = \frac{\partial
\phi}{\partial z}=0 \hskip3mm\mbox{at } z=\pm 1.\no{MHDeqn_bc}
\end{eqnarray}
There are two non-dimensional parameters, the Reynolds number
$\Rey$ and the magnetic interaction parameter $N$, defined in
the previous section. To specify the mean pressure gradient
we assume that the volume flux $Q$ per span width remains
constant.

\subsection{Dynamic subgrid scale model}

In  LES only large and moderate length scales of the flow are
resolved. A spatial filtering procedure is applied to the governing
equations (\ref{MHDeqn}--\ref{MHDeqn_bc}), which results in the
following equations for filtered fields (denoted by overbar):
\begin{eqnarray}
\label{MHDflt} \hskip-15mm& & \frac{\partial \bar{v}_i}{\partial t}
+ \bar{v}_j \frac{\partial \bar{v}_i}{\partial x_j} = -
\frac{\partial \bar{p}}{\partial x_i} + \frac{1}{Re}\nabla^2
\bar{v}_i - \frac{\partial \tau^a_{ij}}{\partial x_j} + N \left(
-\nabla \bar{\phi} \times \bm{e} + \left(\bar{\bm{v}}\times \bm{e}
\right)\times \bm{e}
\right)_i,\no{MHDflt}\\
\hskip-15mm& & \nabla \cdot\bar{\bm{v}}  =  0,\\
\hskip-15mm& & \nabla^2 \bar{\phi} =
\nabla \cdot\left(\bar{\bm{v}} \times \bm{e}\right),\\
\label{MHDflt_bc}
\hskip-15mm& & \bar{v}_x = \bar{v}_y = \bar{v}_z =
\frac{\partial \bar{\phi}}{\partial z}=0 \hskip3mm\mbox{at } z=\pm 1.
\no{MHDflt_bc}
\end{eqnarray}

Here $\tau^a_{ij}$ is the traceless part of the sub-grid scale (SGS)
stress tensor defined as
\begin{equation}
\label{tau_sgs}
\tau^a_{ij} = \tau_{ij} - \frac{\delta_{ij}}{3}\tau_{kk},
\mbox{ with }
\tau_{ij} = \overline{v_i v_j} - \bar{v}_i\bar{v}_j.
\end{equation}

To provide a closure for the filtered equations, the tensor
$\tau^a_{ij}$ has to be modeled
in terms of the resolved fields. The  Smagorinsky model is used in
the present study. It is based on the eddy-viscosity hypothesis and
assumes that
\begin{equation}
\label{tau_smag}
\tau^a_{ij} = -2\nu_T \bar{S}_{ij} =
-2 C_S \bar{\Delta}^2|\bar{S}|\bar{S}_{ij},
\end{equation}
where $\bar{\Delta}$ denotes the filter width and $C_S$  is the
Smagorinsky constant. $|\bar{S}|=\left(2 \bar{S}_{ij}
\bar{S}_{ij}\right)^{1/2}$ is the magnitude of the  strain-rate
tensor 
\begin{equation}
\bar{S}_{ij}=\frac{1}{2}
\left(
\frac{\partial\bar{v}_i}{\partial x_j} +
\frac{\partial\bar{v}_j}{\partial x_i}
\right).
\end{equation}
of the filtered velocity field.

In the classical Smagorinsky eddy-viscosity model, $C_S$ is a
constant parameter usually adjusted for a particular flow
configuration. Germano \etal \cite{Germano:1991} proposed an
extension to this model, the so-called dynamic procedure, later
optimized with a least-square method suggested by
Lilly.\cite{Lilly:1992} Within this approach, the parameter $C_S$ is
determined on the basis of the velocity field and, thus, is a
function of space and, in general, time. Following the assumption of
scale self-similarity of formula (\ref{tau_smag}), the computed
velocity field is filtered once again with the so-called ``test''
filter width $\widehat{\bar{\Delta}}$ larger than the ``grid''
filter width $\bar{\Delta}$.  The spectral cut-off in Fourier space
is used in our computations as the filtering operation.  No test
filtering is performed in the wall-normal direction. The parameter
$C(z)$ is determined as
\begin{equation}
\label{eq:CSdyn} C(z) = \frac{<\bar{M}_{ij}
\bar{L}_{ij}>_{x,y}}{<\bar{M}_{ij} \bar{M}_{ij}>_{x,y}}
\end{equation}
where $\bar{L}_{ij}$ and $\bar{M}_{ij}$ are defined as
\begin{equation}
\label{eq:Lij}
\bar{L}_{ij} = \widehat{\bar{v}_{i} \bar{v}_{j}}
- \hat{\bar{v}}_i \hat{\bar{v}}_j
\end{equation}
and
\begin{equation}
\label{eq:Mij}
\bar{M}_{ij} = 2 \bar{\Delta}^2 \widehat{|\bar{S}|\bar{S}_{ij}}
- 2 \widehat{\bar{\Delta}}^2 \widehat{|\bar{S}|} \widehat{\bar{S}}_{ij}.
\end{equation}
Here symbol~$\widehat{...}$~stands for the test filtering, and the
spatial averaging over the homogeneous ($x,y$)-directions denoted as
$<...>_{x,y}$ is applied to avoid numerical instability. In our
simulations, the ratio $\widehat{\bar{\Delta}} / \bar{\Delta}=2$ is
used. Further details on the dynamic procedure, its application to
classical channel flow, as well as the possible impact of the filter
width, can be found in Refs. \onlinecite{Germano:1991, Lilly:1992,
Piomelli:1993}.

\subsection{Numerical method}

The evolution of the flow is found as a numerical solution of the
full (\ref{MHDeqn})--(\ref{MHDeqn_bc}) or filtered
(\ref{MHDflt})--(\ref{MHDflt_bc}) equations (for DNS and LES parts
of our study, correspondingly). We use a pseudo-spectral method,
where the flow field is represented by velocity potentials complying
with the incompressibility constraint. A detailed description of the
algorithm and the corresponding flow solver are given
elsewhere.\cite{Krasnov:2004, Krasnov:2007} Briefly, the method
applies a Fourier expansion in the horizontal directions, where
periodical boundary conditions are imposed, and a Chebyshev
polynomial expansion in the vertical direction between insulating
walls with no-slip conditions. Nonlinear terms are calculated
through Fast Fourier Transforms (FFT). The algorithm is parallelized
using domain decomposition in the streamwise direction. FFTs are
performed locally by transposition of the data array across
the processors.

The modifications made for the present study concern the Lorentz
force and the time-stepping method (see also Ref.
\onlinecite{Krasnov:2007}). The Lorentz force term is changed to the
case of spanwise orientation of the magnetic field and is now
treated as an explicit term in the temporal discretization.
Furthermore, the new time-stepping scheme uses three time levels for
the approximation of the time derivative and is second-order
accurate. We have also implemented de-aliasing following the $2/3$
rule.\cite{Canuto:1988} At last, the code now includes the dynamic
and conventional Smagorinsky procedures of computation of the SGS
stress term $\tau^a_{ij}$, so that it can be used both for DNS and
LES calculations.


For the conventional Smagorinsky model, the parameter $C_s$ in
equation (\ref{tau_smag}) is estimated as
\begin{equation}
\label{eq:delta1} C_s = C^2 (1-\exp(-z^+/25))^2,
\end{equation}
where the value $C = 0.1$ found optimal for LES in wall-bounded
flows is used\cite{Moin:1982}. The other term is the van Driest damping
term\cite{vanDriest:1956} for  $C_S$ near the walls, which is based
on  the wall distance $z^+$ in friction units. To identify the grid
filter $\bar{\Delta}$ in that case, we apply the commonly used
definition
\begin{equation}
\label{eq:delta2}
\bar{\Delta}^3 = \bar{\Delta}_x \bar{\Delta}_y \bar{\Delta}_z.
\end{equation}
Here the homogeneous filter widths
$\bar{\Delta}_x$, $\bar{\Delta}_y$ in the horizontal directions
and the variable filter width $\bar{\Delta}_z$ in the wall-normal
are defined as
\begin{equation}
\label{eq:delta3}
\bar{\Delta}_x = \tfrac{3}{2} L_x/N_x,
\hskip5mm
\bar{\Delta}_y = \tfrac{3}{2} L_y/N_y,
\hskip5mm
\bar{\Delta}_z^k = \tfrac{3}{2} |z_{k-1} - z_{k+1}|/2,
\end{equation}
where the collocation points in the Chebyshev direction are denoted
by $z_k$. The $3/2$ prefactor accounts for the effective number of
collocation points due to de-aliasing by the $2/3$-rule.


\subsection{Procedure and parameters of numerical experiments}

The parameters of the numerical experiments, such as the dimensions
of the computational domain, numerical resolution, and Reynolds
numbers are listed in Table \ref{table:param1}. The simulations
were conducted for two values of the constant volume flux identified by
the values of the Reynolds number $\Rey$ based on the half-channel
width and centerline velocity of the laminar parabolic profile.
The values of the flux-based Reynolds number $\Rey_q$ are also
given in table \ref{table:param1}.
The
Hartmann number $Ha$ varied between zero and a value close to the
threshold above which the three-dimensional turbulence could not be
sustained.  The flows at higher $Ha$ were characterized by
intermittency with long periods of nearly two-dimensional behavior
interrupted by three-dimensional bursts.  This phenomenon is still
under investigation and will not be discussed in the present paper.
We only note that the corresponding threshold values of $Ha$ and $N$
seem to depend on the Reynolds number. Continuous turbulence disappears
beyond $N\approx 0.09$ for $Re=10000$ and beyond $N\approx 0.08$ for
$Re=20000$. For $Re=3000$ studied by Lee \& Choi\cite{Lee:Choi:2001}
the turbulence is already completely suppressed for $N=0.05$. The
intermittent dynamics is missing at this value of $Re$  since the
laminar state is linearly stable.

The computations were conducted (at given $\Rey$ and $Ha$) for
sufficiently long periods of time until the integral characteristics
showed that the statistically steady state was reached. In order
to accelerate the evolution of DNS fields, the LES solutions at
the same parameters were used as initial conditions. The computations
of statistically steady flows were continued for several (not less
than 20) convective units $L/U$ and the flow statistics were
collected and averaged.

The choice of numerical resolution and domain sizes used for DNS
calculations was verified through a comparison with the results by
Moser \etal \cite{Moser:1999} for non-magnetic turbulent channel
flow. The values of $Re_\tau$  in our simulations at $Ha=0$ and both
$\Rey$ numbers are very close to those in Ref.
\onlinecite{Moser:1999} so a detailed comparison between the two
studies could be made. In particular, we have analyzed mean velocity
profiles and rms values of turbulent fluctuations and found very
good qualitative and quantitative agreement.

As an additional verification, we examined two-point correlations
and energy spectra of velocity fluctuations for both $\Rey$. The
two-point correlations computed at $z=0$ in stream- and spanwise
directions  showed rapid fall-off to nearly zero values at the
distance of the half-domain size. The energy spectra were observed
to span over $4.5 - 5$ decimal orders with significant decay at high
wavenumbers and without noticeable pile-up at the smallest scales.
This demonstrated that the small scales were sufficiently resolved.

The flow solver was also verified in LES computations.
In the first series of simulations we examined the effect of
LES resolution on convergence of integral flow parameters, such
as $Re_\tau$ and friction coefficient $C_f$. For the test case
we used our DNS at $256^3$ points and $Re_\tau \approx 180$
(corresponding to the turbulent channel flow studied by Kim
\etal \cite{Kim:Moin:1987}) and two LES calculations at $32^3$
and $64^3$ points employing the dynamic model. The comparison
indicated that the LES at $64^3$ points almost saturated at
the target $Re_\tau$ ($178.4$ vs. $180$), whereas the value is
clearly underpredicted on the coarser grid ($171.8$ vs. $180$).
The latter contributed to approximately $8\%$ difference in
the friction coefficient $C_f$. Similar dependence of accuracy
on numerical resolution was observed earlier by Piomelli \etal
\cite{Piomelli:1988} with the classical Smagorinsky model at
the same $Re_\tau$.

We also tested the dynamic model through comparison with
the results of channel flow LES at high Reynolds number by
Piomelli.\cite{Piomelli:1993} Simulations targeting the LES
results at $Re_\tau = 1050$ were conducted for different
resolutions ($64^3$ and $128^3$ points) close to those used
in Ref. \onlinecite{Piomelli:1993}. The analysis of profiles
of mean velocity and rms of turbulent fluctuations as well as
the model parameters such as the Smagorinsky constant and SGS
stresses $\tau_{ij}$ confirmed proper performance of our LES
model.

\section{DNS vs. LES comparison}
\label{sec:comparison} In this section we report the results of
\emph{a-posteriori} verification of the dynamic and classical
Smagorinsky models conducted via comparison with the data of
high-resolution DNS.  We also include results of deliberately
under-resolved DNS (hereafter UDNS) conducted at the same resolution
as the LES runs (see Table \ref{table:param1}) but without any SGS
terms appearing in the equations. Comparison with UDNS is commonly
used in studies of LES models in order to assess  the significance
of SGS contributions. The assessment becomes even more important
in the MHD case because of the possibility indicated by earlier
studies\cite{Knaepen1:2004,Vorobev:2005,Vorobev:2007} that at
high $Ha$ the MHD flow transformation depopulates the small length
scales and reduces their effect on the flow, thus rendering the LES
model unnecessary. In such case, the results of UDNS would approach
fully-resolved DNS data.

The time-averaged values of the key integral quantities of the flow
are listed in table \ref{table:param2}. They include the centerline
velocity $U_{cl}$, the friction Reynolds number
\begin{equation}\label{retau}
\Rey_{\tau} = \Rey \, u_{\tau},
\end{equation}
where the non-dimensional wall friction velocity $u_{\tau}$
is calculated as
\begin{equation}\label{utau}
u_{\tau}^2 = \Rey^{-1}\left(\partial u/\partial z\right)_{z = -1},
\end{equation}
the friction coefficient
\begin{equation}\label{cfri}
C_f=2(u_{\tau}/U_q)^2,
\end{equation}
where $U_q$ is the flux (mean) velocity, and the volume-averaged
dissipation rate at the resolved scales
\begin{equation}\label{dissrate}
\epsilon=\langle2\Rey^{-1}S_{ij}S_{ij}\rangle_v.
\end{equation}
The dissipation rate can be used to assess the importance of the
model component in LES since the difference between the DNS and LES
values may serve as an approximate measure of the dissipation
provided by the SGS closure. One can see that, despite the
relatively high numerical resolution of our LES runs, the
subgrid-scale dissipation constitutes a significant fraction of the
total, especially at the resolution of $64^3$ at $\Rey=20000$. The
under-resolved DNS produce significantly stronger dissipation, which
can be viewed as a result of energy pile-up at the scales near the
grid cut-off scale.  As expected, the difference between UDNS and
DNS decreases with $Ha$, but remains large even at the strongest
magnetic fields used in our study.

The data for the centerline velocity $U_{cl}$ show fairly good
agreement between the DNS and LES results in all cases and do not
allow us to differentiate between the dynamic and classical models.
Clear differentiation can, however, be made on the basis of computed
$\Rey_{\tau}$ and $C_f$.  With the possible exception of the case
$\Rey=20000$, $Ha=0$, the dynamic model consistently provides values
that are closer to the DNS data.  The improvement furnished by the
dynamic mechanism is particularly significant in the experiments
conducted at the lower resolution $64^3$ at $\Rey=20000$ (the rows
marked as DSM64 and SM64 in table \ref{table:param2}). The values of
all these coefficients are consistently overpredicted by UDNS.
Similarly to the dissipation rate, the error decreases with Ha, but
remains significant.

\begin{figure}[t]
\centerline{
\includegraphics[width=0.48\textwidth]{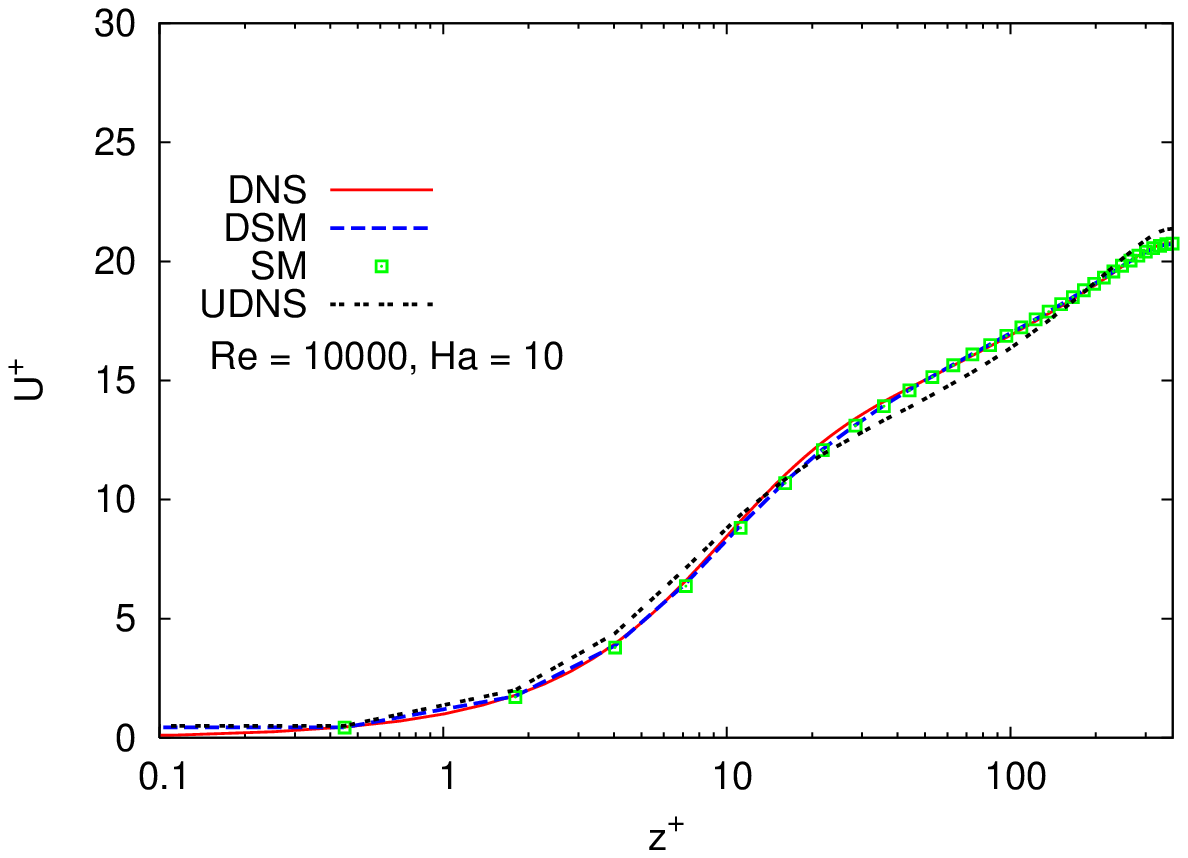}
\includegraphics[width=0.48\textwidth]{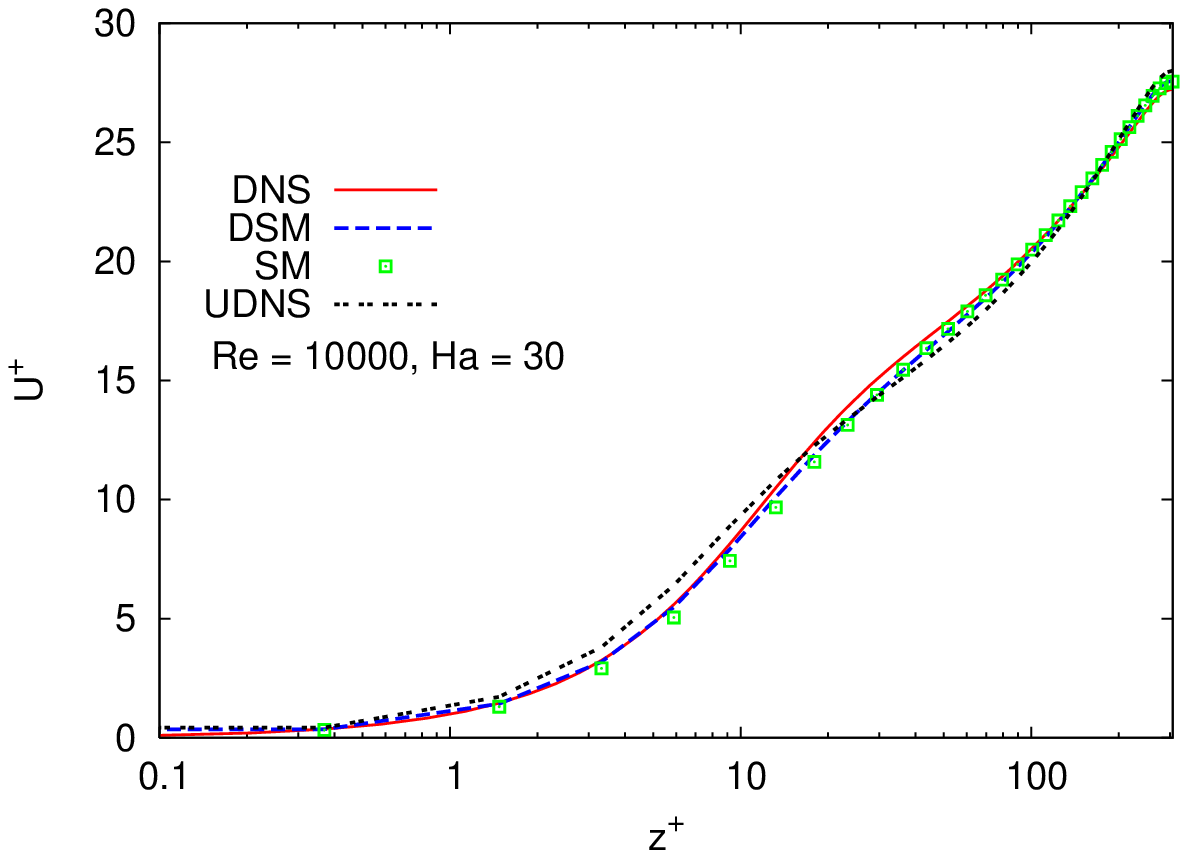}
}
\caption{Comparison between the LES and filtered DNS results. Mean
velocity profiles in wall coordinates based on $u_{\tau}$ and $\nu$
for $Re=10000$ at $Ha=10$ (left) and $Ha=30$ (right). The results
of unresolved DNS (indicated as UDNS) are shown for comparison.}
\label{fig3:re10_prof}
\end{figure}

\begin{figure}[t]
\centerline{
\includegraphics[width=0.48\textwidth]{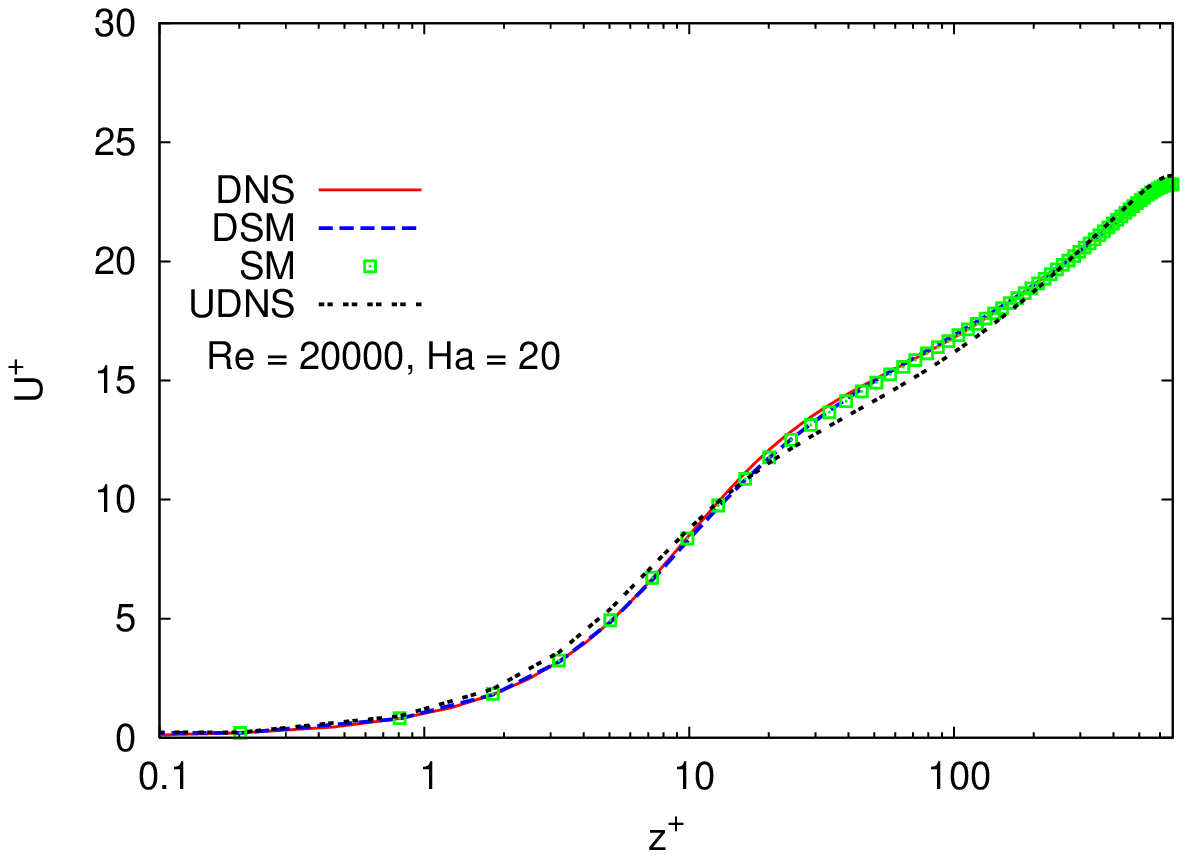}
\includegraphics[width=0.48\textwidth]{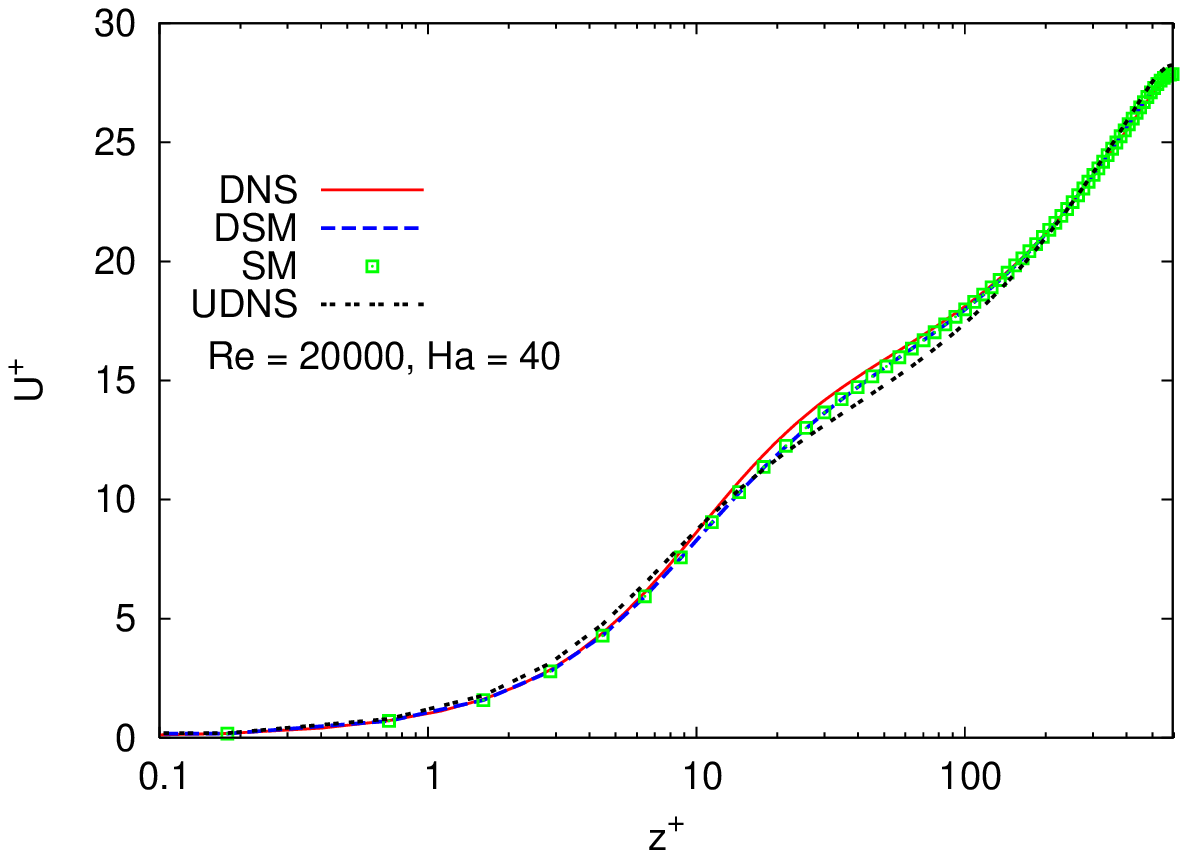}
}
\centerline{
\includegraphics[width=0.48\textwidth]{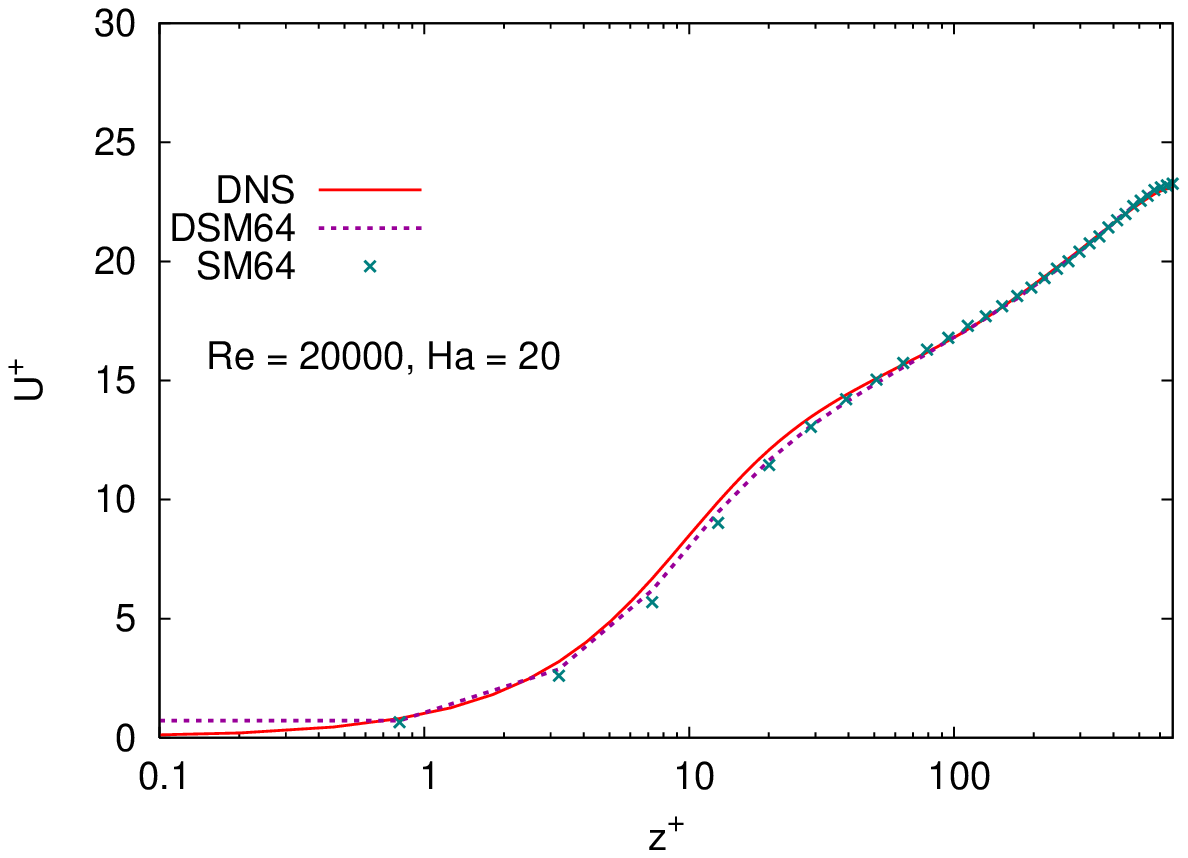}
\includegraphics[width=0.48\textwidth]{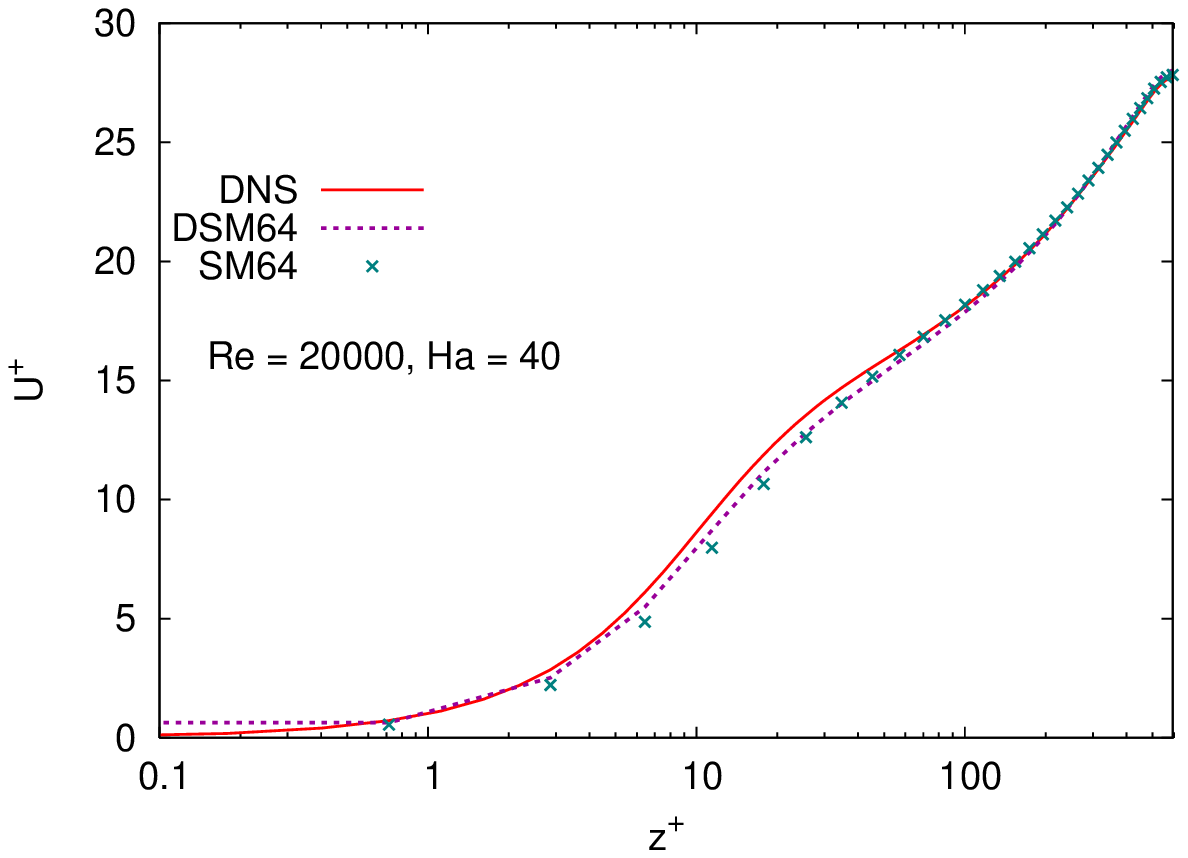}
}
\caption{Comparison between the LES and filtered DNS results. Mean
velocity profiles in wall coordinates based on $u_{\tau}$ and $\nu$
for $Re=20000$ at $Ha=20$ (left) and $Ha=40$ (right). The results
of unresolved DNS (UDNS on top figures) are shown for comparison.}
\label{fig3:re20_prof}
\end{figure}

The time-averaged mean velocity profiles are shown in figures
\ref{fig3:re10_prof} and \ref{fig3:re20_prof} for $\Rey=10000$ \&
$20000$ and two values of the Hartmann number. Wall units are used
and, for the sake of comparison, the DNS velocity field is filtered
to the LES resolution using the spectral cut-off operation. One can
see that both the dynamic and the classical models accurately
reproduce the DNS profile. The difference between the two models
becomes more visible if the resolution is lowered  to $64^3$. The
results presented by the bottom pictures of figure 3 show that the
classical model poorly reproduces the buffer region, especially for
the case of higher $Ha$.  As illustrated by the top pictures of
figure \ref{fig3:re20_prof}, the UDNS fails to reproduce the entire
velocity profile in the whole range of $\Rey$ and $Ha$ numbers. At
$\Rey=20000$, the $64^3$ simulation using the dynamic model produces
better results than UDNS at the resolution $128^3$.

\begin{figure}[t]
\centerline{
\includegraphics[width=0.48\textwidth]{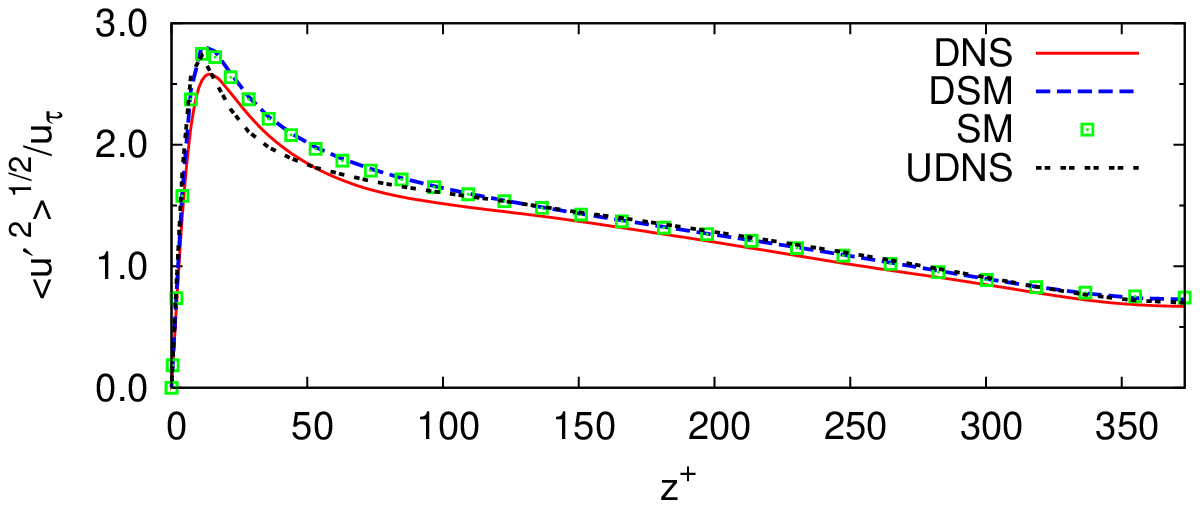}
\includegraphics[width=0.48\textwidth]{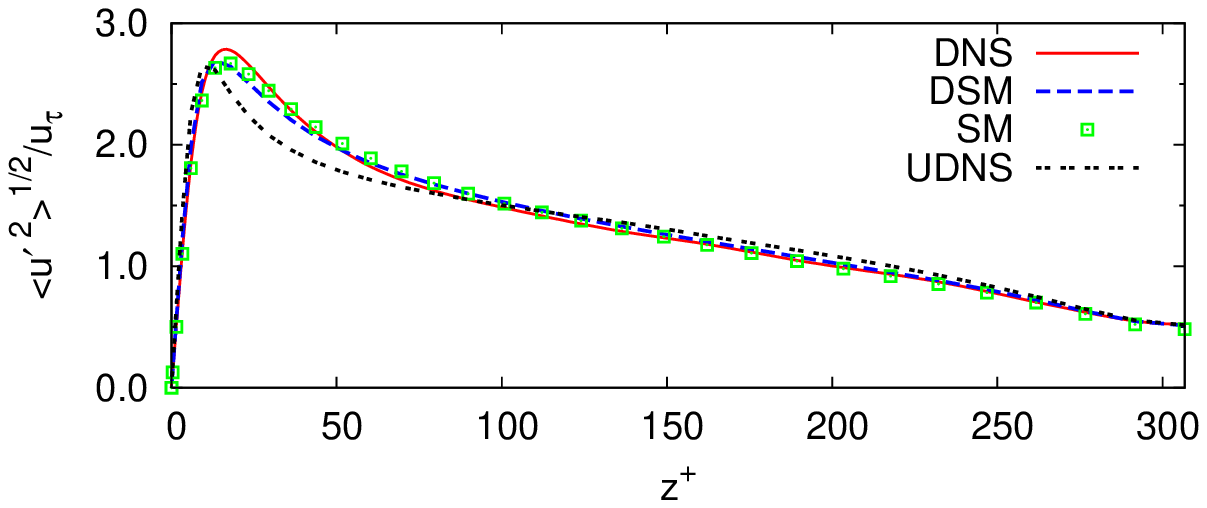}
}
\centerline{
\includegraphics[width=0.48\textwidth]{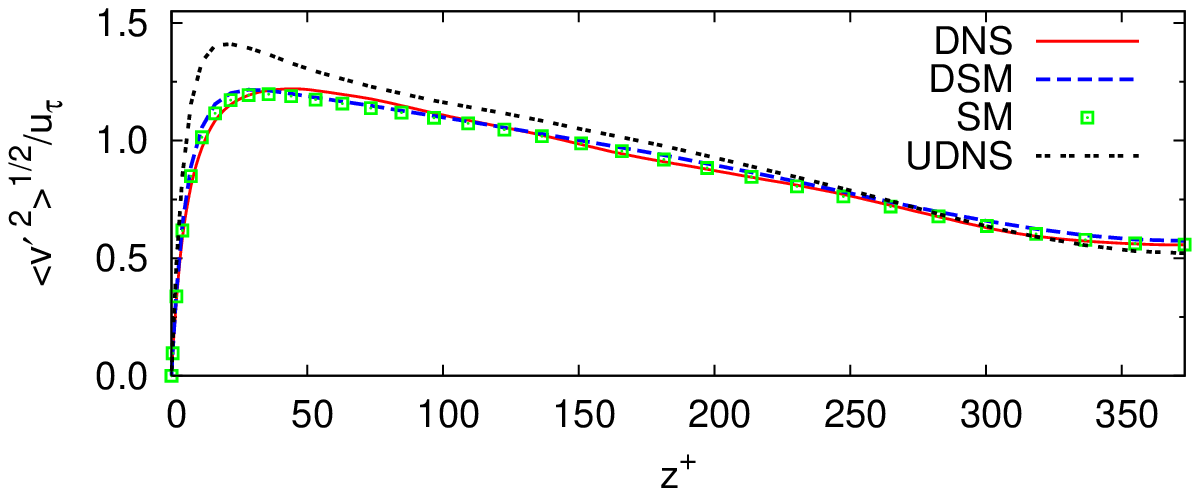}
\includegraphics[width=0.48\textwidth]{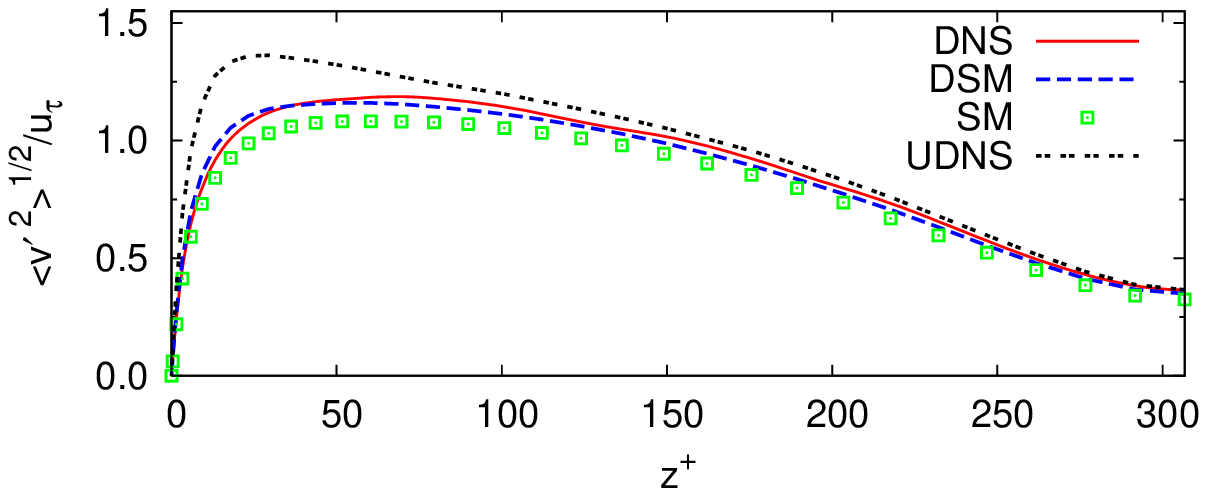}
}
\centerline{
\includegraphics[width=0.48\textwidth]{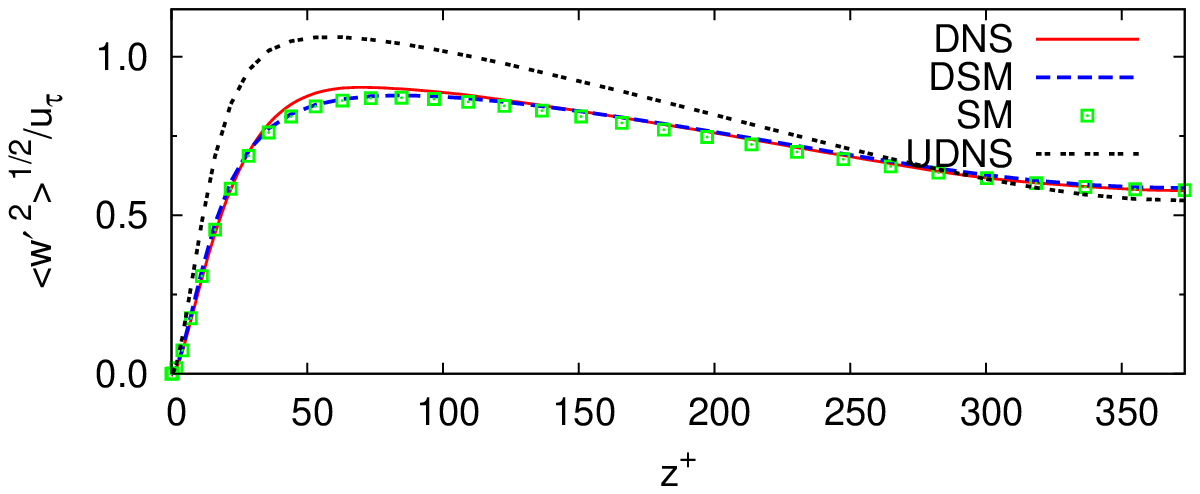}
\includegraphics[width=0.48\textwidth]{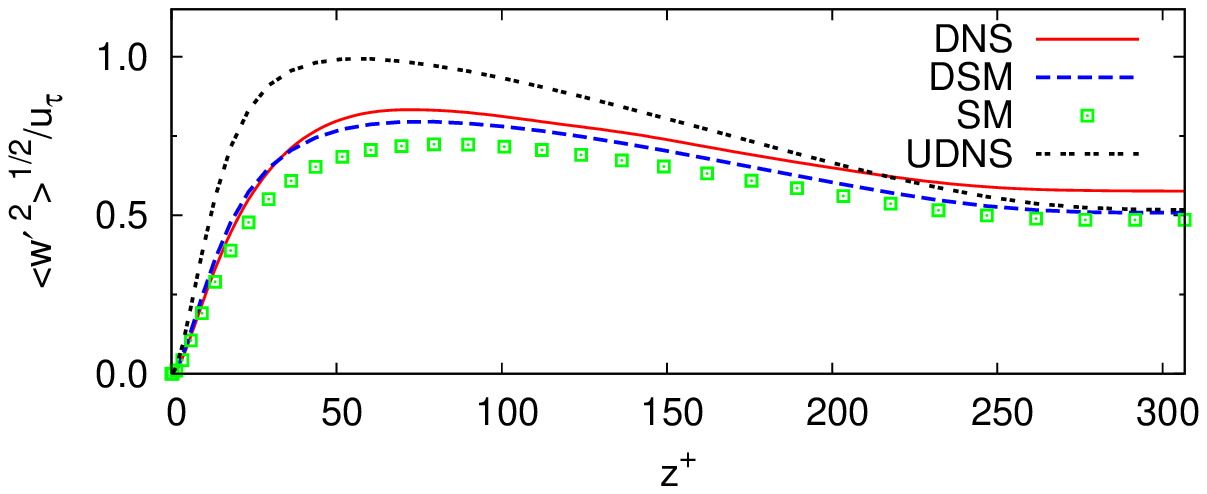}
}
\caption{Comparison between the LES and filtered DNS results.
Horizontally and time-averaged turbulence intensities in wall
coordinates for $Re=10000$ at $Ha=10$ (left) and $Ha=30$ (right).
Root mean square of velocity fluctuations scaled by DNS wall
shear velocity (\ref{utau}) are shown. The results of unresolved
DNS are also shown (UDNS).} \label{fig3:re10_rms}
\end{figure}

\begin{figure}[t]
\centerline{
\includegraphics[width=0.48\textwidth]{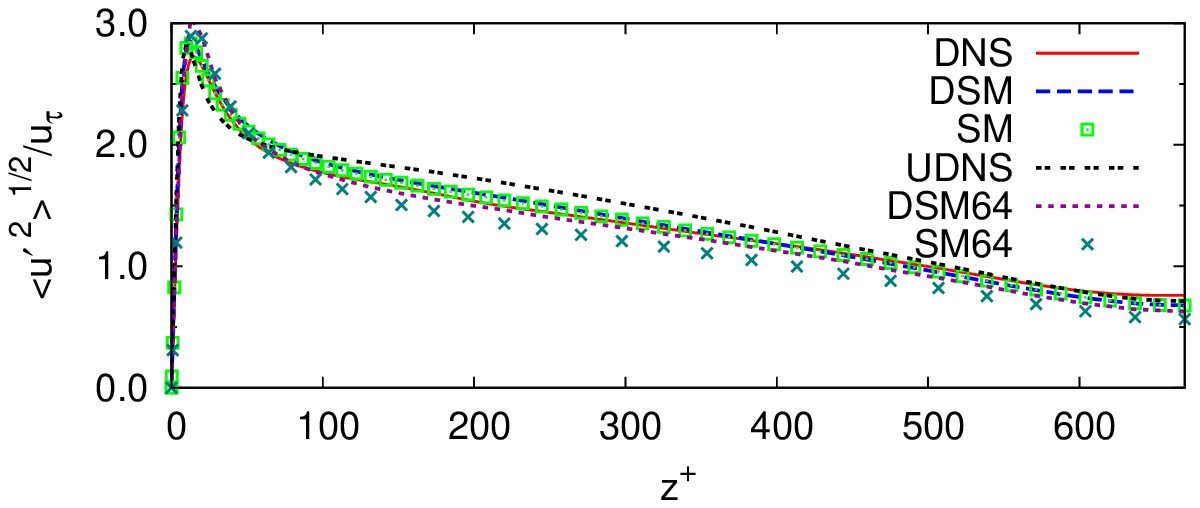}
\includegraphics[width=0.48\textwidth]{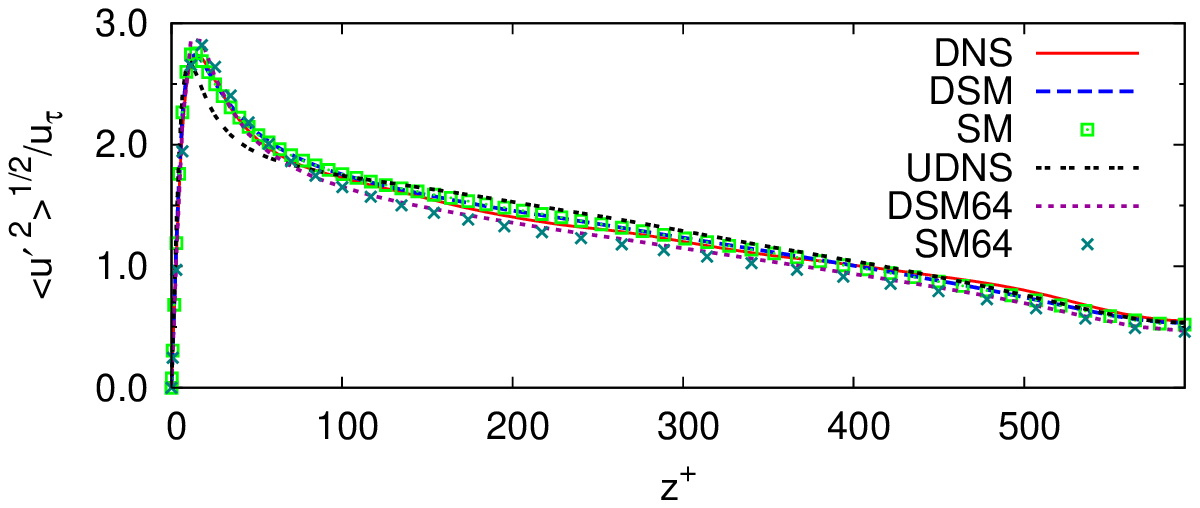}
}
\centerline{
\includegraphics[width=0.48\textwidth]{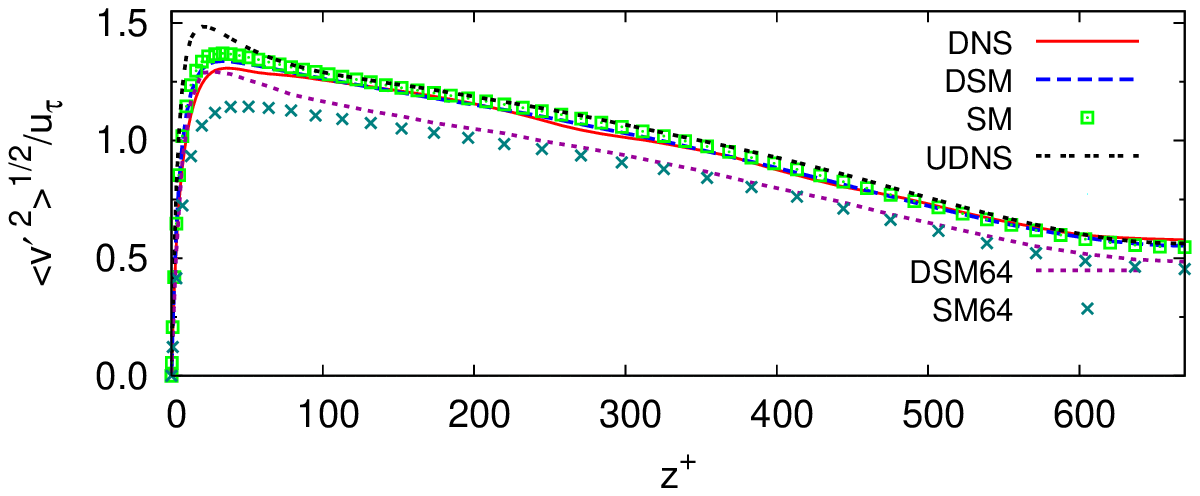}
\includegraphics[width=0.48\textwidth]{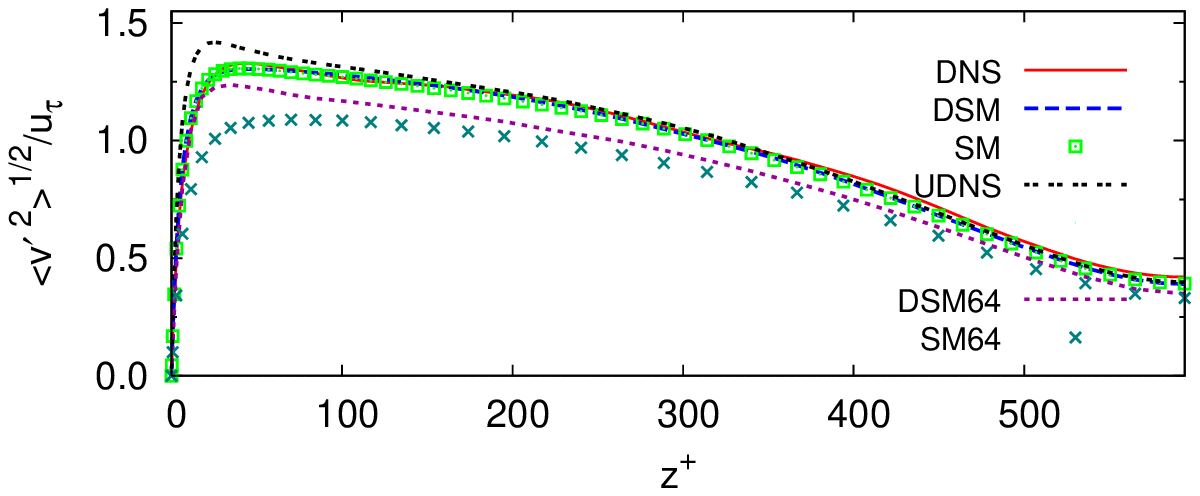}
}
\centerline{
\includegraphics[width=0.48\textwidth]{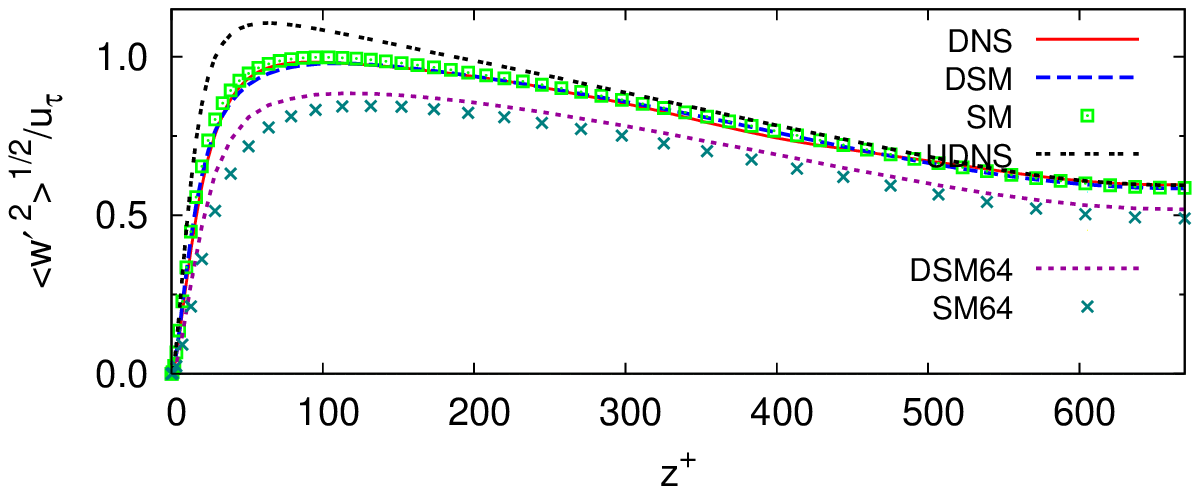}
\includegraphics[width=0.48\textwidth]{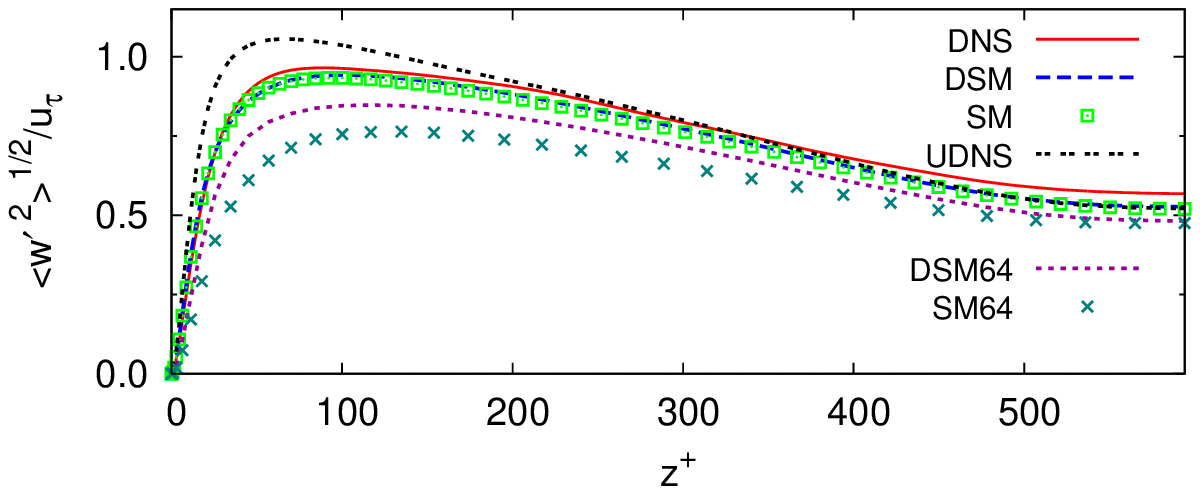}
}
\caption{Comparison between the LES and filtered DNS results.
Horizontally and time-averaged turbulence intensities in wall
coordinates for $Re=20000$ at $Ha=20$ (left) and $Ha=40$ (right).
Root mean square of velocity fluctuations scaled by DNS wall
shear velocity (\ref{utau}) are shown. The results of unresolved
DNS are also shown (UDNS).} \label{fig3:re20_rms}
\end{figure}

The typical results for the DNS -- LES comparison of the turbulence
intensities are shown in figures \ref{fig3:re10_rms} and
\ref{fig3:re20_rms}. Horizontally- and time-averaged profiles of the
rms of fluctuations of streamwise ($u^{\prime}$), spanwise
($v^{\prime}$), and wall-normal ($w^{\prime}$) velocity components are
plotted for $\Rey=10000$ at $Ha=10$ and $30$ (fig.
\ref{fig3:re10_rms}) and $\Rey=20000$ at $Ha=20$ and $40$ (fig.
\ref{fig3:re20_rms}). The scaling by the DNS wall shear velocity
is applied to all data. One can see that the intensity of the
streamwise fluctuations is well reproduced by both LES models.
The situation is different in the cases of spanwise and wall-normal
components. Here, the classical Smagorinsky model underestimates the
fluctuation intensities at high Hartmann number, while the dynamic
model remains reasonably accurate. The results for $\Rey=20000$ at
resolution $128^3$ demonstrate better agreement between the DNS and
LES without clear prevalence of any model. This situation changes
for the lower  resolution, in which case the dynamic model shows
better performance than its classical counterpart. The
under-resolved DNS gives reasonably accurate results for the
streamwise intensities but severely overestimates intensity of
fluctuations of the spanwise and wall-normal components.

\begin{figure}[t]
\centerline{
\includegraphics[width=0.48\textwidth]{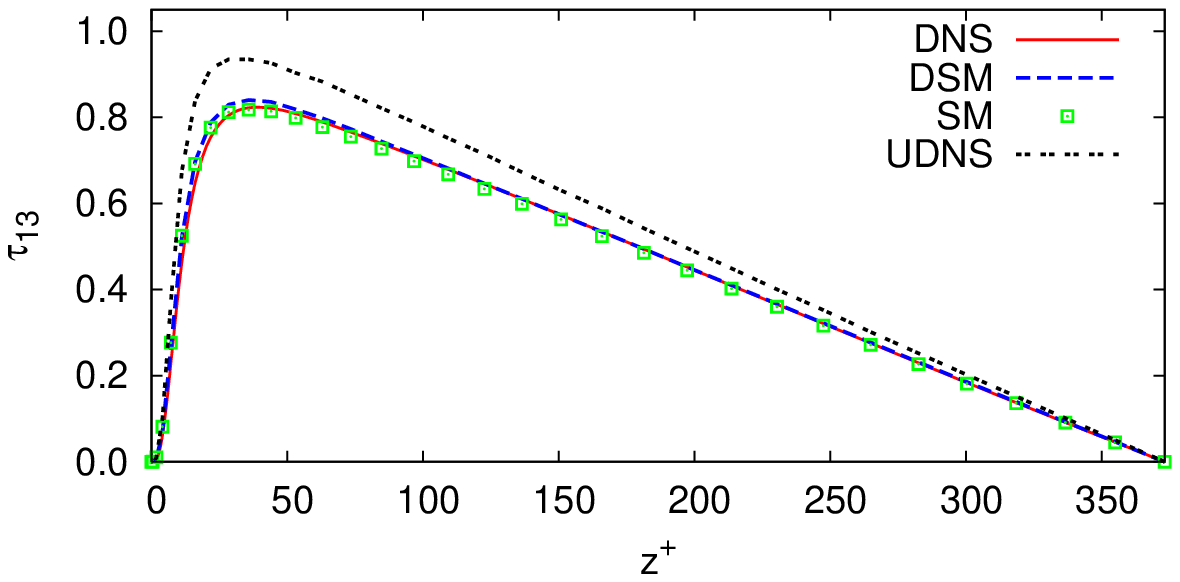}
\includegraphics[width=0.48\textwidth]{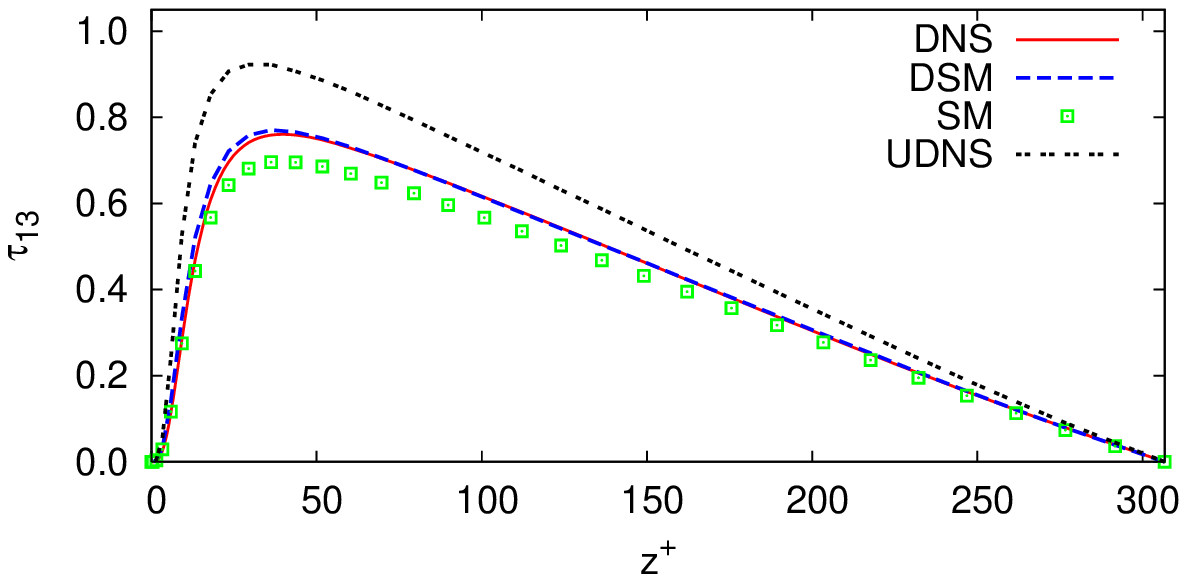}
}
\centerline{
\includegraphics[width=0.48\textwidth]{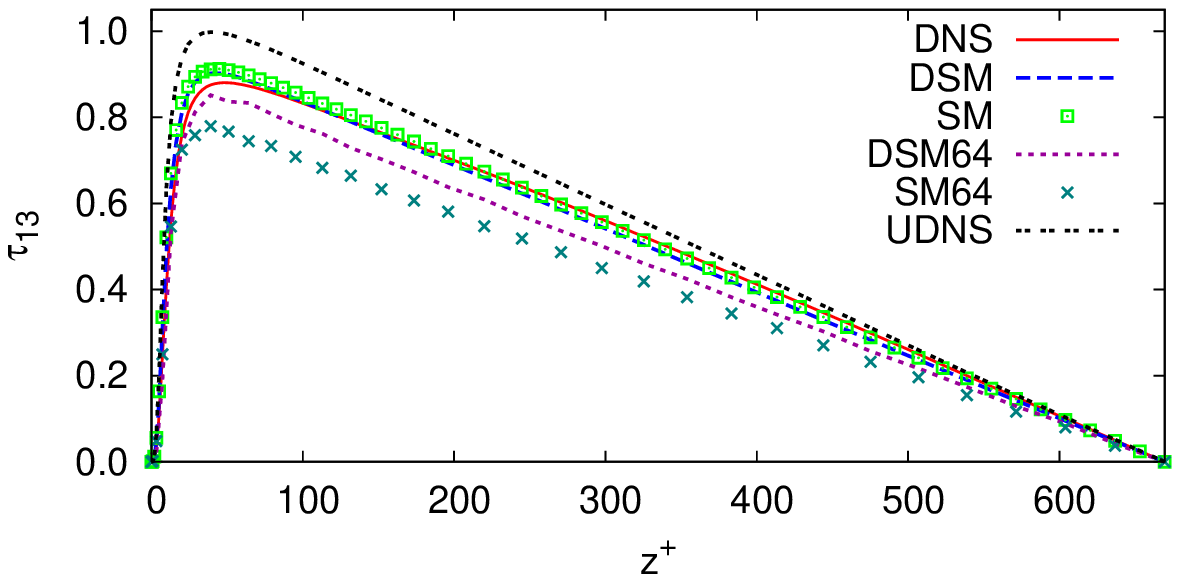}
\includegraphics[width=0.48\textwidth]{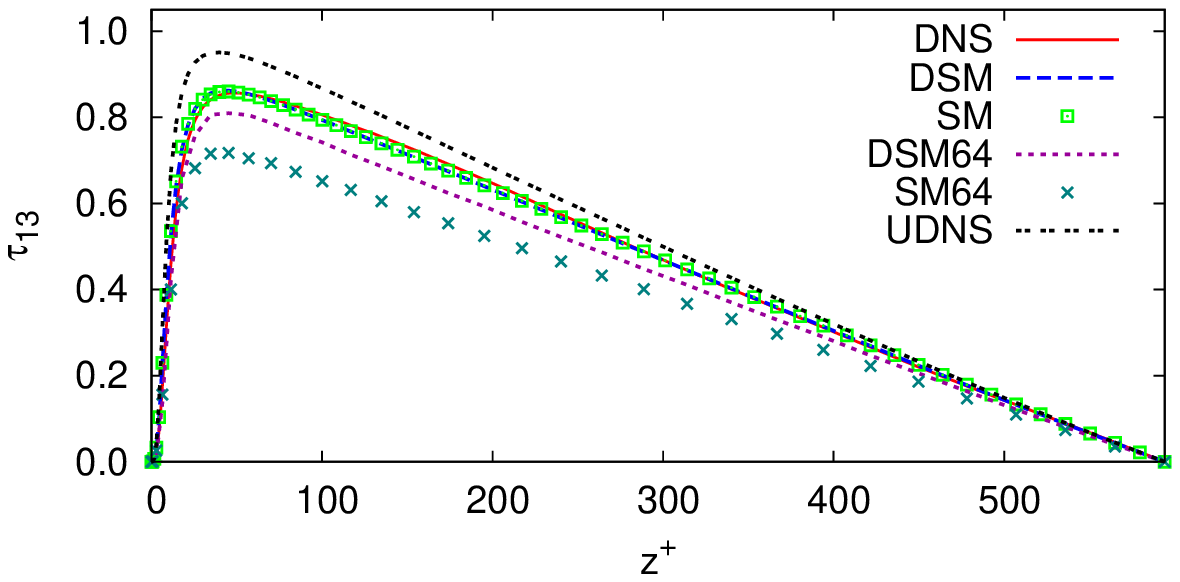}
}
\caption{Comparison between the LES and DNS results. Horizontally and
time-averaged Reynolds shear stress $\tau_{13}$ in wall coordinates
for $Re=10000$ (top) at $Ha=10$ (left) and  $Ha=30$ (right) and for
$Re=20000$ at $Ha=20$ (left) and $Ha=40$ (right). Unfiltered DNS,
UDNS, and full LES ($\tau_\textrm{\tiny resolved}$ +
$\tau_\textrm{\tiny SGS}$) stresses are shown. Normalization is by
$u^2_{\tau}$ obtained for each case from DNS.}
\label{fig3:re10_tau13}
\end{figure}

The profiles of time-averaged Reynolds shear stresses
$\tau_{13}=\langle u^{\prime}w^{\prime}\rangle$ are shown in figure
\ref{fig3:re10_tau13}.  The unfiltered velocity field is used to
compute $\tau_{13}$ in DNS and UDNS, while the sum of computed
$\tau_{resolved}$ and modeled $\tau_{SGS}$ components represents the
LES solution.  The data for $\Rey=10000$ and $Ha=10$, $30$ and for
$\Rey=20000$ and $Ha=20$, $40$ are presented. In each case, the
curves are normalized by $u^2_{\tau}$ taken from the DNS solution.
The comparison between the DNS and LES curves shows higher accuracy
of the dynamic model, which becomes obvious when the data for higher
Hartmann numbers or for lower numerical resolution (at $\Rey=20000$)
are considered.  It is clear that the conventional Smagorinsky model
underestimates the shear stress,  whereas UDNS yields significant
over-estimation.

\begin{figure}[t]
\centerline{
\includegraphics[width=0.48\textwidth]{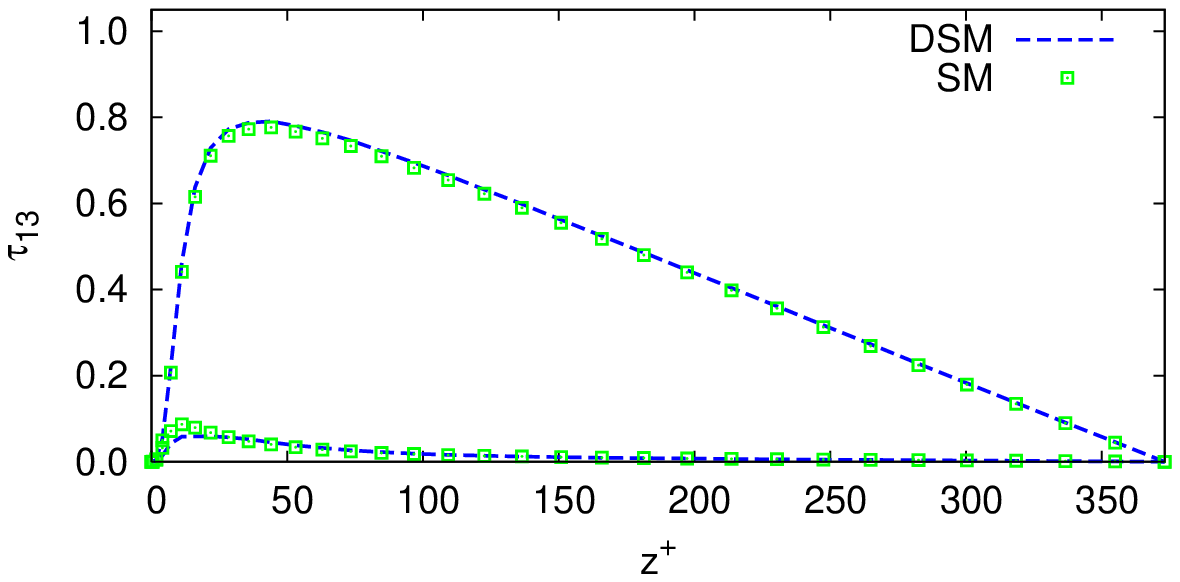}
\includegraphics[width=0.48\textwidth]{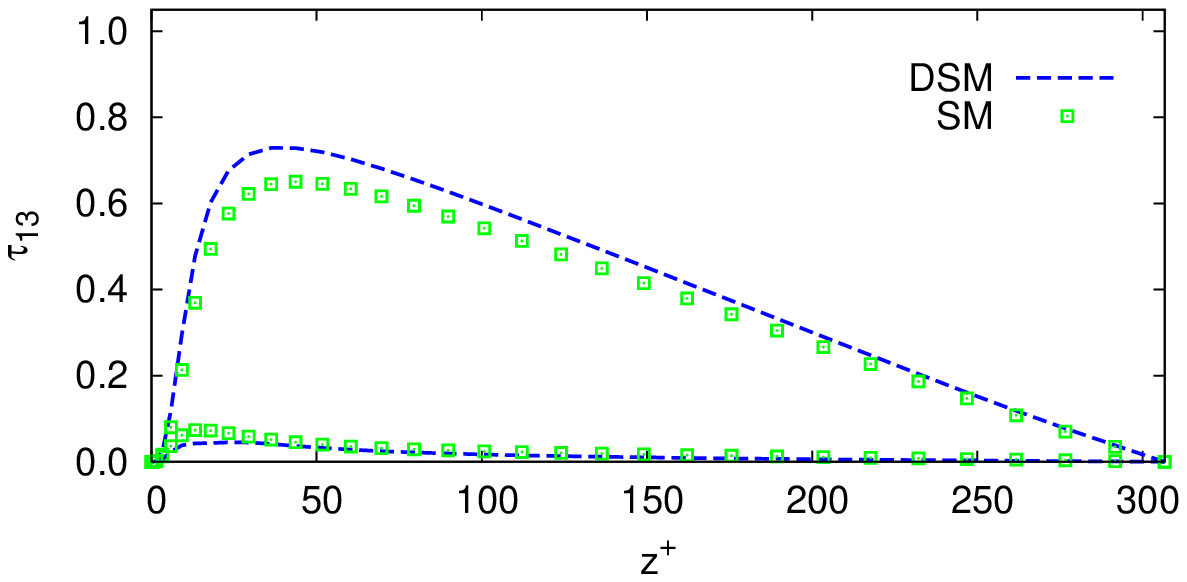}
}
\centerline{
\includegraphics[width=0.48\textwidth]{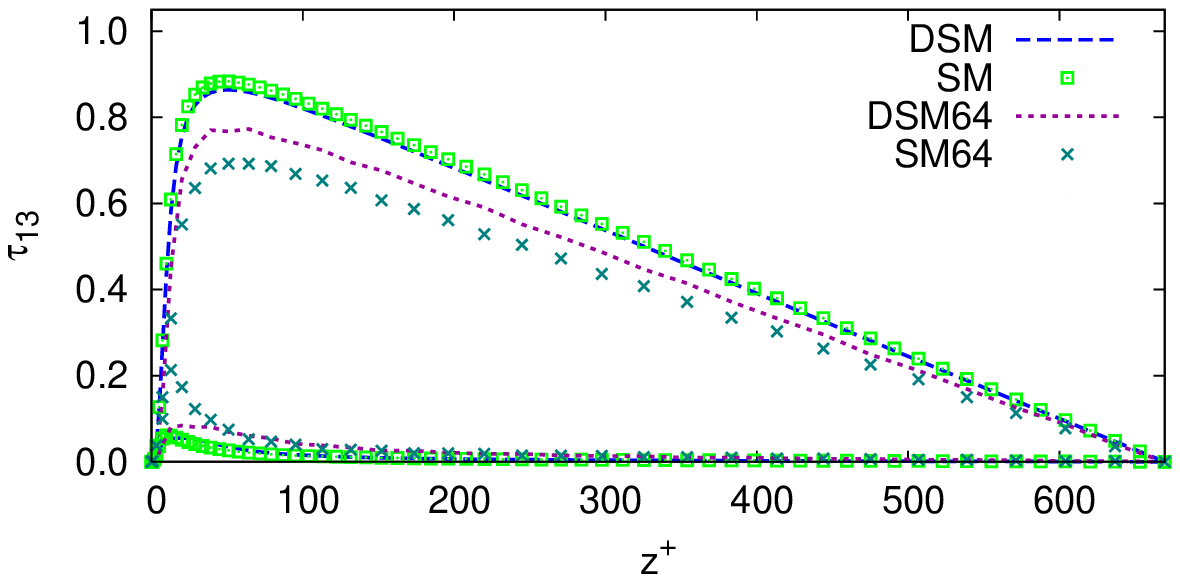}
\includegraphics[width=0.48\textwidth]{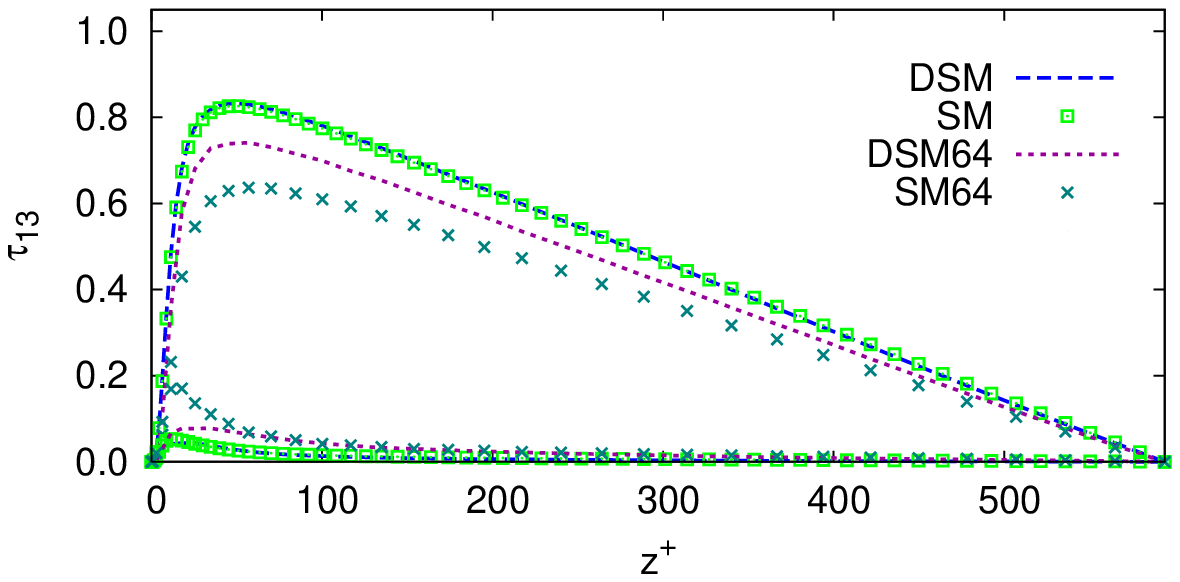}
} \caption{Separate contributions of the resolved
$\tau_\textrm{\tiny resolved}$ (upper curves) and subgrid
$\tau_\textrm{\tiny SGS}$ (lower curves) stresses in LES
simulations. Horizontally and time-averaged Reynolds shear stress
$\tau_{13}$ in wall coordinates for $Re=10000$ (top) at $Ha=10$
(left) and  $Ha=30$ (right) and for $Re=20000$ at $Ha=20$ (left) and
$Ha=40$ (right). Normalization is by $u^2_{\tau}$ obtained for each
case from DNS.} \label{fig3:re10_tau13a}
\end{figure}

Figure \ref{fig3:re10_tau13a} shows separate contributions from the
resolved $\tau_{resolved}$ and the modelled $\tau_{SGS}$ stresses.
Apart from presenting the relative roles played by the resolved and
modeled parts of the turbulent momentum transfer, the curves
illustrate difference between the dynamic and classical Smagorinsky
models. The two models generate close results in the high-resolution
LES. In the case of the low-resolution LES, the contribution of the
modeled stresses is higher for the classical model. Correspondingly,
the resolved stresses are lower than for the dynamic model.
Considering the effect of the magnetic field, it is interesting
to note that stronger field does not result in preferential
suppression of the SGS stress component. Both components are
visibly reduced (this phenomenon is discussed in the following
section).

We can now formulate conclusions regarding the performance of the
classical and dynamic Smagorinsky models. The models perform
significantly better than simple underresolved DNS. Both models are
reasonably accurate in reproducing the mean flow.  The same is true
for the intensities of turbulent fluctuations, although the
classical model shows a tendency toward underestimation of the
normal and spanwise components at high Hartmann numbers. The real
difference between the models becomes visible when we consider the
characteristics of the momentum transfer in the wall-normal
direction, such as the Reynolds shear stress $\tau_{13}$, wall shear
Reynolds number $\Rey_{\tau}$, or the friction coefficient $C_f$.
For all these characteristics, the dynamic model consistently
generates significantly more accurate results.

\section{Flow transformation caused by the magnetic field}
\label{sec:transform} This section presents the results of a
systematic study of the effects of the spanwise magnetic field on
the flow properties. Since the accuracy of the dynamic Smagorinsky
model has been proven for the set of parameters under consideration
in the previous section, the data from DNS and dynamic LES will be
used interchangeably. Table \ref{table:param2} summarizes the
parameters and computed integral characteristics of all numerical
experiments.

\subsection{Integral characteristics}
As already noted by Lee \& Choi\cite{Lee:Choi:2001} and confirmed by
the results of our computations presented below, the spanwise
magnetic field tends to suppress the turbulent fluctuations and
turbulent momentum transfer in the wall-normal direction. The
integral effect is a reduced friction drag. This also leads to a
smaller slope of the velocity profile near the wall. The centerline
velocity should increase to maintain the constant volumetric flux.

\begin{figure}[t]
\scriptsize{
\parbox{0.22\linewidth}{(a)}\parbox{0.76\linewidth}{(b)}
\centerline{
\includegraphics[width=0.49\textwidth]{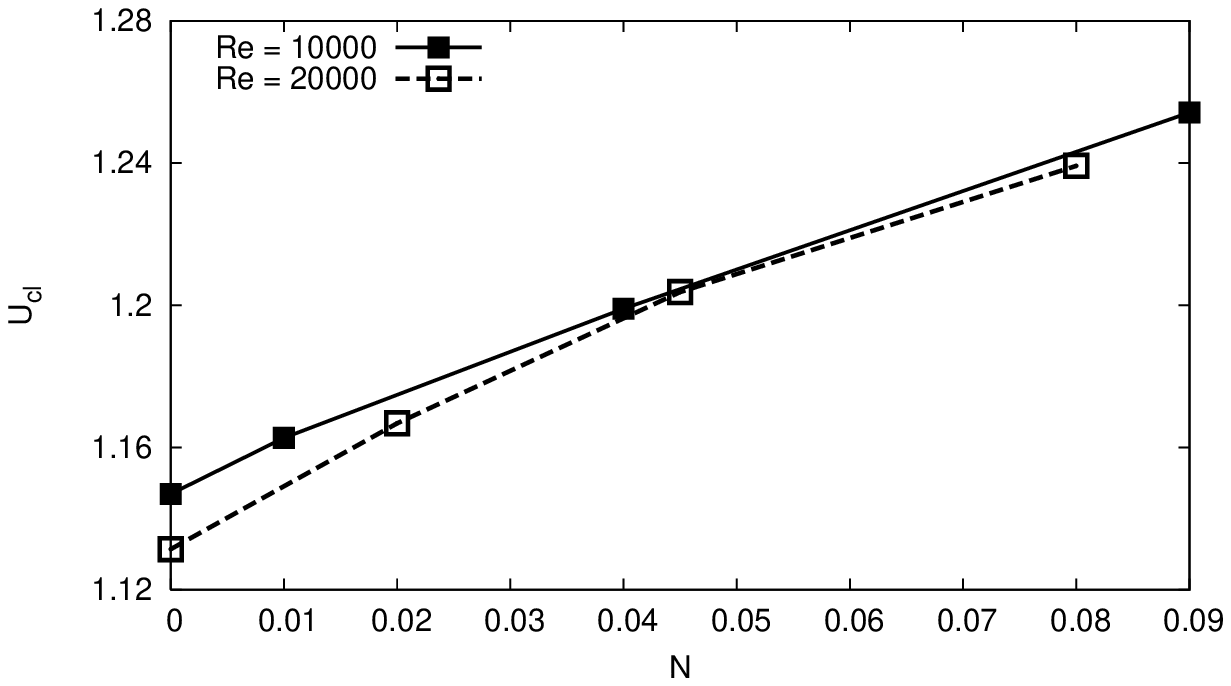}
\includegraphics[width=0.49\textwidth]{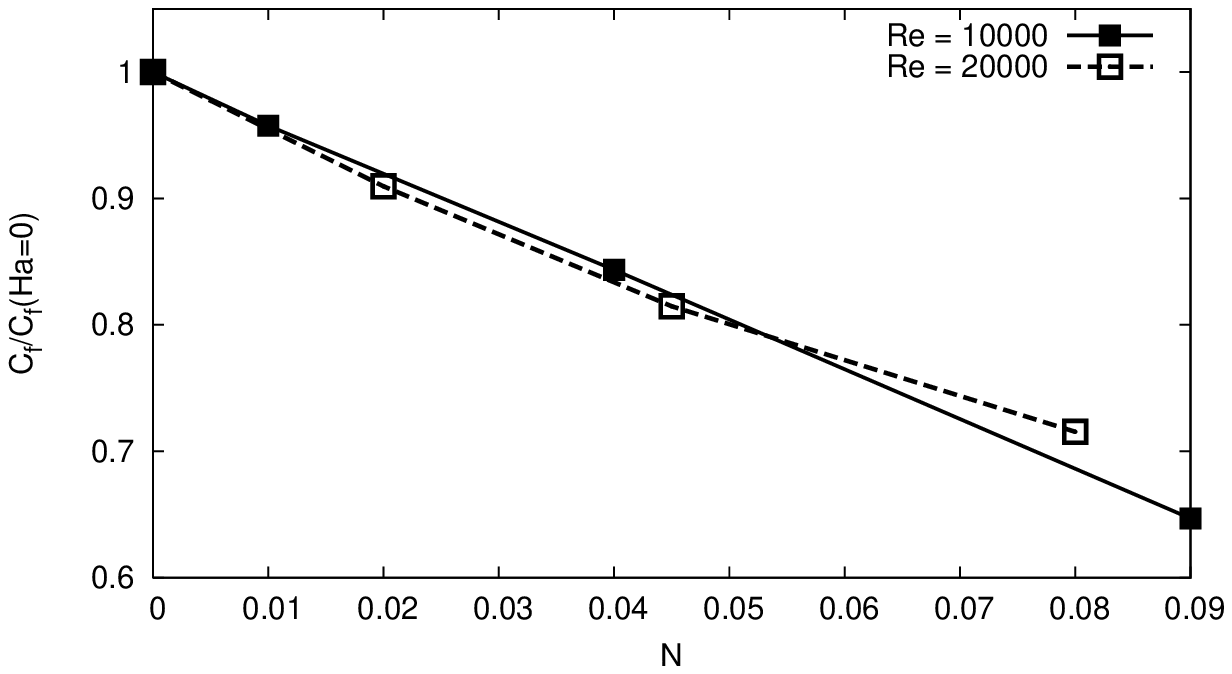}
} } \caption{Effect of the magnetic field on the flow properties:
centerline velocity $U_{cl}$ \emph{(a)} and relative friction
coefficient $C_f/C_f(Ha = 0)$ \emph{(b)} versus magnetic interaction
parameter $N = Ha^2/Re$. Results of DNS for $Re=10000$ and
$Re=20000$.} \label{fig4:ucl_cf}
\end{figure}

While all this can be inferred from table II, it is nonetheless
instructive to present this information graphically. Figs.
\ref{fig4:ucl_cf}(a,b) show the centerline velocity and the
relative friction
coefficient $C_f/C_f(Ha = 0)$  as functions of the magnetic
interaction parameter $N$ for the two different Reynolds numbers. We
see that the centerline velocity increases monotonously with $N$ and
that the data sets for the different Reynolds numbers are fairly
close.

The same observation applies for the normalized friction
coefficient, which decreases almost linearly with $N$. The friction
coefficient is reduced about 30 percent before the sustained
turbulence is replaced by intermittent dynamics.

We finally note that by comparison with the laminar state the drag
reduction is rather modest. The laminar friction coefficient is
$C_f=0.00090$ for $Re=10000$ and $C_f=0.00045$ for $Re=20000$, i.e.
it is almost an order of magnitude smaller than for the turbulent
states obtained in our simulations.

\subsection{Mean profiles}

Figure \ref{fig4:re20_prof} shows the profiles of time-averaged
mean velocity $U=\langle u \rangle$ as obtained in DNS. In figures
\ref{fig4:re20_prof}(a,b), the global coordinates are used and the
velocity is scaled with the flux velocity $U_q$. One can see that
the mean flow profile changes significantly as the magnetic field
becomes stronger (the Hartmann number grows). The transformation
resembles a `transition' towards the laminar profile, with the
centerline velocity growing and the gradient near the wall decreasing.
As shown below, this does not mean actual laminarization -- the flow
remains fully turbulent. Similar behavior was observed in the duct
flow experiment of Ref. \onlinecite{Votsish:Kolesnikov:1977}.

In general, there is no reason why the logarithmic layer behavior
should not be observed in the MHD channel flow with spanwise
magnetic field. The Lorentz force does not directly appear in the
equation for wall-normal transport of mean momentum. The dimensional
arguments leading to the log-layer solution should remain valid
provided, of course, the flow retains the pattern of inner, outer,
and overlap sub-layers. On the other hand, the magnetic field
affects the solution indirectly, by transforming the turbulent
fluctuations and, thus, the turbulent momentum transport by
$\tau_{13}$.

Our attempt to identify the log-layer behavior is illustrated in
figure \ref{fig4:re20_prof}. The DNS mean velocity profiles are
plotted in wall units in figures \ref{fig4:re20_prof}(c,d). It is
difficult to make a definite conclusion but one may convince
oneself upon observation that the profiles contain intervals of
nearly logarithmic behavior.  This conclusion would be wrong as
illustrated in figures \ref{fig4:re20_prof}(e,f), where the
 profiles of compensated velocity gradient $\gamma\equiv z^+du^+/dz^+$ are shown.
 Such profiles were used in the past (for example, in Ref.
\onlinecite{Moser:1999}) to assess the agreement between the
computed data and the log-law, according to which the inverse von
K\'arm\'an  constant $\kappa$ equals $\gamma$.  One can see in figure
\ref{fig4:re20_prof} that the logarithmic layer is absent in the
flows affected by the magnetic field.  This is observed for all
non-zero Hartmann numbers considered and for both  values of the
Reynolds number.  We also tried to identify possible power law
behavior $u^+\sim (z^+)^n$ suggested in Ref.
\onlinecite{Barenblatt:1997}.  After considering the compensated
profiles $\beta\equiv z^+(u^+)^{-1}du^+/dz^+$, we found a
situation similar to that of the log-layer.  $\beta$ was nearly
constant in an extended range of $z^+$ for non-magnetic flows (see
Ref. \onlinecite{Moser:1999} for similar results) but not for the
flows in the presence of  the magnetic field.

The effect of the magnetic field on the turbulent velocity
fluctuations is illustrated in Figure \ref{fig4:re20_rms}.
Time-averaged profiles of root mean square fluctuations for each
velocity component and of full (resolved plus SGS) turbulent shear
stress $\tau_{13}$ are shown as found in dynamic Smagorinsky LES.
For the purpose of comparison of the absolute magnitudes, the common
velocity scale, namely the flux velocity $U_q$ is used for
normalization. The main conclusion is in agreement with the DNS
results of Lee and Choi\cite{Lee:Choi:2001} and with the
experimental data.\cite{Votsish:Kolesnikov:1977} The magnetic field
suppresses the turbulent fluctuations. The degree of suppression
grows monotonically with the strength of the magnetic field and is
somewhat larger for the wall-normal velocity component than for the
other two. An important consequence is the substantial reduction of
the turbulent shear stress illustrated by figures
\ref{fig4:re20_rms}(g,h).  The resulting decrease of the turbulent
momentum transfer in the wall-normal direction is, obviously, a
reason for the transformation of the mean flow profile shown in
figure \ref{fig4:re20_prof} and for the drag reduction.
Interestingly, the suppression of turbulent fluctuations is not
accompanied by any noticeable change of the global spatial structure
of the flow.  In particular, the dimension of the viscous sublayer
and locations of the maxima of the fluctuation energy and turbulent
stress are unaffected by the magnetic field.

\subsection{Coherent structures}
In this part of the paper, we analyze the effect of the spanwise
magnetic field on the flow structures. The study of such structures
in the homogeneous MHD turbulence\cite{Zikanov:1998, Vorobev:2005,
Knaepen1:2004, Kassinos:Knaepen:Wray:2006} proved quite fruitful. It
was found that a sufficiently strong magnetic field transforms the
velocity and vorticity  field in a unique and clearly visible
way.  The main feature of the transformation is growth of the
typical size of the coherent structures, strong in the direction of
the magnetic field and weaker, but noticeable in the transverse
directions. In general, the structures become larger, slower, and
less intense.

The situation in the channel flow is more complex, in particular,
due to the obvious difference between the core flow and the boundary
layers. In the core, where the mean shear is weak, a transformation
similar to that in homogeneous turbulence can be expected.  On the
contrary, in the boundary layers, the evolution of the coherent
structures is dominated by the mean shear and the effect of the
magnetic field can be quite different.  The results of our
calculations, illustrated in figure \ref{fig7:re10_lam2}, largely
meet these expectations.  To identify the coherent structures, we
analyzed the fields of streamwise vorticity $(\bnabla \times
\textbf{u})_x$, fluctuations of streamwise velocity $u^{\prime}$,
and the intermediate eigenvalue $\lambda_2$ of the tensor
$S_{ik}S_{kj}+\Omega_{ik}\Omega_{kj}$, where $S_{ik}\equiv
(v_{i,k}+v_{k,i})/2$ and $\Omega_{ik}\equiv (v_{i,k}-v_{k,i})/2$.

The contour plots of the streamwise vorticity component in the plane
perpendicular to the direction of the channel are shown in figures
\ref{fig7:re10_lam2}(a,b).  The vorticity field is scaled by its
root mean square value in each case. Near the walls, the regions of
localized strong vorticity become somewhat larger in the presence of
the magnetic field.  Similar behavior was observed by Lee and
Choi,\cite{Lee:Choi:2001} although in their case, the transformation
was more pronounced due to lower Reynolds number and numerical
resolution.  The effect of the magnetic field is stronger in the
middle part of the channel, where the structures seem to be
suppressed to a larger degree than near the walls.  In general, the
vorticity magnitude decreases as can be seen from comparison of the
rms values $\langle \omega_x^2 \rangle^{1/2}=1.643$ at $Ha=0$ and
$1.022$ at $Ha=30$.

The connected regions of negative $\lambda_2$ eigenvalue are often
considered as indicators of coherent vortical
structures\cite{Jeong:Hussein:1995} especially in the regions of
strong mean shear. Our analysis of the $\lambda_2$ fields led to the
same conclusions as based on the vorticity field.  As an
illustration, the figures \ref{fig7:re10_lam2}(c,d) show the
isosurfaces $\lambda_2=-0.1\lambda_{2,rms}$ in the middle part of
the channel. One can see that the structures decrease in number,
increase in size, and, seemingly, become elongated in the direction
of the magnetic field.  The latter observation will be confirmed
below by quantitative characteristics of the flow anisotropy.

The effect of the magnetic field on near-wall streaks is illustrated
in figures \ref{fig7:re10_lam2}(e,f), where the contours of
streamwise velocity perturbations at $z=0.95$ are shown.  The fields
are normalized by their rms values averaged over the horizontal
plane. This should not obscure the fact that, as illustrated in
figure \ref{fig4:re20_rms}a, the amplitude of the streaks is lower
at $Ha=30$.  As can be seen in figures \ref{fig7:re10_lam2}(e,f),
they also become more stable (less susceptible to small-scale
perturbations) and wider in the spanwise direction. These
observations are also in agreement with the earlier results by Lee
and Choi.\cite{Lee:Choi:2001}

\subsection{Anisotropy}
Discussing the flow anisotropy we should clearly separate the following two
types: the anisotropy of dimensionality referring to variation of
the typical length scale with direction, and the anisotropy of
componentality, which means imparity between the velocity components
and can be viewed as anisotropy of the Reynolds stress tensor (see
Ref. \onlinecite{Kassinos:2001} for a discussion of terminology).
Both types are present in the channel flow.  The focus of our study
is, however, on the dimensionality anisotropy, since only this
type is affected by the magnetic field directly.

The question of anisotropy in the combined presence of mean shear
and magnetic field is non-trivial.  We deal with two competing
effects, each leading to establishment of its own anisotropy.  The
mean shear promotes elongation and alignment of flow structures in
the streamwise direction, while the magnetic field produces similar
(in  effect but not in mechanism) action along the magnetic field
lines, i.e. in the spanwise direction.  The situation was considered
earlier for the case of homogeneous
turbulence.\cite{Kassinos:Knaepen:Wray:2006}  The DNS computations
in a periodic box confirmed the predictions based on an analysis of
the typical time scales.  The two relevant scales are the typical
shear time $\tau_S=S^{-1}$, where $S$ is the mean shear rate, and
the Joule damping time $\tau_J=\rho/\sigma B^2$.  It was
found\cite{Kassinos:Knaepen:Wray:2006} that at low magnetic Reynolds
number the type of the anisotropy is determined by the ratio
$M\equiv \tau_S/\tau_J=\rho^{-1}B^2\sigma S^{-1}$.  The flow
structures elongate with the mean shear at $M\ll 1$ and with the
magnetic field at $M\gg 1$.

We evaluated the parameter $M$ as a function of the wall-normal
coordinate $z$ using the local magnitude of the derivative of the
mean velocity $U$.  In our units, the parameter is defined as
$M=N\left(d U/dz \right)^{-1}$.  It was found that the
condition $M\ll 1$ is satisfied in a larger part of the channel in
all our experiments.  At the highest interaction parameter $N=0.09$
corresponding to $\Rey=10000$ and $Ha=30$, $M$ was below 0.1, 0.25,
and 0.5 at $|z|\gtrsim 0.85$, 0.5, and 0.1, respectively.  This
means that the anisotropy properties are predominantly controlled by
the mean shear with exception of the region in the middle of the
channel.  The visualizations of the flow structures presented above
provide certain support to this conclusion.  A more quantitative
assessment is
provided in what follows.

The typical results for the two-point correlations are presented in
figure \ref{fig7:re10_corr}. The correlations were calculated for
the streamwise velocity component as
\begin{equation}
\label{eq:corr} C=\frac{\langle
u(\textbf{x}+\textbf{r})u(\textbf{x}) \rangle}{\langle
u(\textbf{x})^2\rangle},
\end{equation}
where $\textbf{x}$ stands for the physical coordinates and
$\textbf{r} = (r_x,r_y,0)$
is the horizontal
displacement vector, and the averaging is done over a horizontal
plane.  No time averaging is applied to the correlation
coefficients.  The results for the flow without magnetic field and
for the flow with the highest magnetic interaction parameter
$N=0.09$ are shown in figures \ref{fig7:re10_corr}(a,c) and
\ref{fig7:re10_corr}(b,d), respectively.  In the middle of the
channel (see figures \ref{fig7:re10_corr}(a,b)), the effect of the
magnetic field is to extend the correlations in the spanwise
direction. The effect of the magnetic field is not very strong and
concerns primarily the weak correlation ranges. In the area of
strong mean shear near the walls (see figures
\ref{fig7:re10_corr}(c,d)), the effect of the magnetic field is also
visible.  The degree of correlation increases in the streamwise and
spanwise directions. This is in agreement with the conclusion made
on the basis of visualization of the same flow fields in figures
\ref{fig7:re10_lam2}(e,f), and which will be further confirmed
below. It can also be noticed that the ratio of the typical
correlation lengths in the two directions is not affected by the
magnetic field.

The anisotropy of dimensionality can be viewed as anisotropy of the
velocity gradients.  This viewpoint seems particularly attractive in
the case of the low-$\Rey_m$ MHD turbulence, where the mechanism of
anisotropy generation is the preferential suppression of flow modes
with strong gradients along the magnetic field lines.  A natural way
to evaluate such anisotropy is to use the ratios of the mean square
velocity gradients
\begin{equation}
\label{eq:Gij} G_{ij}=\frac{\langle \left( \partial v_i/\partial y
\right)^2 \rangle \left(1+\delta_{i2}\right)}{\langle \left(
\partial v_i/\partial x_j \right)^2 \rangle
\left(1+\delta_{ij}\right)}.
\end{equation}
Here $y$ is the coordinate in the direction of the
magnetic field. The factors in the numerator and denominator are
introduced so that each coefficient $G_{ij}$ varies between 0 in a
purely two-dimensional flow, which is uniform in the $y$-direction
and 1 in a perfectly isotropic flow.

The coefficients (\ref{eq:Gij}) were successfully used in the
numerical studies of turbulence in a periodic box.\cite{Schumann:1976,
Zikanov:1998, Vorobev:2005} One of the reasons is that in such flows
the degree of anisotropy is nearly scale-independent in a wide range
of intermediate and small scales.\cite{Vorobev:2005} The dominant
contribution by the intermediate scales makes the coefficients
(\ref{eq:Gij}) convenient and fairly accurate measures of anisotropy
in this range,\cite{Vorobev:2005} which, importantly for the LES
modeling, includes the typical scales of filtering. This property
was employed in Ref. \onlinecite{Vorobev:2007}, where a simple
anisotropy correction of the classical Smagorinsky model based on
computed values of $G_{ij}$ was proposed and verified.

The channel flow is a more complex and less straightforward subject
of the analysis in terms of $G_{ij}$ than the flow in a periodic
cube.  The major complicating factor is the presence of the mean
shear that creates the anisotropy of a different orientation.  It
also leads to spatial non-uniformity. The situation in the core flow
is clearly different from that near the walls.  Furthermore, one may
expect  the behavior of the coefficients to vary depending on the
choice of the velocity component $v_i$. Keeping these difficulties
in mind, we conduct the analysis of the $G_{ij}$ coefficients in the
rest of this section.

The coefficients are calculated as functions of $z$ using horizontal
and time-averaging  as a substitute to the volume averaging employed
in the periodic box flows. The values of the mean square of velocity
gradients $\langle\left( \partial v_i/\partial x_j \right)^2\rangle$
obtained in DNS and LES differ significantly from each other, which
indicates substantial contribution of the subgrid scales. The same
effect was illustrated in table \ref{table:param2} on the example of
the dissipation rate. The agreement between the LES and filtered DNS
is quite good.  Unlike the case of periodic box flows, the ratios of
the gradients, i.e. the coefficients (\ref{eq:Gij}) are also
noticeably different between DNS and LES, especially for the
coefficients that include $x$-derivatives.  This can be attributed
to the scale-dependent character of the anisotropy generated by mean
shear. Only the DNS data are presented and discussed below.

The results for $\Rey=10000$ are shown in figure
\ref{fig8:re10_gij}. Similar behavior was detected in the runs with
$\Rey=20000$. The ratios of the spanwise and normal derivatives are
plotted in figures \ref{fig8:re10_gij}(a-c) (the top row).  One can
clearly see the effect of the magnetic field in the middle of the
channel.  The perturbations are nearly isotropic in the $y-z$ plane
at $Ha=0$. As the Hartmann number increases, the growing suppression
of the spanwise derivatives by the magnetic field results in
consistent decrease of the coefficients. The effect of the magnetic
field becomes negligible near the walls, where the behavior of all
coefficients is nearly identically dominated by the growing normal
derivatives.

The streamwise and normal derivatives, ratios of which are plotted
in figures \ref{fig8:re10_gij}(d-f) (the middle row), show
anisotropy created by the mean shear at $Ha=0$.  According to the
commonly known picture, the flow structures are characterized by
weaker gradients (slower variation) in the streamwise than in the
spanwise direction. It is interesting that the magnetic field
enhances this anisotropy.  The effect of growing Hartmann number is
obvious and unambiguous.  It is also quite remarkable since neither
streamwise nor normal derivatives are directly affected by the
magnetic field.

The last row of figure \ref{fig8:re10_gij} shows the ratios of
streamwise and spanwise derivatives.  Here, again, we see the
coefficients decreasing near the wall, which is a consequence of the
dominance of streaky structures.  Regarding the effect of the
magnetic field, one could expect that, due to direct suppression of
the spanwise derivatives, the coefficients would grow with the
Hartmann number.  The figures \ref{fig8:re10_gij}(g-i) prove such
expectations wrong.  First, the coefficients do not change
significantly at all.  Second, they, in fact, decrease with $Ha$.
This occurs for all three velocity components throughout the channel
except in the very central part of it for the coefficients based on
$u$ and $w$.  We see that the magnetic suppression of spanwise
derivatives is accompanied by stronger reduction of the streamwise
ones.

An alternative measure of the deviations from local isotropy (or of
anisotropy) is the skewness of the transverse derivative of the
streamwise velocity fluctuations which is defined as
\begin{equation}
S_3(z)=\frac{\langle(\partial u/\partial z)^3\rangle_{}}
{\langle(\partial u/\partial z)^2\rangle_{}^{3/2}}\,,
\label{sz}
\end{equation}
where the statistical averages are taken in $x,y$-planes at fixed
height $z$. This ratio has been succesfully applied to measure
the return of shear flow turbulence to a state of local isotropy
in the presence of a large-scale source of anisotropy, the sustained
mean flow. \cite{Warhaft:2002,Gualtieri:2002,Schumacher:2003}
In a perfectly isotropic flow, all odd-order moments have to be
zero. Note that a non-vanishing skewness profile has to reverse
sign at the midplane for symmetry reasons in our case.

A positive derivative skewness over the half-channel ($z\in [-1,0]$)
is observed in Fig. \ref{fig8:re10_skew} for all four cases. The
profiles indicate that positive gradients are more likely than
negative ones. In other words, the streamwise velocity fluctuations
increase preferentially with growing $z$. It can be observed
that magnitude in both non-magnetic cases decreases slightly
with increasing Reynolds number, thus indicating that a return
to local isotropy which states that a growing range of small
scales is less affected by the presence of the global shear.
The rapid increase of all profiles close to the wall is
a fingerprint of the streamwise streaky structures that are
formed. The figures also display that a less rapid increase
of the skewness profile is in line with less rapid decrease
of the local mean shear or the thickening of the viscous layer.
All profiles decrease then to zero at the midplane.

The strongest suppression of the turbulent fluctuations and
the preferential alignment with the mean flow becomes clearly
visible in the case of $Re=20000$ and $Ha=40$ in comparison to
both, the $Ha=0$ case at the same Reynolds number and the $Ha=30$
case at the lower Reynolds number of 10000. The findings are
consistent with the more pronounced streamwise structures which
are displayed in Fig. \ref{fig7:re10_lam2}(e) compared to (f).
We can conclude that the drag reduction goes in line with a larger
amplitude of the derivative skewness. The spanwise magnetic field
reduces the degree of local isotropy which is also consistent with
the trends of the profiles in Fig. \ref{fig8:re10_gij}.

We now can summarize the results of our study of the dimensionality
anisotropy in the MHD channel flow.  The picture of strong
anisotropy induced by the magnetic field in simulations of
homogeneous (periodic box) turbulence\cite{Schumann:1976,
Zikanov:1998, Knaepen1:2004, Vorobev:2005} cannot be directly
translated to the channel case. The anisotropy properties remain
predominantly determined by the mean shear, especially in the wall
regions.  The main effect of the magnetic field is to reduce the
velocity gradients in both horizontal directions.  We can view this
as another manifestation of the phenomenon already observed in the
visualization of the coherent structures, namely growth of typical
spanwise and streamwise length scales.

\section{Comparison with drag reduction by stratification and polymer additives}
\label{sec:drag}

The results in Table \ref{table:param2} show that the spanwise
magnetic field results in noticeable reduction of the friction
drag.
It is interesting to relate the present mechanism of drag reduction
to other known mechanisms. In this section, we do so for  two
cases: channel flow in the presence of stable stratification and
channel flow with minute amounts of long flexible polymer chains added.
In particular, we seek to identify the similarities and differences
in the momentum flux balance for both types of flows and to
rationalize the resulting differences in the mean profiles.

The turbulent channel with stable density stratification in the
wall-normal direction was a subject of a detailed LES
study\cite{Armenio:2002}.   The strong effect on the momentum
transfer and the mean flow was observed, which, albeit caused by a
different mechanism, is  similar to the effect of the magnetic field
detected in the present study.  In particular, increase of the
Richardson number led to growing suppression of turbulent
fluctuations and reduction of wall-normal turbulent momentum flux
$\tau_{13}$.  The transformation of the mean velocity profile
visually similar to that shown in our figures
\ref{fig4:re20_prof}a,b was found.  In contrast to  the case with
the magnetic field, the stratification was reported to retain the
logarithmic layer behavior. On the other hand, the details of the
transformation of the log-layer, such as increased slope, reduced
intercept, and widened zone of the core flow, were consistent with
our observations.

The phenomenon of polymer drag reduction has been observed
in the late 1940ies by Toms\cite{Toms:1949} and studied well in
experiments\cite{Virk:1975,Warholic:1999} and
DNS.\cite{Sureshkumar:1997,Dimitropoulos:2005,Peters:2007} Dilute
polymer solutions are modeled as two-component fluids, where
stresses of the Newtonian solvent and polymer component are
additive. The simplest additional macroscopic stress field is given
by
\begin{equation}
\tau_{ij}^{(p)}({\bf x},t)=\frac{\eta_p}{\lambda} (f
C_{ij}-\delta_{ij})\,.
\end{equation}
Here, $\eta_p$ is the dynamic polymer viscosity (depending on
polymer concentration), $\lambda$ the characteristic relaxation time
of the macromolecule chains, $f=f(|{\bf R}|)$ the dimensionless
Peterlin function which reflects the finite extensibility of the
chains. The components of the end-to-end vector of an ensemble of
polymer chains, ${\bf R}$, are combined to a macroscopic dimensionless
conformation tensor $C_{ij}$ by a dyadic product. The streamwise
component of the Reynolds equation obtained after an integration
with respect to $z$ and horizontal averaging is given by
\begin{equation}
-\langle u^{\prime} w^{\prime}\rangle(z) +\nu
\frac{dU}{dz}+\frac{\nu_p}{\lambda}\langle f C_{xz}\rangle(z)=
\frac{\tau_w}{\rho}\, \label{balance}
\end{equation}
in dimensional form,
where $\tau_w$ is the wall-shear stress. The polymer chains are
assumed to be significantly extended beyond their equilibrium
extension. Following Benzi {\it et al.},\cite{Benzi:2006} the
unknown polymer stress term can be closed to
\begin{equation}
\langle f C_{xz}\rangle(z)\approx\lambda \frac{dU}{dz} \langle
C_{zz}\rangle(z)\,,
\end{equation}
which gives
\begin{equation}
-\langle u^{\prime} w^{\prime}\rangle(z) +[\nu +\nu_p \langle C_{zz}\rangle(z)]
\frac{dU}{dz}\approx\frac{\tau_w}{\rho}\,.
\end{equation}
DNS in Ref. \onlinecite{Benzi:2006} also suggest  that $\langle
C_{zz}\rangle(z)\sim z$ close to the wall and $\langle
C_{zz}\rangle(z)\sim const$ in the logarithmic layer. This causes a
linearly increasing effective viscosity in the viscous layer and a
constant but larger one in the logarithmic layer. Drag reduction
seems to be associated with the growth of the viscous layer and
transition to the logarithmic scaling at larger $z^+$ with nearly
the same von K\'arm\'an constant $\kappa$.

In contrast, the Reynolds stress balance for the magnetic case
remains unchanged in comparison to the pure hydrodynamic case, i.e.,
no additional stress term appears. The spanwise magnetic field
provides a turbulent kinetic energy sink that reduces the turbulent
velocity fluctuations over the entire width of the channel (see
figures \ref{fig4:re20_rms}(g,h)) and does not contribute as a
$z$-dependent effective viscosity. This seems to be the only feature
separating the MHD case from other flows with drag reduction
effects.  It can explain why we do not observe growth of the viscous
layer as for turbulent polymer solutions. The absence of such growth
is clearly seen, for example, in the Reynolds stress profiles in
Figs. \ref{fig4:re20_rms}(g,h), which display no shift of the
position of the maximum.

The suppression of turbulent energy and the associated reduction of
the turbulent momentum transfer in the wall-normal direction also
explains the steepening of the mean flow profiles seen in figures
\ref{fig4:re20_prof}(a,b). At last, the weakening of the wall-normal
momentum transfer can be identified as the mechanism leading to the
drag reduction in our flow.

\section{Conclusions}
\label{sec:conclusion} In this paper, we presented the results of a
detailed investigation of a turbulent channel flow affected by a
uniform spanwise magnetic field.  The case of low magnetic Reynolds
number and electrically insulating walls was considered.  Numerical
simulations, both DNS and LES, were conducted for two relatively
large values of the hydrodynamic Reynolds number and  Hartmann
numbers in the range from zero to the threshold values above which
statistically steady three-dimensional turbulence could not
sustained.

A-posteriori comparison between the results of DNS and LES
computations demonstrated the ability of the dynamic Smagorinsky
model to accurately reproduce the effect of the MHD flow
transformation on the SGS stresses.  This is in contrast with
relatively poor performance of the channel-optimized classical
Smagorinsky model and in agreement with the earlier
studies\cite{Knaepen1:2004,Vorobev:2005,Kobayashi:2006,Sarris:Kassinos:Carati:2007,Vorobev:2007}
that showed the suitability of the dynamic model for the MHD
homogeneous turbulence and Hartmann flow.

We conducted a thorough investigation of the flow transformation
caused by the magnetic field.  Some results confirmed and extended
the conclusions made earlier on the basis of the low-$\Rey$
DNS\cite{Lee:Choi:2001} and
experiments\cite{Votsish:Kolesnikov:1977}, while the others were
entirely new.  In particular, we found that the key effect of the
magnetic field is the suppression of turbulent fluctuations.  The
important results are the reduction of the turbulent momentum
transfer in the wall-normal direction and decrease of the friction
drag coefficient.  Another consequence is the transformation of the
mean flow profile, which becomes steeper, acquires higher centerline
velocity, and resembles a laminar rather than turbulent channel flow
profile.

\begin{figure}[t]
\scriptsize{
\parbox{0.22\linewidth}{(a)}\parbox{0.76\linewidth}{(b)}
\centerline{
\includegraphics[width=0.48\textwidth]{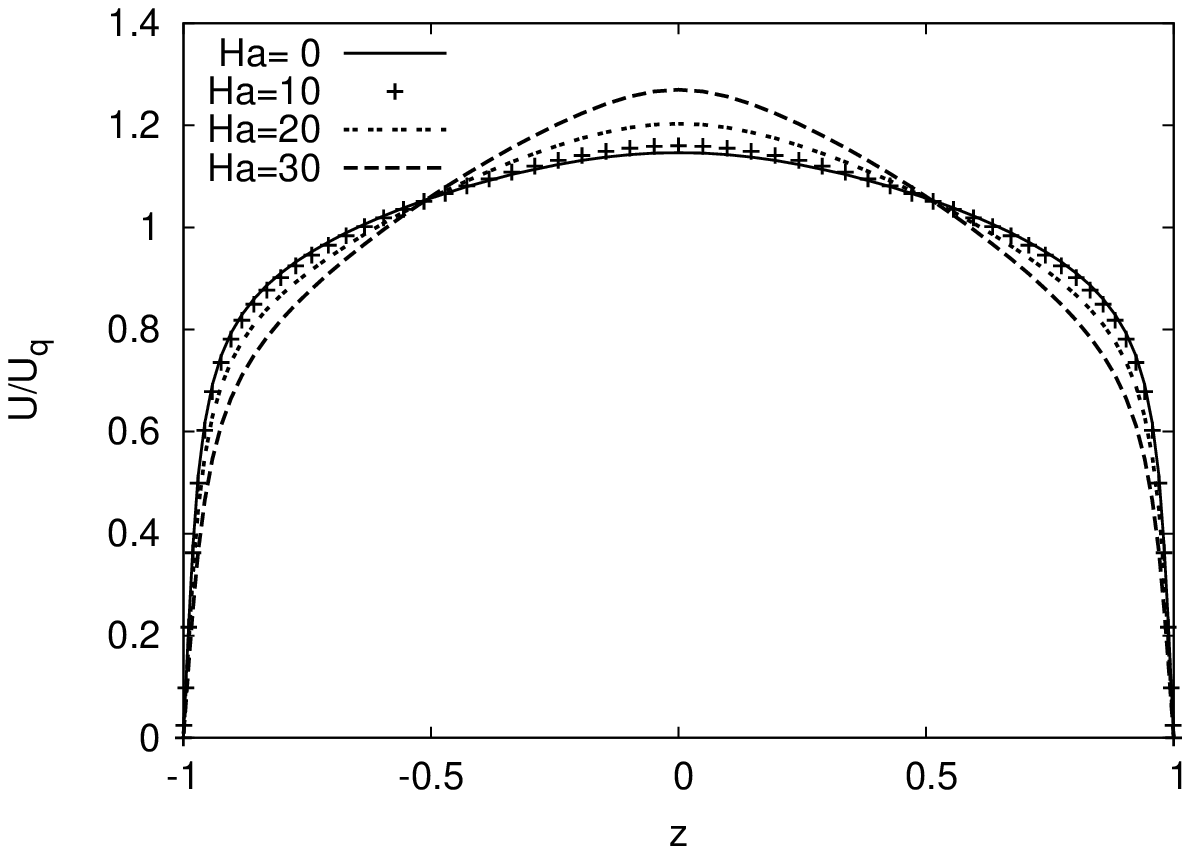}
\includegraphics[width=0.48\textwidth]{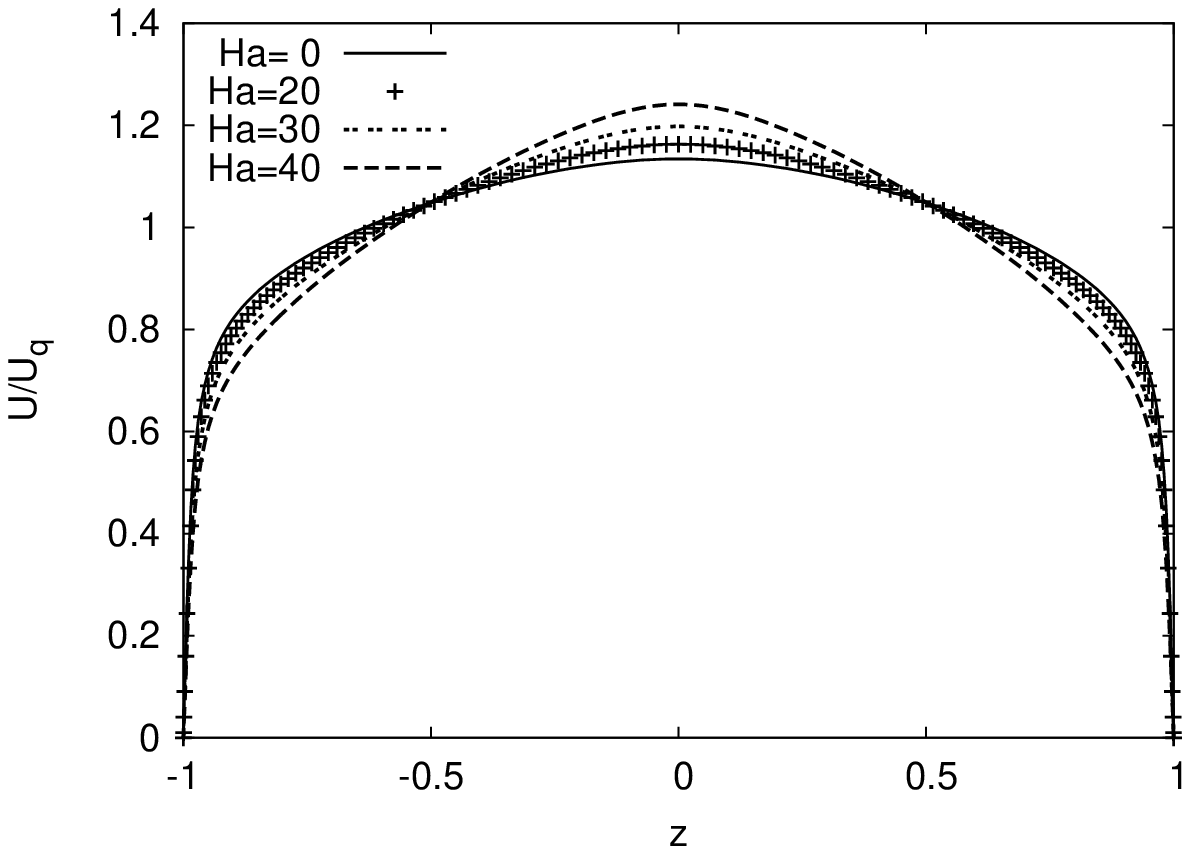}
}
\parbox{0.22\linewidth}{(c)}\parbox{0.76\linewidth}{(d)}
\centerline{
\includegraphics[width=0.48\textwidth]{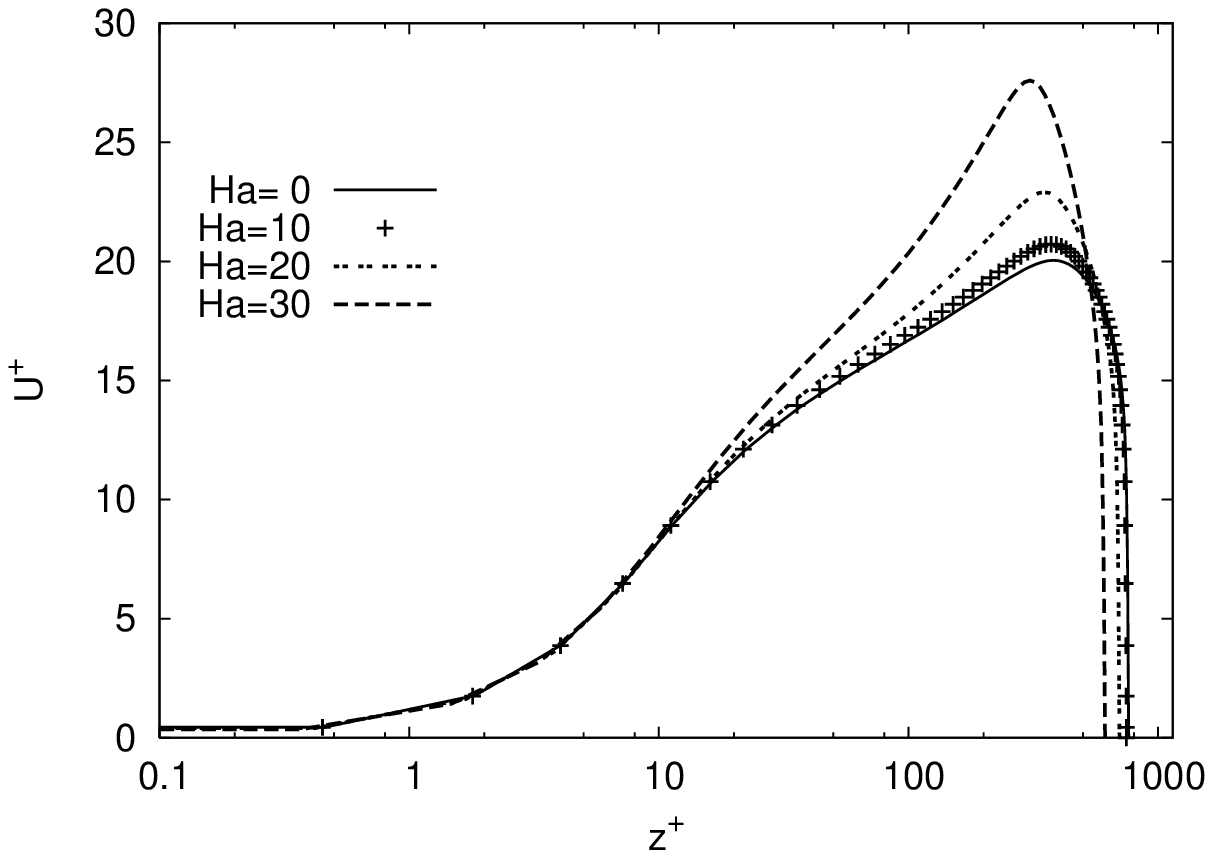}
\includegraphics[width=0.48\textwidth]{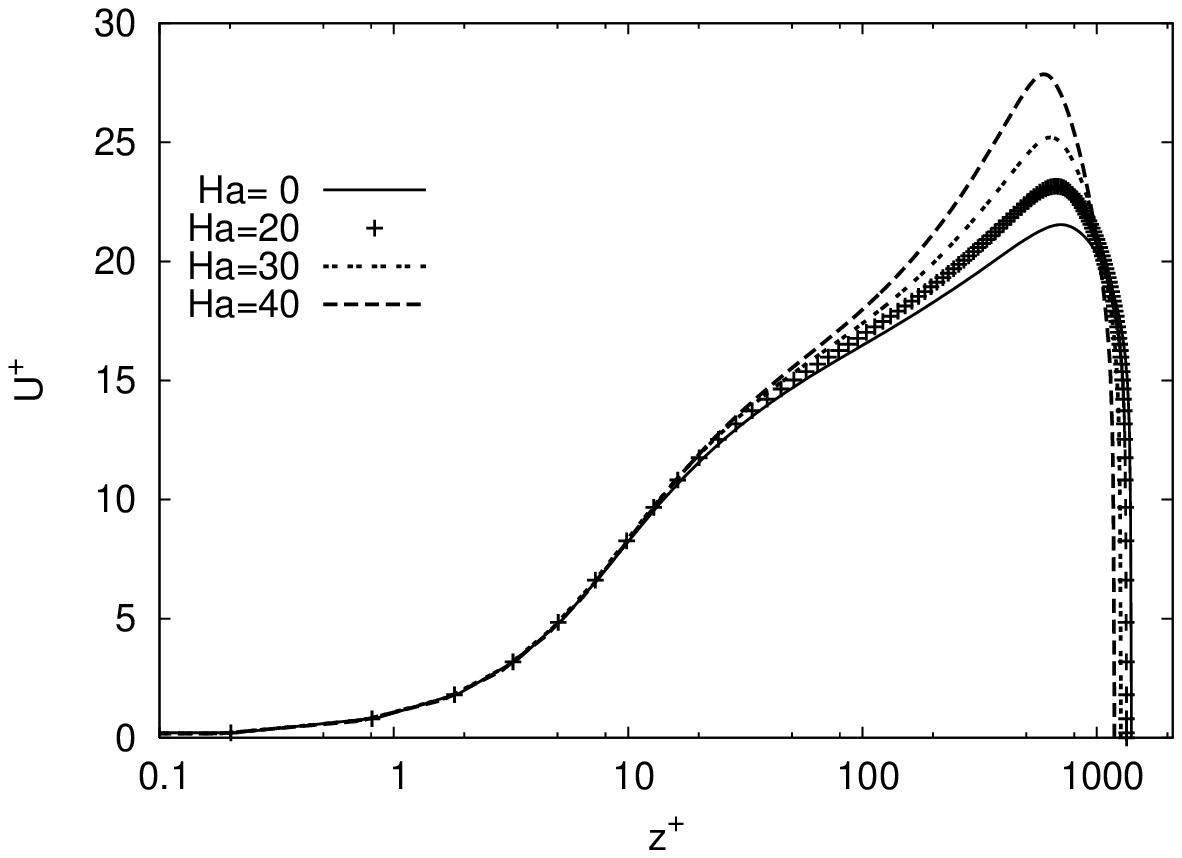}
}
\parbox{0.22\linewidth}{(e)}\parbox{0.76\linewidth}{(f)}
\centerline{
\includegraphics[width=0.48\textwidth]{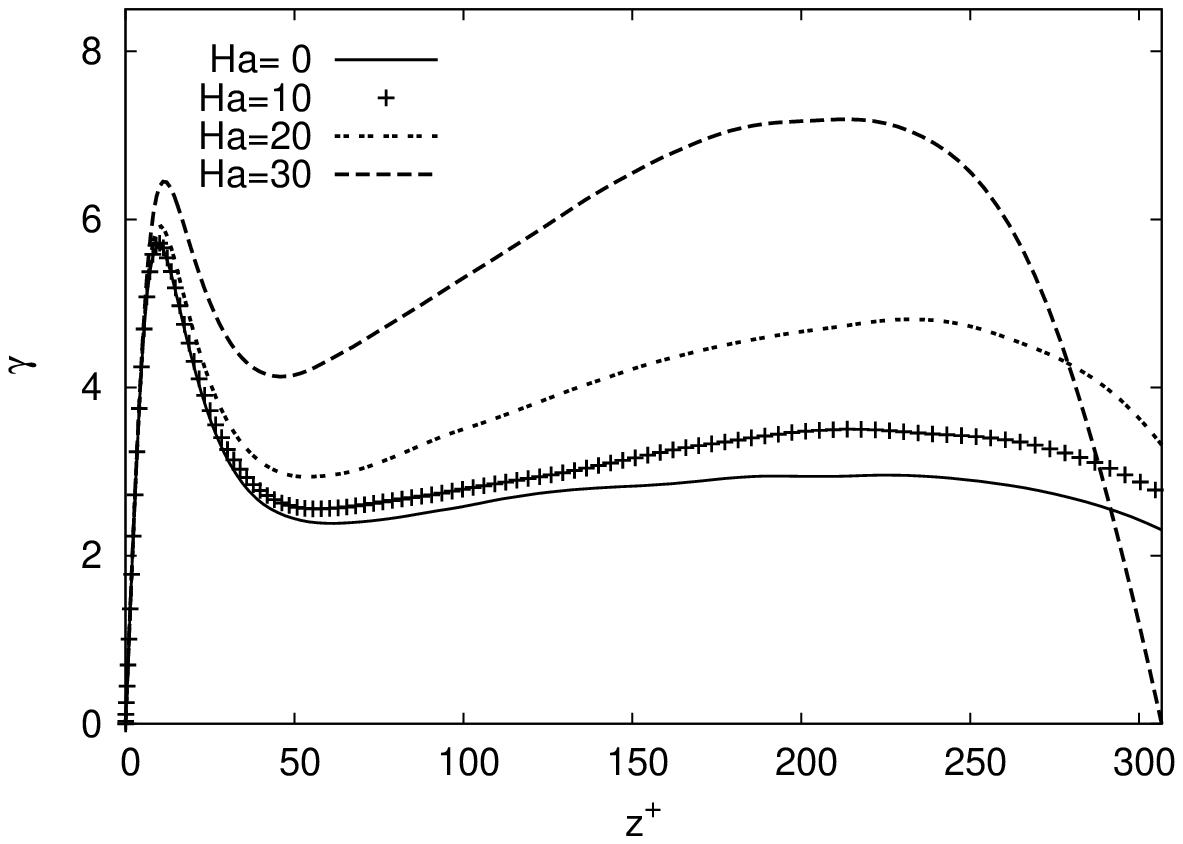}
\includegraphics[width=0.48\textwidth]{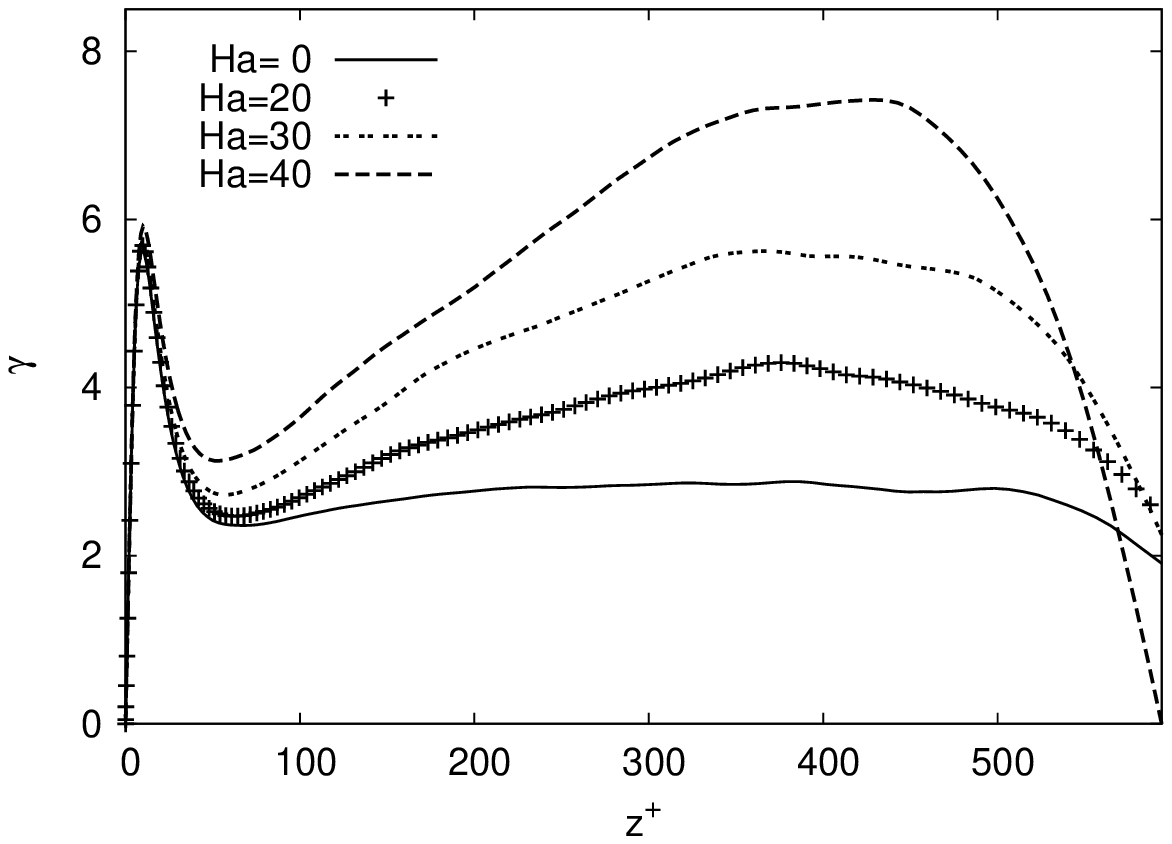}
}
}
\caption{Effect of the magnetic field on the flow properties.
Results of DNS are shown for $\Rey=10000$ (left column) and
$\Rey=20000$ (right column). \emph{(a,b)}, Mean velocity profiles
in global units. \emph{(c,d)} Mean velocity profiles in wall
coordinates. \emph{(e,f)}, Compensated profiles
$\gamma=z^+dU^+/dz^+$.} \label{fig4:re20_prof}
\end{figure}

\begin{figure}[t]
\scriptsize{
\parbox{0.22\linewidth}{(a)}\parbox{0.76\linewidth}{(b)}
\centerline{
\includegraphics[width=0.49\textwidth]{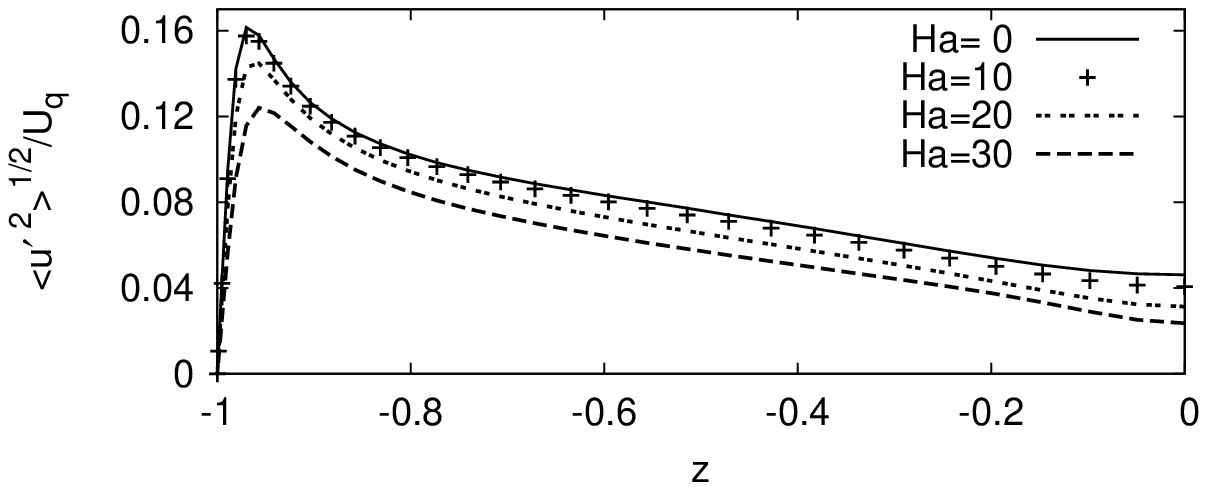}
\includegraphics[width=0.49\textwidth]{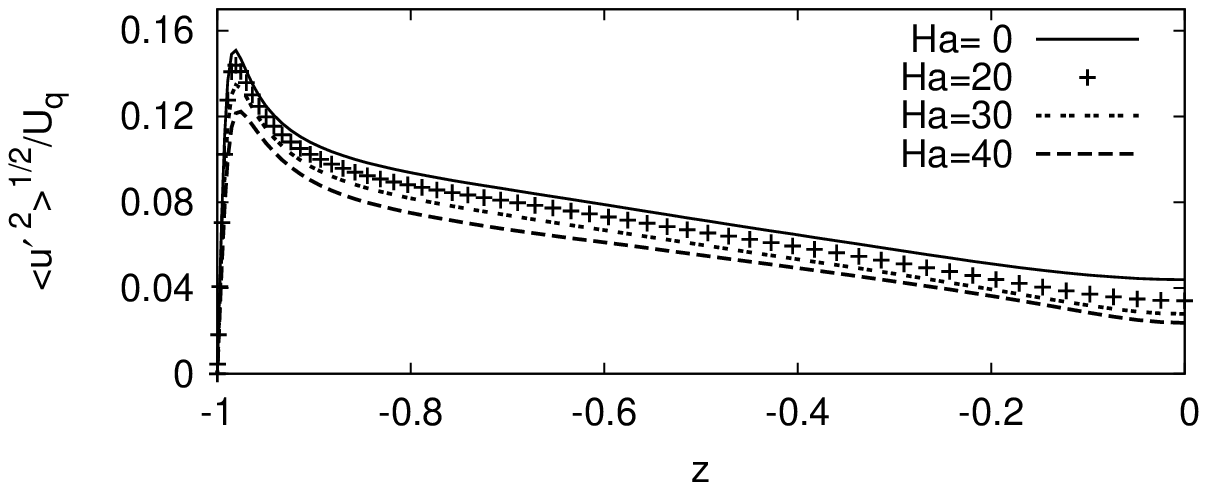}
}
\parbox{0.22\linewidth}{(c)}\parbox{0.76\linewidth}{(d)}
\centerline{
\includegraphics[width=0.49\textwidth]{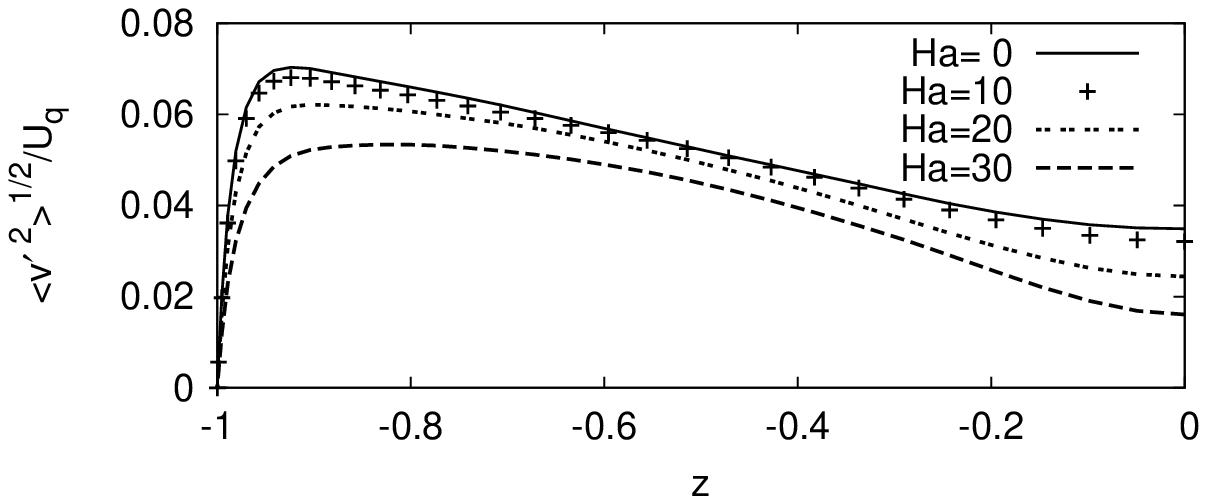}
\includegraphics[width=0.49\textwidth]{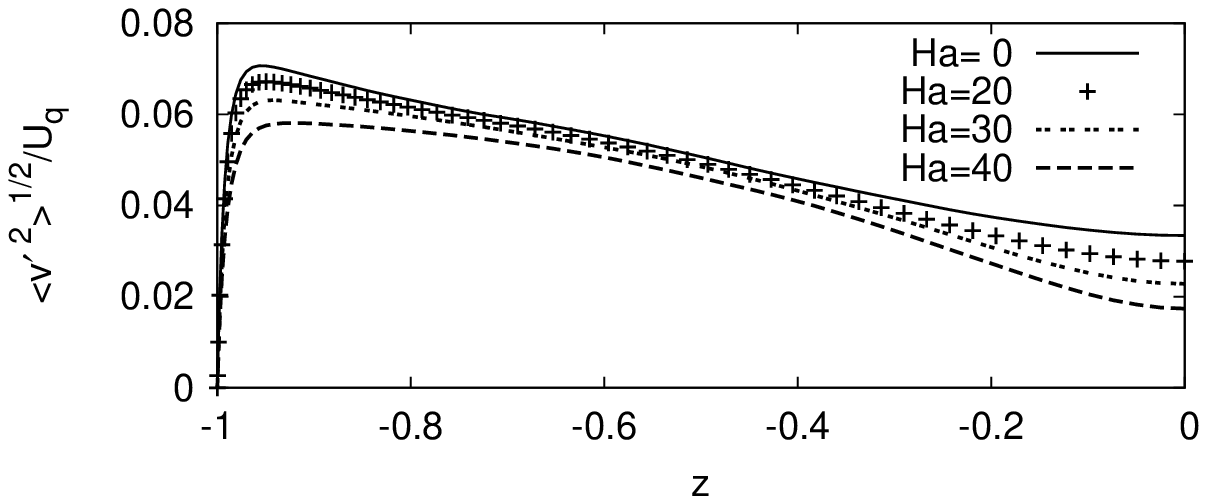}
}
\parbox{0.22\linewidth}{(e)}\parbox{0.76\linewidth}{(f)}
\centerline{
\includegraphics[width=0.49\textwidth]{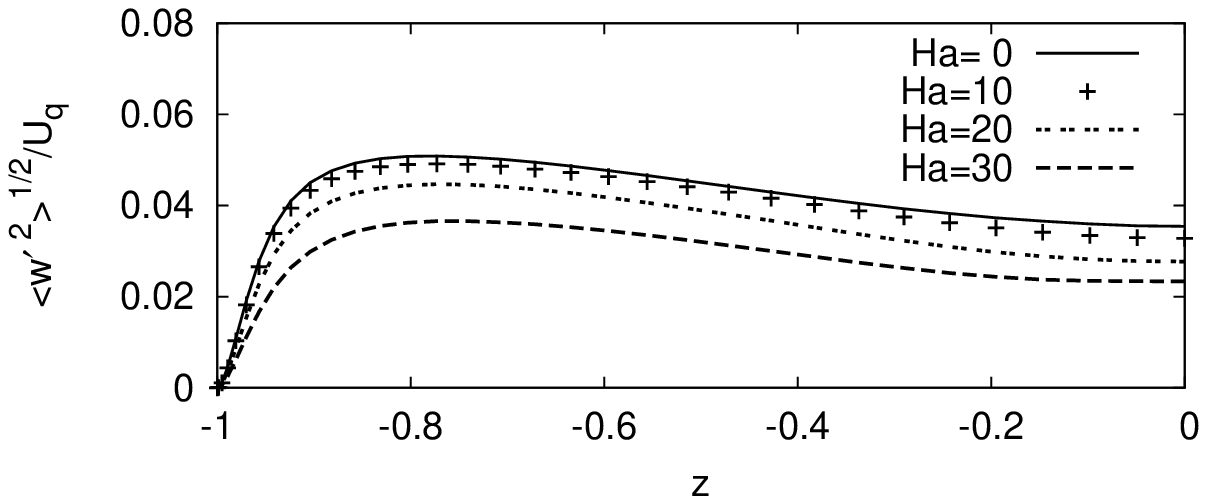}
\includegraphics[width=0.49\textwidth]{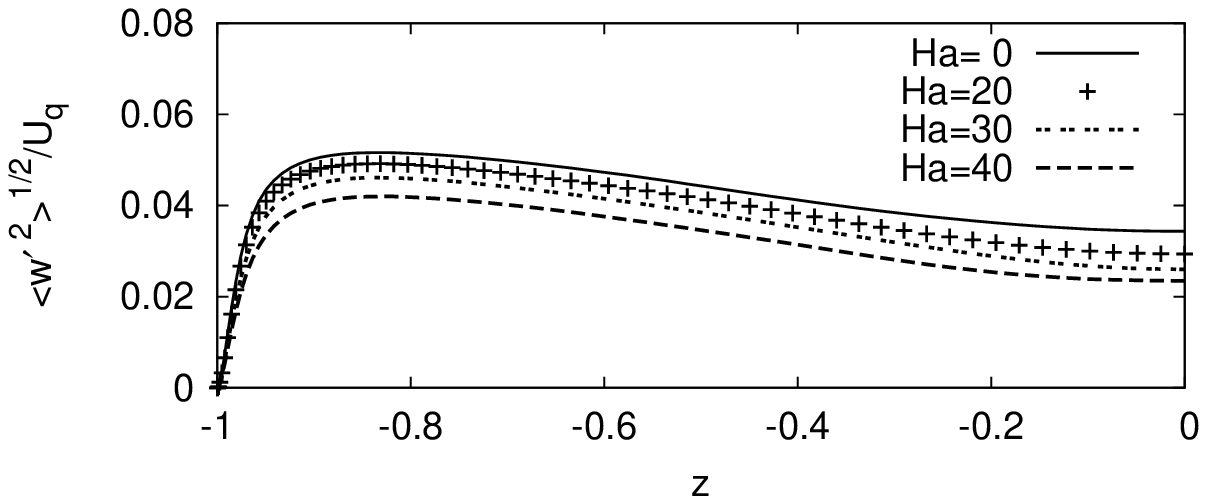}
}
\parbox{0.22\linewidth}{(g)}\parbox{0.76\linewidth}{(h)}
\centerline{
\includegraphics[width=0.49\textwidth]{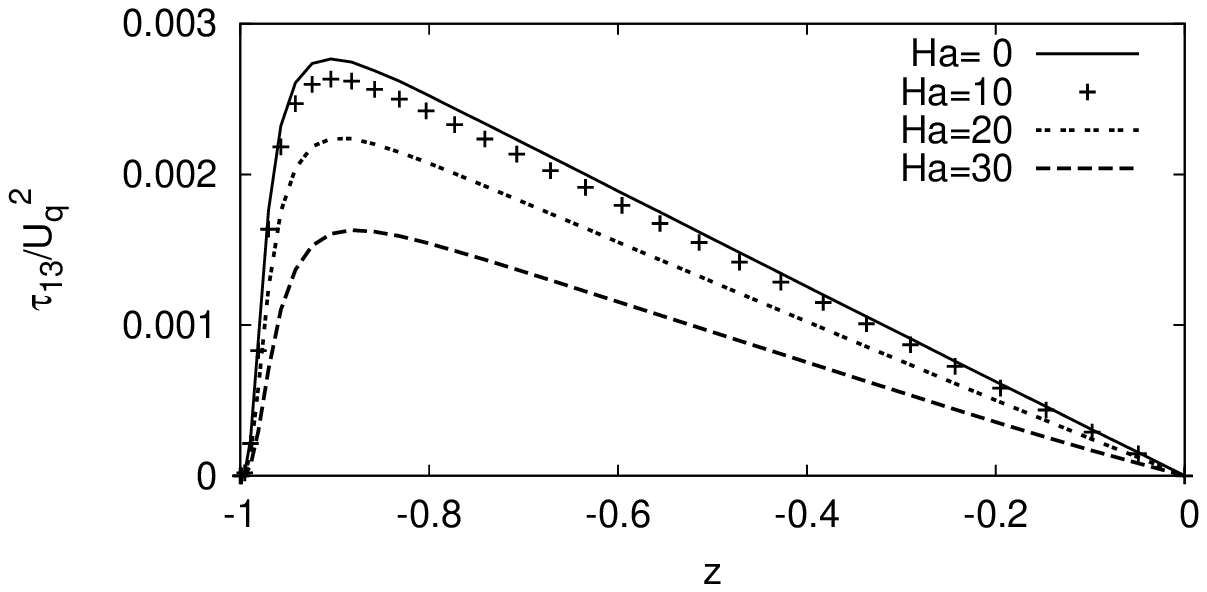}
\includegraphics[width=0.49\textwidth]{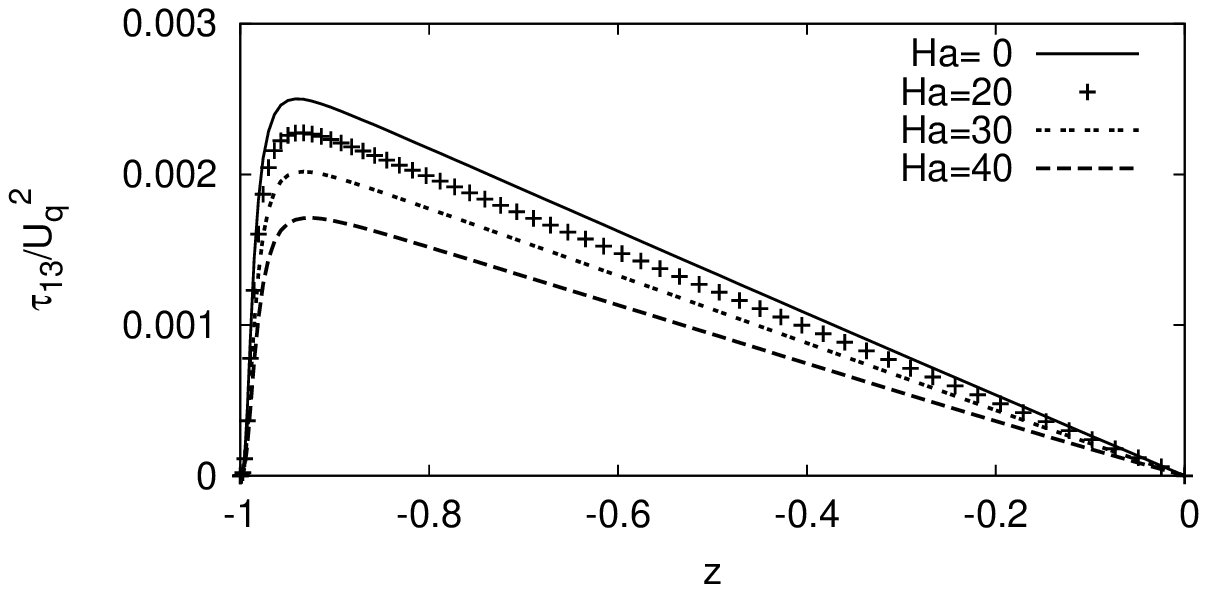}
}
}
\caption{Effect of the magnetic field on the flow properties.
Results of dynamic Smagorinsky LES are shown for $\Rey=10000$ (left
column) and $\Rey=20000$ (right column). \emph{(a-f)}, Horizontally
and time-averaged rms fluctuations of streamwise \emph{(a,b)},
spanwise \emph{(c,d)}, and normal \emph{(e,f)} velocities.
\emph{(g,h)}, horizontally and time-averaged full (resolved and SGS)
turbulent shear stress $\tau_{13}$.} \label{fig4:re20_rms}
\end{figure}

\begin{figure}[t]
\scriptsize{
\parbox{0.22\linewidth}{(a)}\parbox{0.76\linewidth}{(b)}
\centerline{
\includegraphics[width=0.47\textwidth]{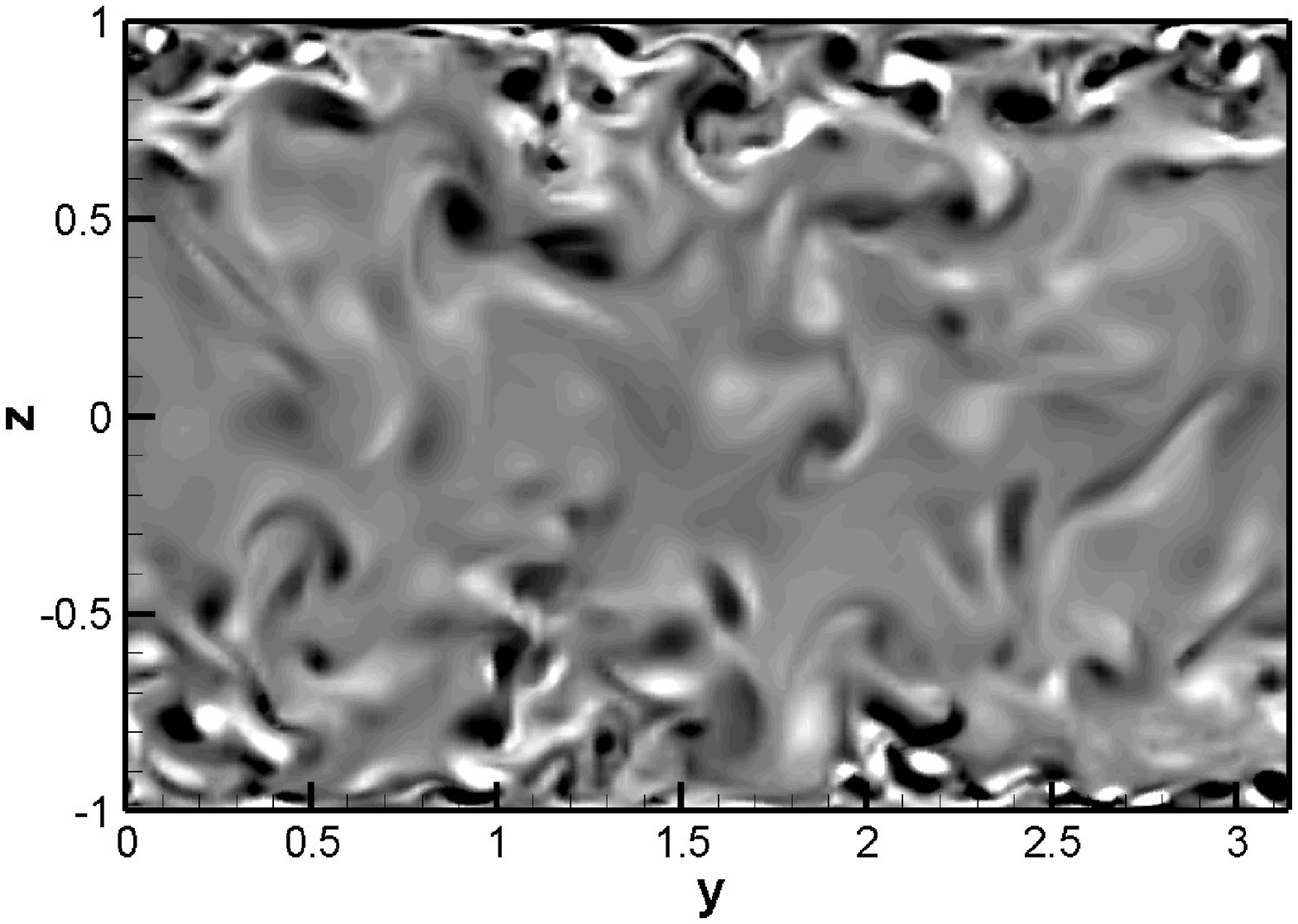}
\includegraphics[width=0.47\textwidth]{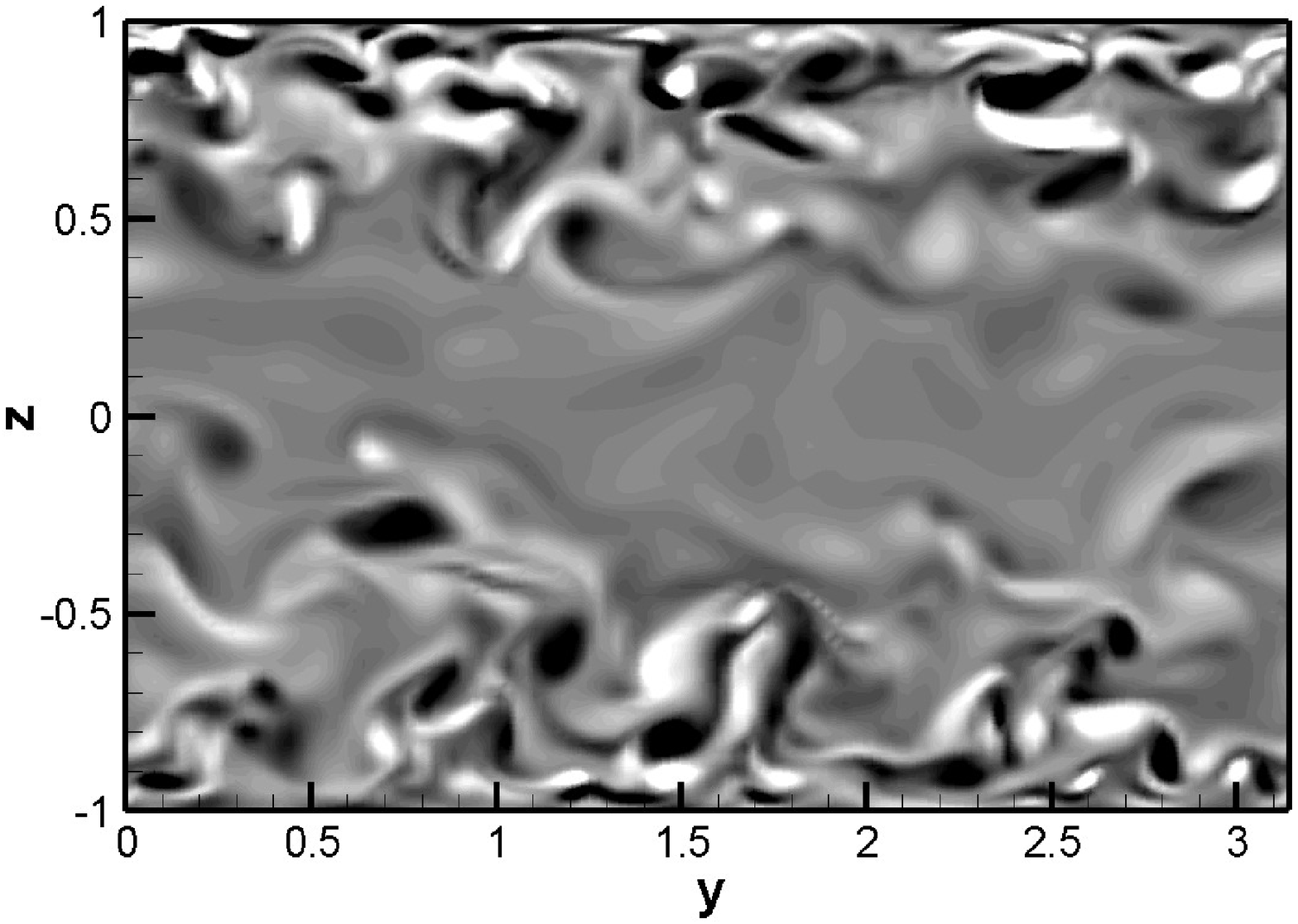}
}
\parbox{0.22\linewidth}{(c)}\parbox{0.76\linewidth}{(d)}
\centerline{
\includegraphics[width=0.49\textwidth]{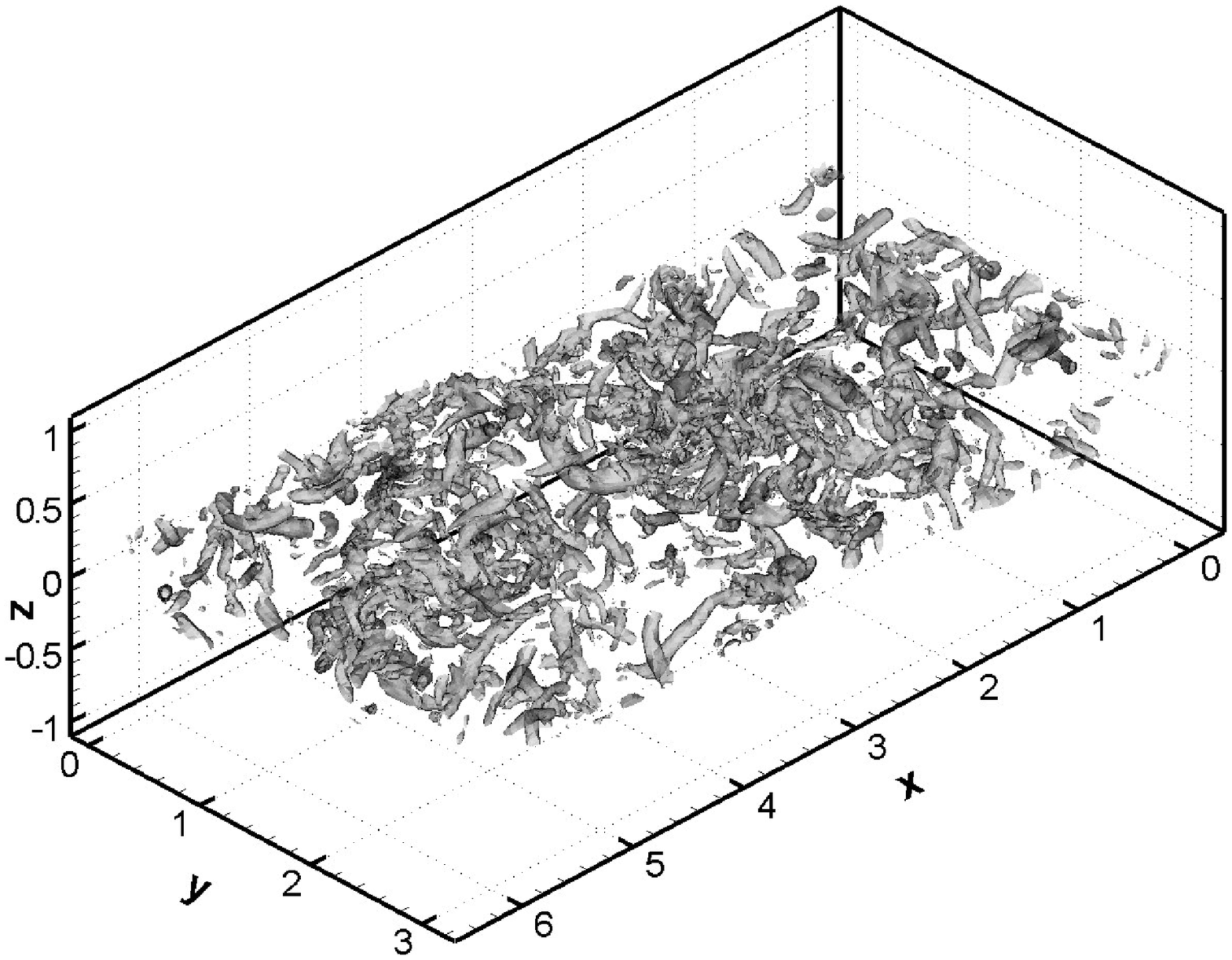}
\includegraphics[width=0.49\textwidth]{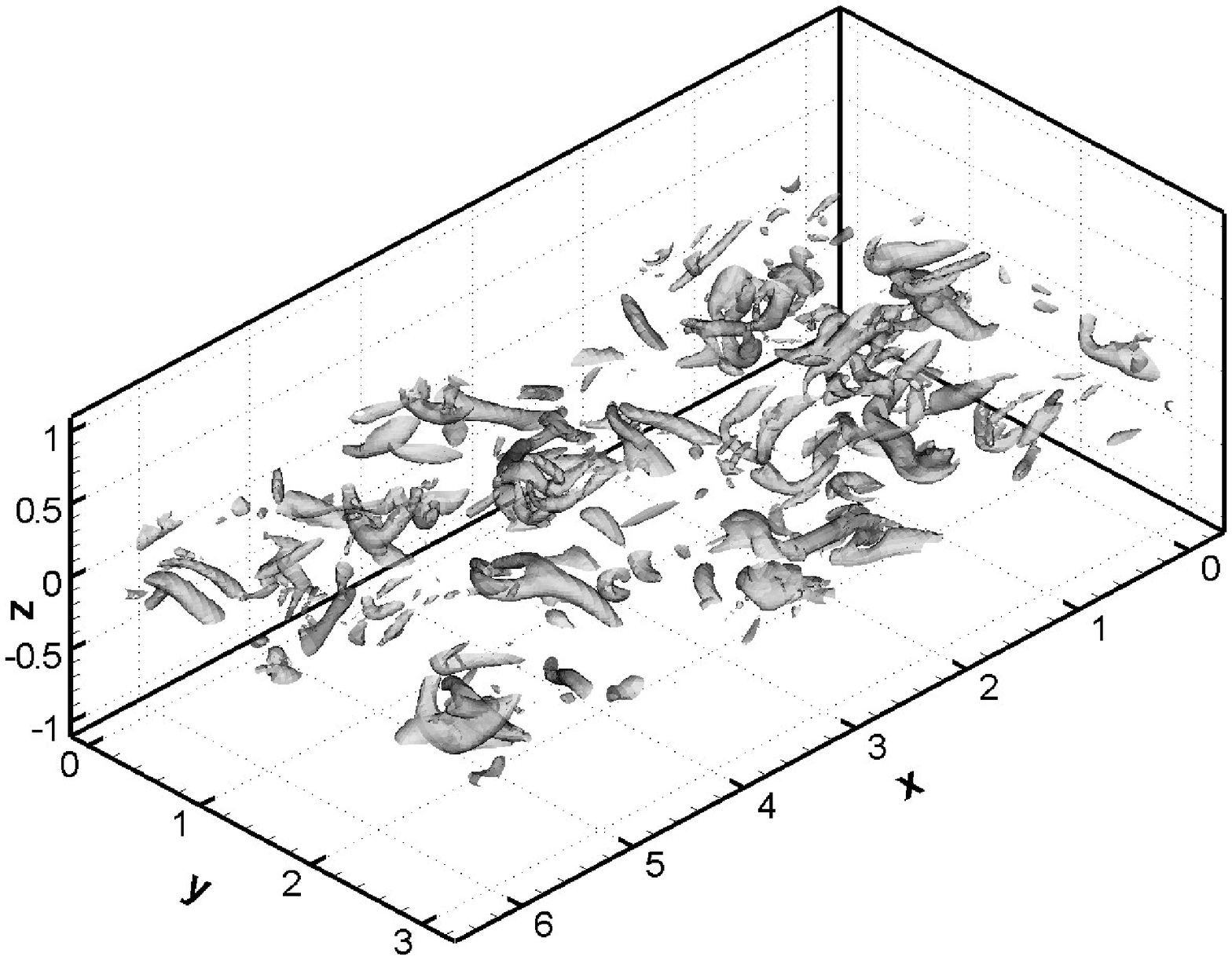}
}
\parbox{0.22\linewidth}{(e)}\parbox{0.76\linewidth}{(f)}
\centerline{
\includegraphics[width=0.48\textwidth,clip=]{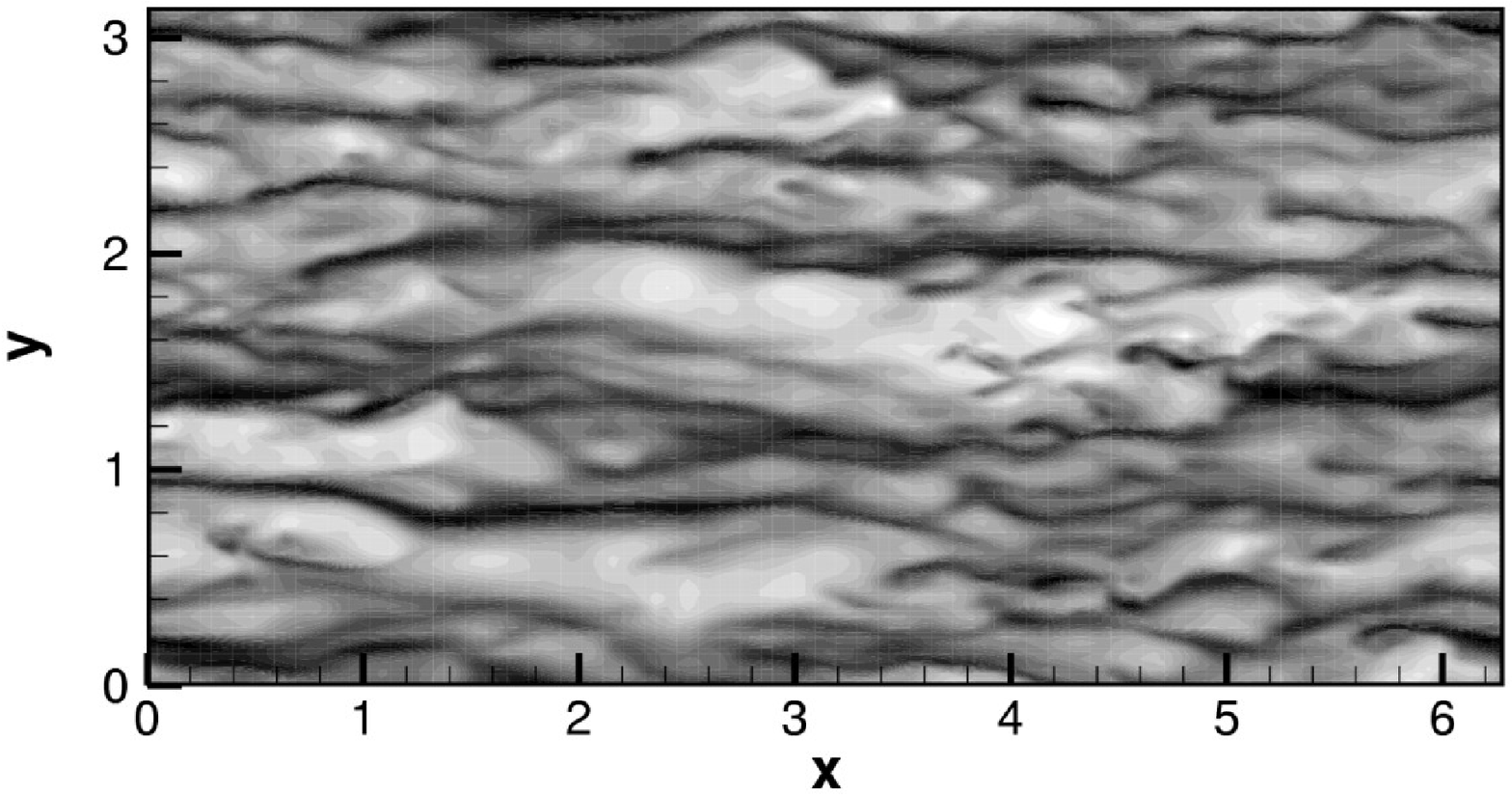}
\includegraphics[width=0.48\textwidth,clip=]{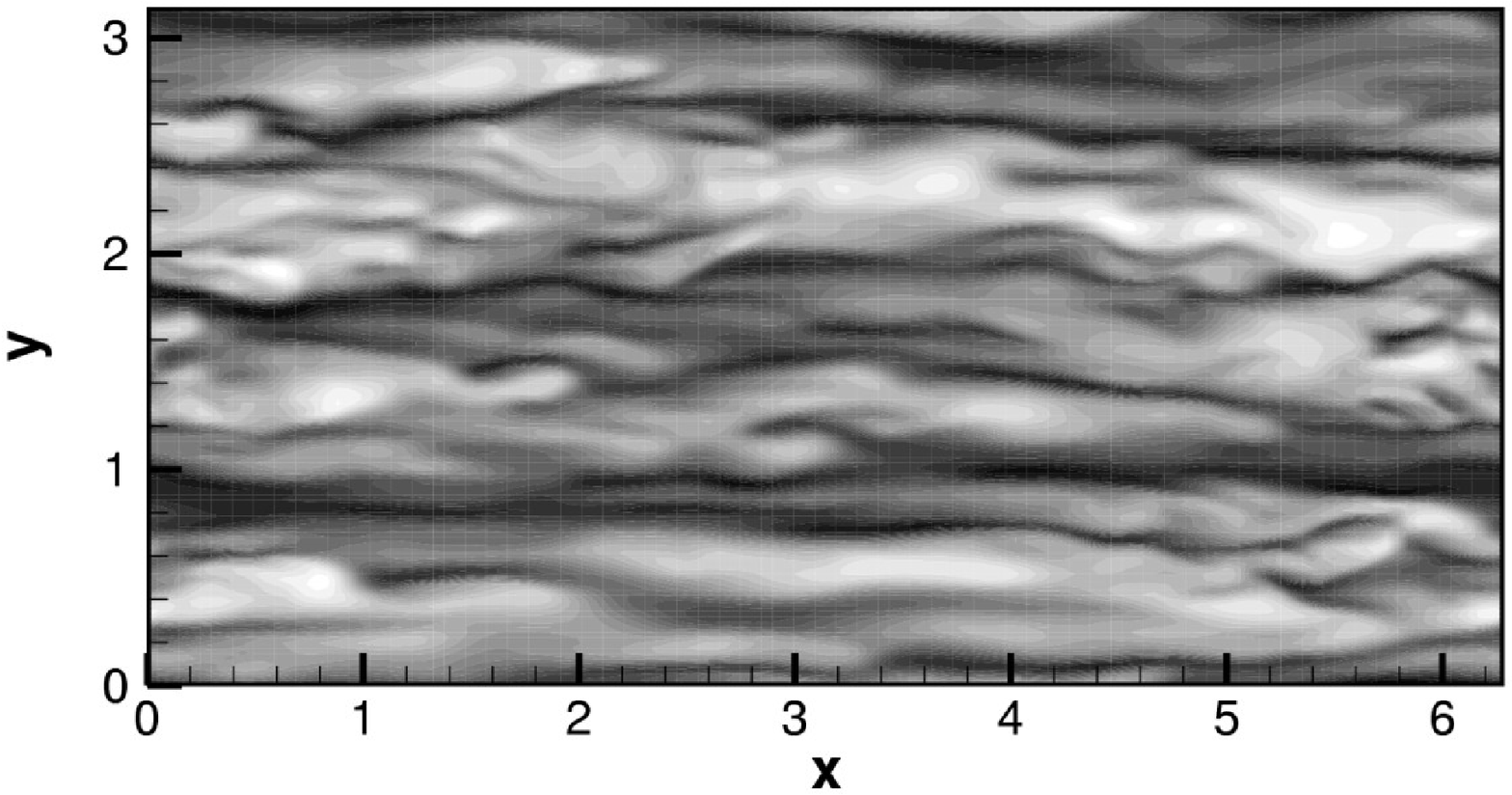}
}
}
\caption{Effect of magnetic field on coherent structures of the
flow.  Results of DNS are shown for $\Rey=10000$ and $Ha=0$
\emph{(a,c,e)} and $Ha=30$ \emph{(b,d,f)}.  All fields are
normalized by the corresponding rms values. \emph{(a,b)}, Contour
plots of streamwise vorticity in the transverse plane. \emph{(c,d)},
Iso-surfaces $\lambda_{2}=-0.1$ shown in the middle of the channel
at ($-0.25 < z < 0.25$). \emph{(e,f)}, Contour plots of the
streamwise velocity fluctuations at $z=0.95$. }
\label{fig7:re10_lam2}
\end{figure}

\begin{figure}[t]
\scriptsize{
\parbox{0.22\linewidth}{(a)}\parbox{0.76\linewidth}{(b)}
\centerline{
\includegraphics[width=0.48\textwidth,clip=]{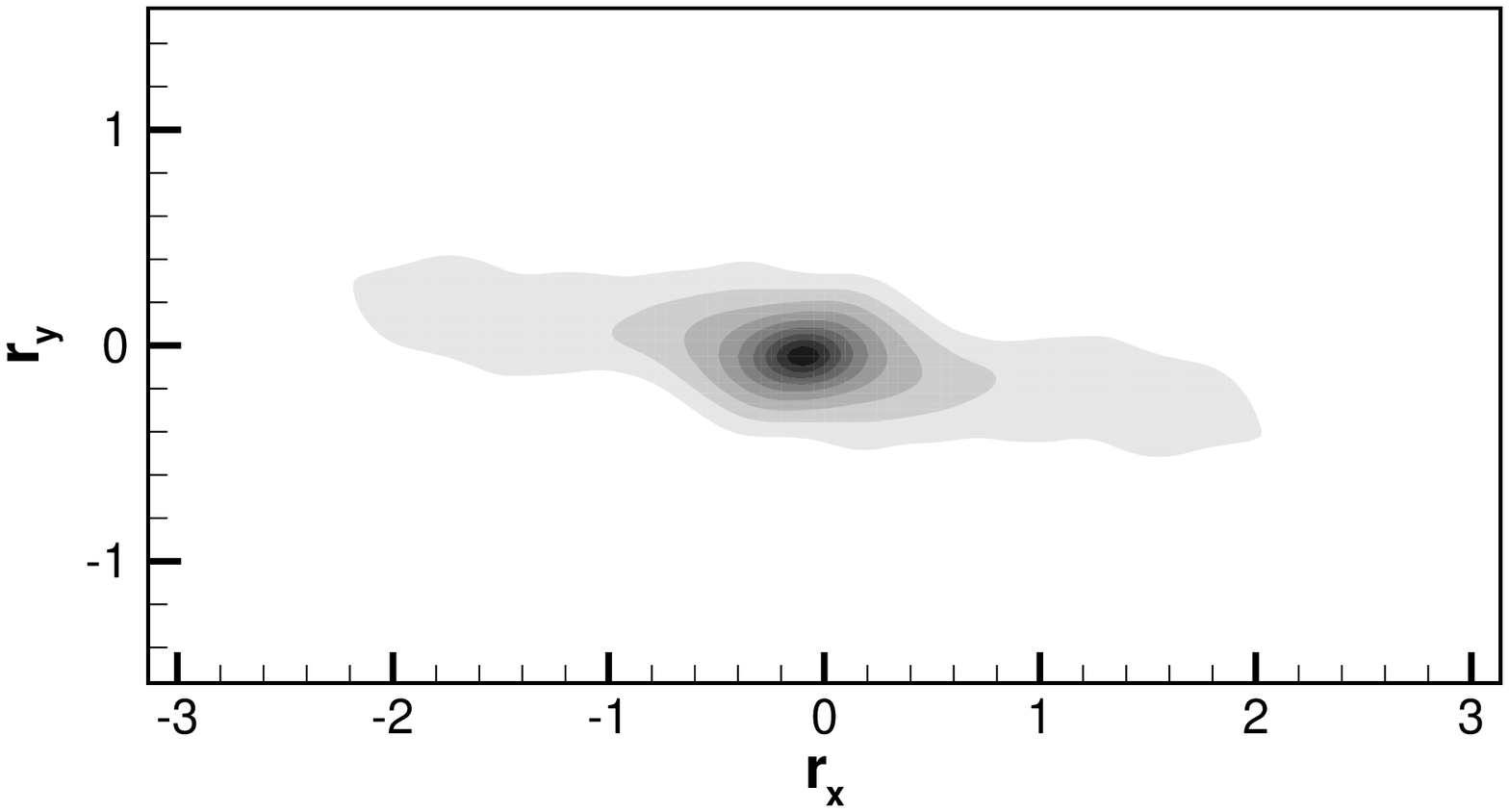}
\includegraphics[width=0.48\textwidth,clip=]{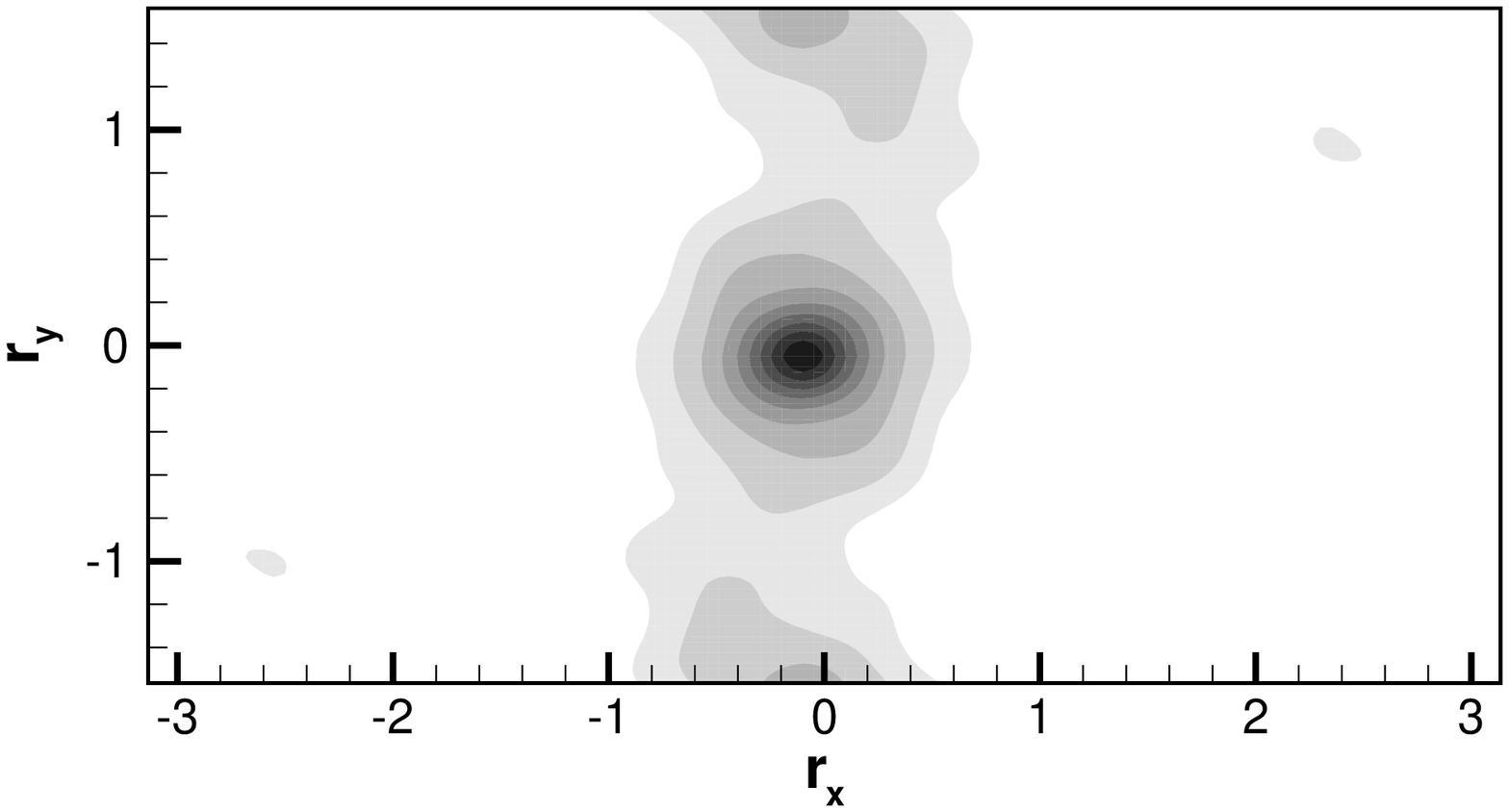}
}
\parbox{0.22\linewidth}{(c)}\parbox{0.76\linewidth}{(d)}
\centerline{
\includegraphics[width=0.48\textwidth,clip=]{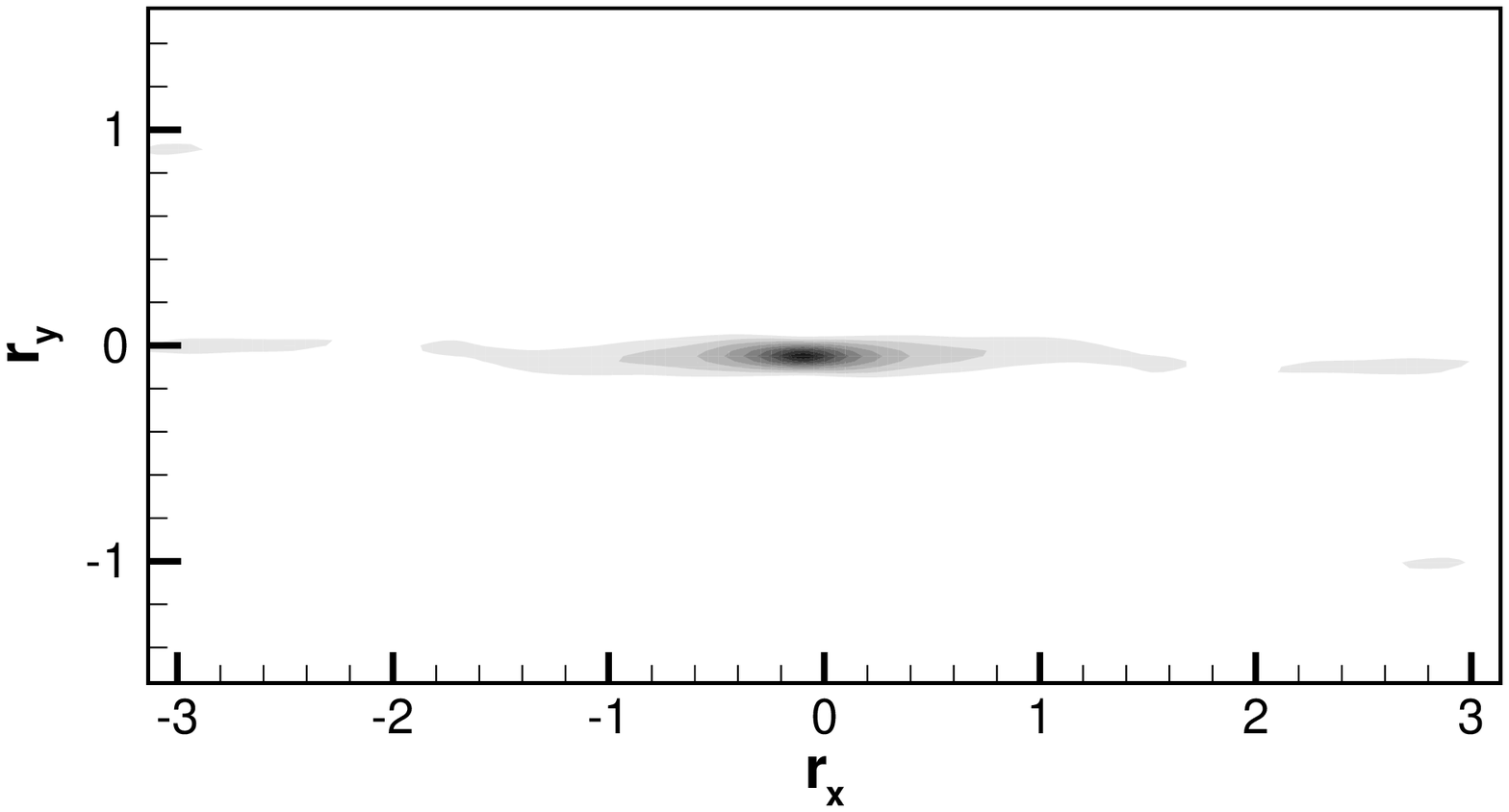}
\includegraphics[width=0.48\textwidth,clip=]{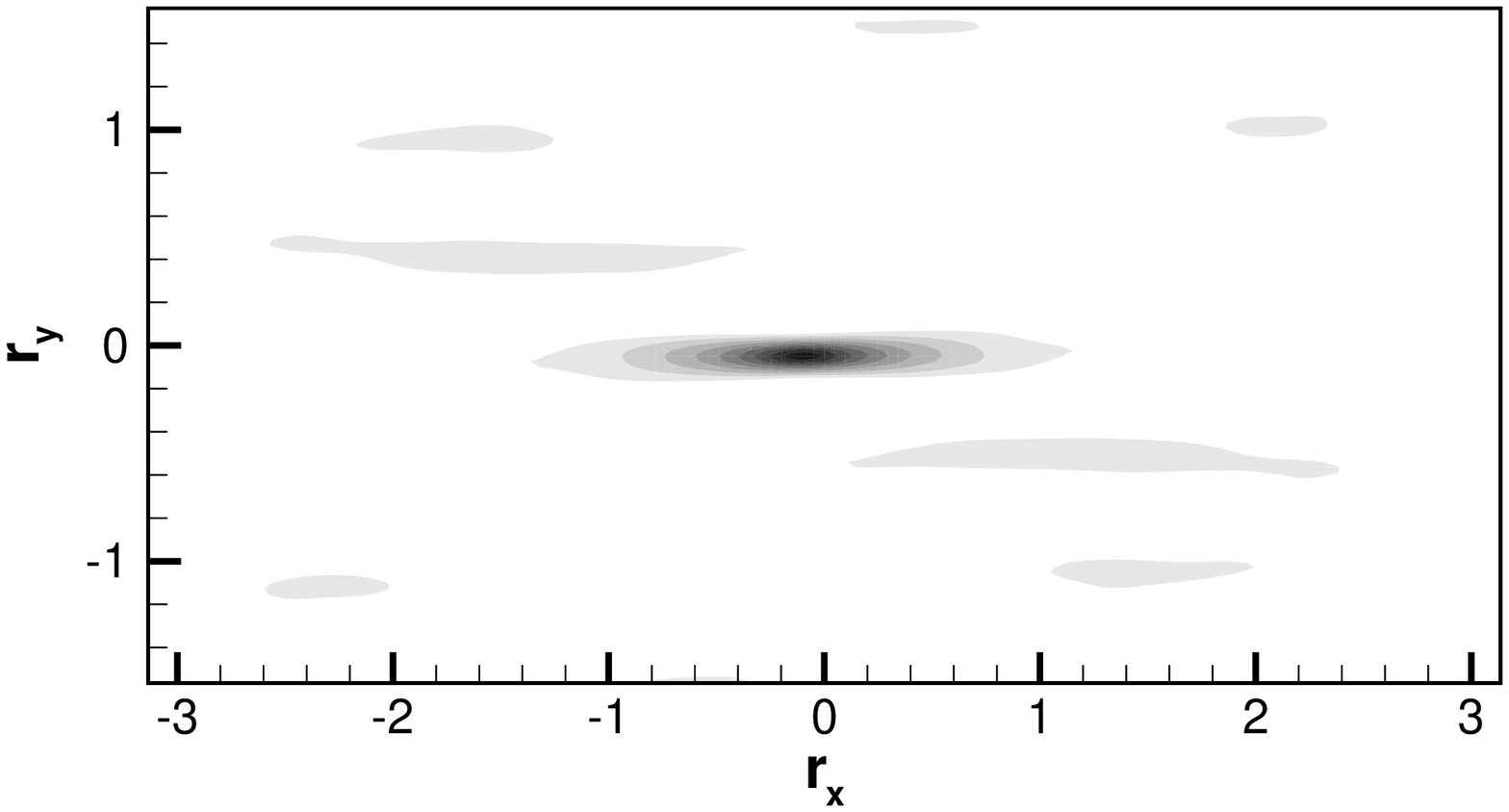}
}
}
\caption{Effect of magnetic field on flow anisotropy. Contour
plots of the correlation coefficient (\ref{eq:corr}) for the
streamwise velocity component computed for the DNS fields at
$\Rey=10000$ $Ha=0$ \emph{(a,c)} and $Ha=30$ \emph{(b,d)} are shown
in the middle of the channel at $z=0$ \emph{(a,b)} and near the wall
at $z=0.95$ \emph{(c,d)}. } \label{fig7:re10_corr}
\end{figure}

\begin{figure}[t]
\scriptsize{
\parbox{0.05\linewidth}{(a)}\parbox{0.27\linewidth}{$ $}
\parbox{0.05\linewidth}{(b)}\parbox{0.28\linewidth}{$ $}
\parbox{0.05\linewidth}{(c)}\parbox{0.16\linewidth}{$ $}
\centerline{
\includegraphics[width=0.32\textwidth,clip=]{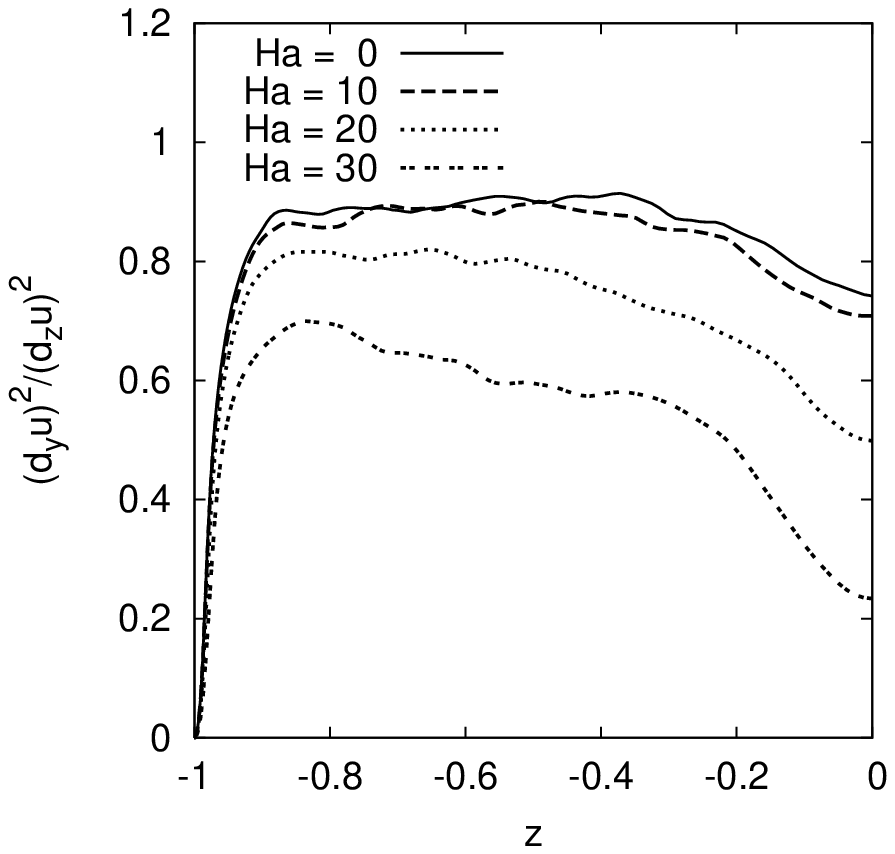}
\includegraphics[width=0.32\textwidth,clip=]{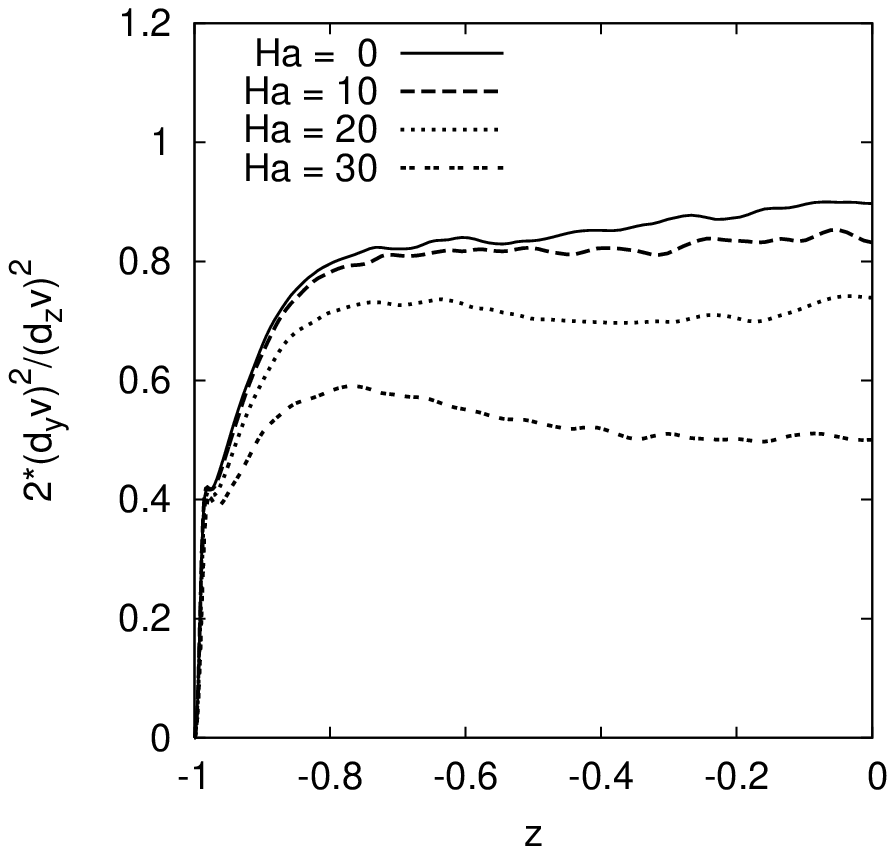}
\includegraphics[width=0.32\textwidth,clip=]{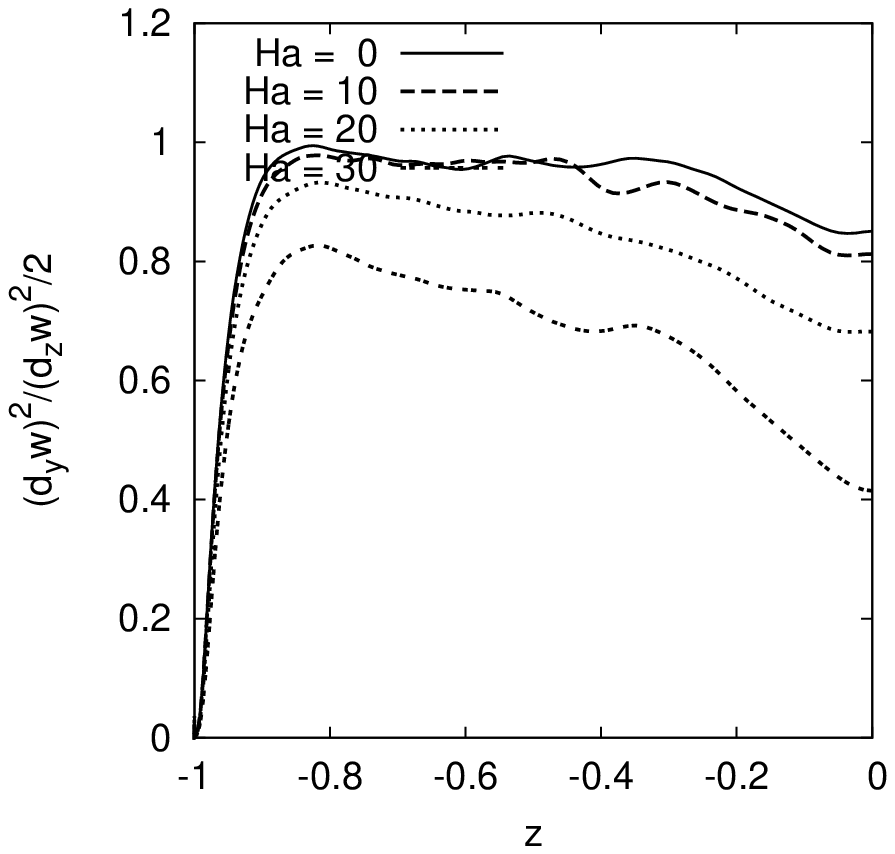}
}
\parbox{0.05\linewidth}{(d)}\parbox{0.27\linewidth}{$ $}
\parbox{0.05\linewidth}{(e)}\parbox{0.28\linewidth}{$ $}
\parbox{0.05\linewidth}{(f)}\parbox{0.16\linewidth}{$ $}
\centerline{
\includegraphics[width=0.32\textwidth,clip=]{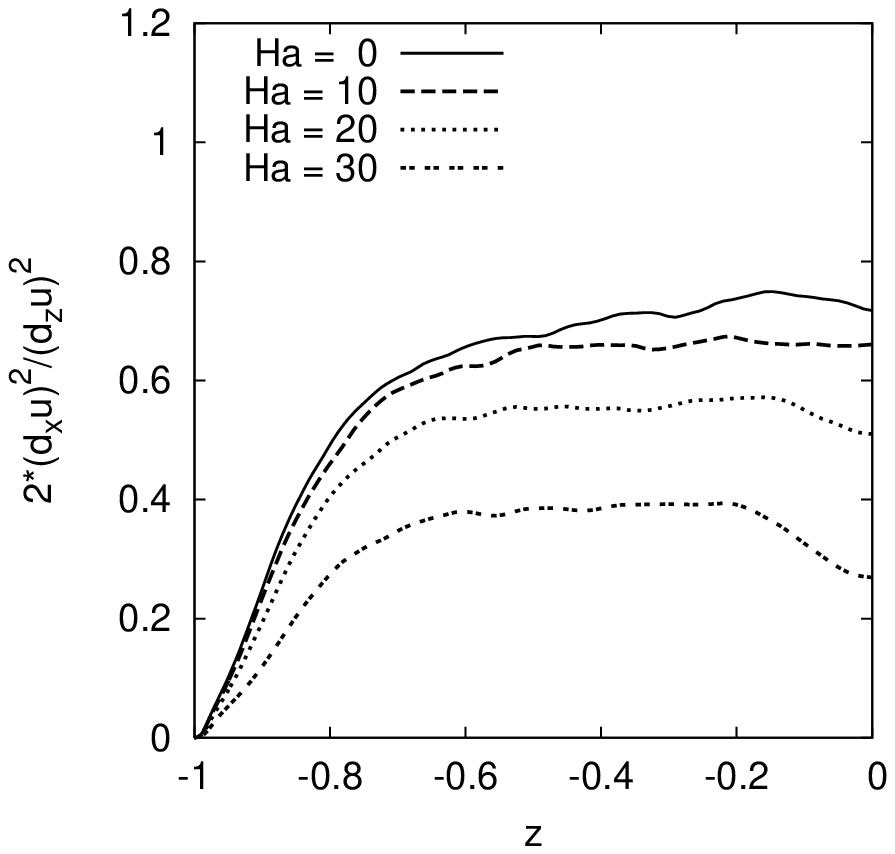}
\includegraphics[width=0.32\textwidth,clip=]{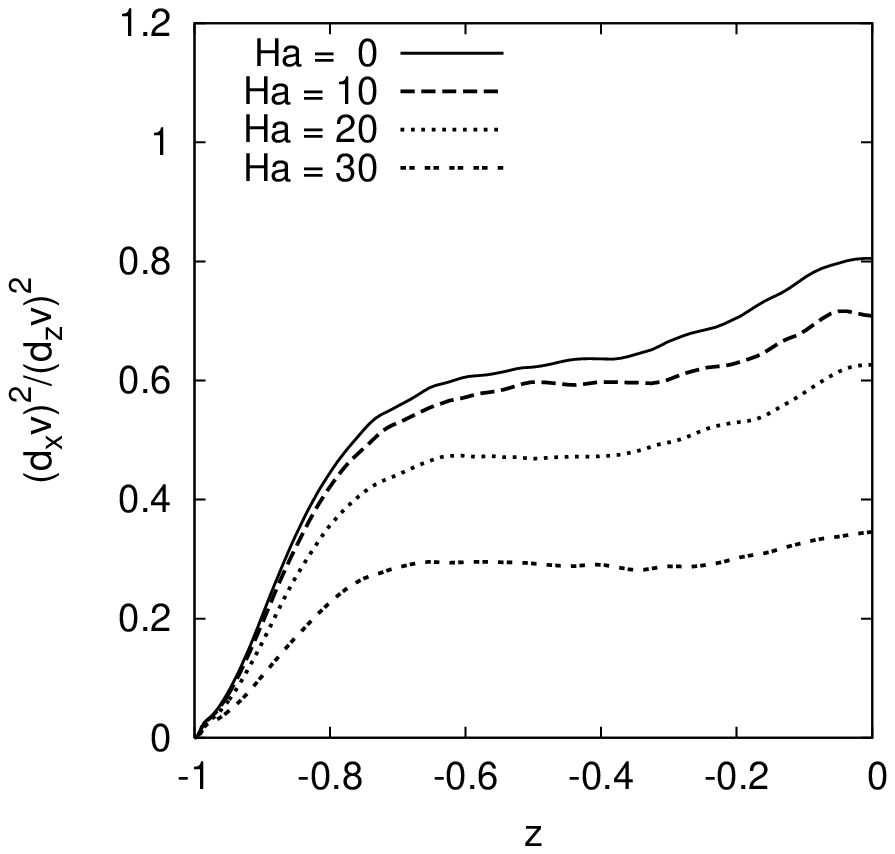}
\includegraphics[width=0.32\textwidth,clip=]{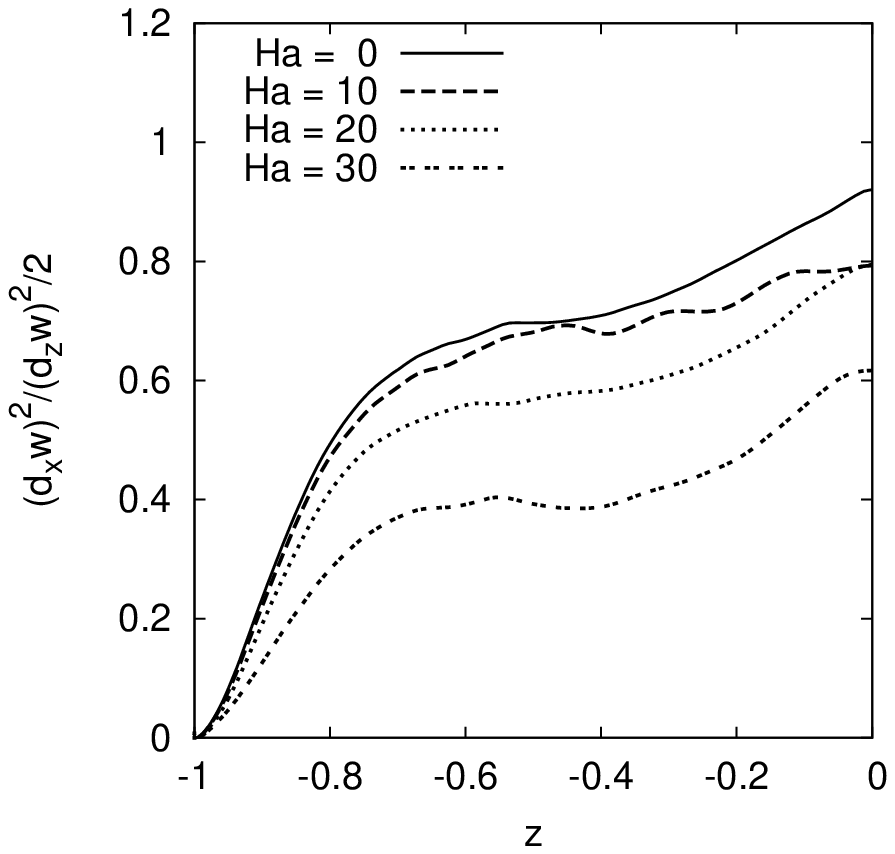}
}
\parbox{0.05\linewidth}{(g)}\parbox{0.27\linewidth}{$ $}
\parbox{0.05\linewidth}{(h)}\parbox{0.28\linewidth}{$ $}
\parbox{0.05\linewidth}{(i)}\parbox{0.16\linewidth}{$ $}
\centerline{
\includegraphics[width=0.32\textwidth,clip=]{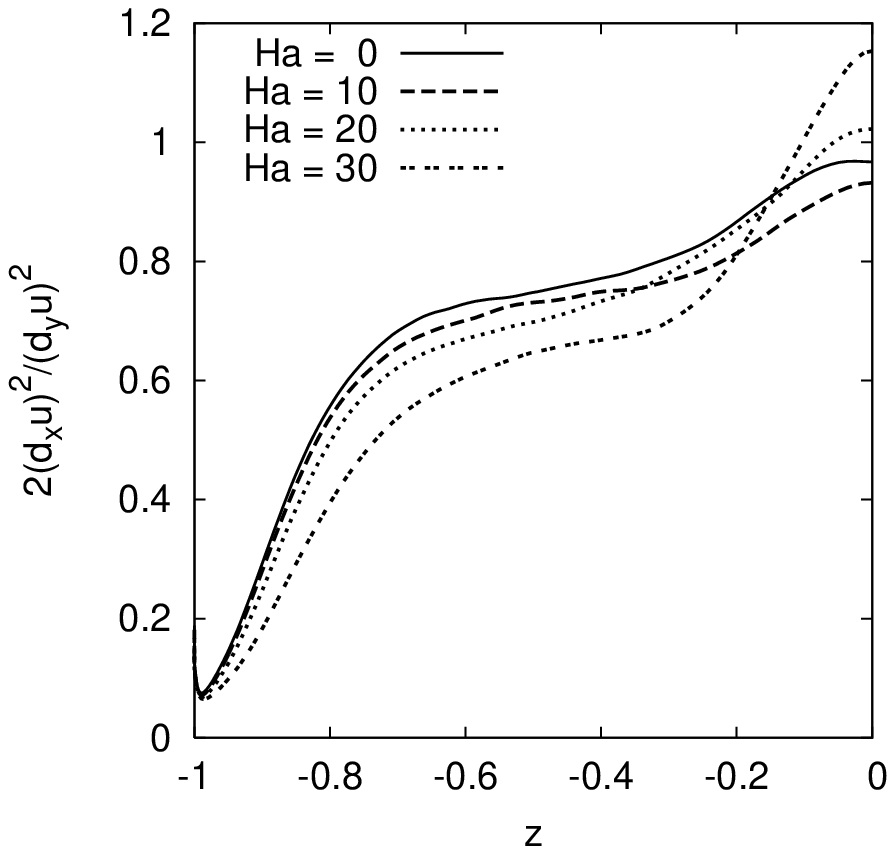}
\includegraphics[width=0.32\textwidth,clip=]{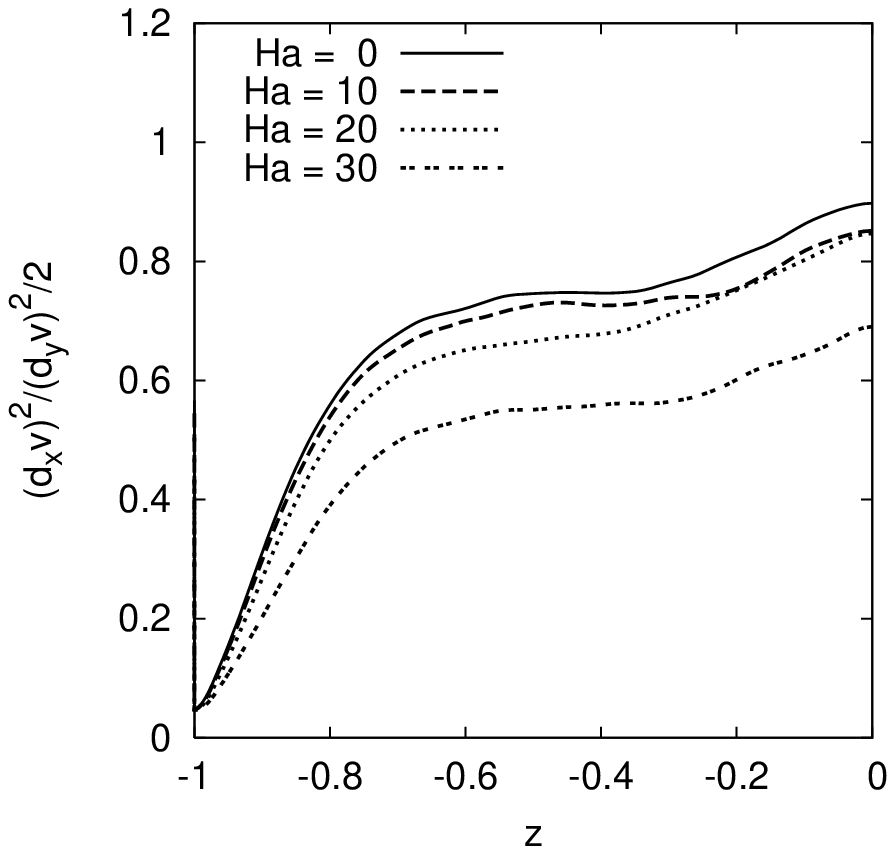}
\includegraphics[width=0.32\textwidth,clip=]{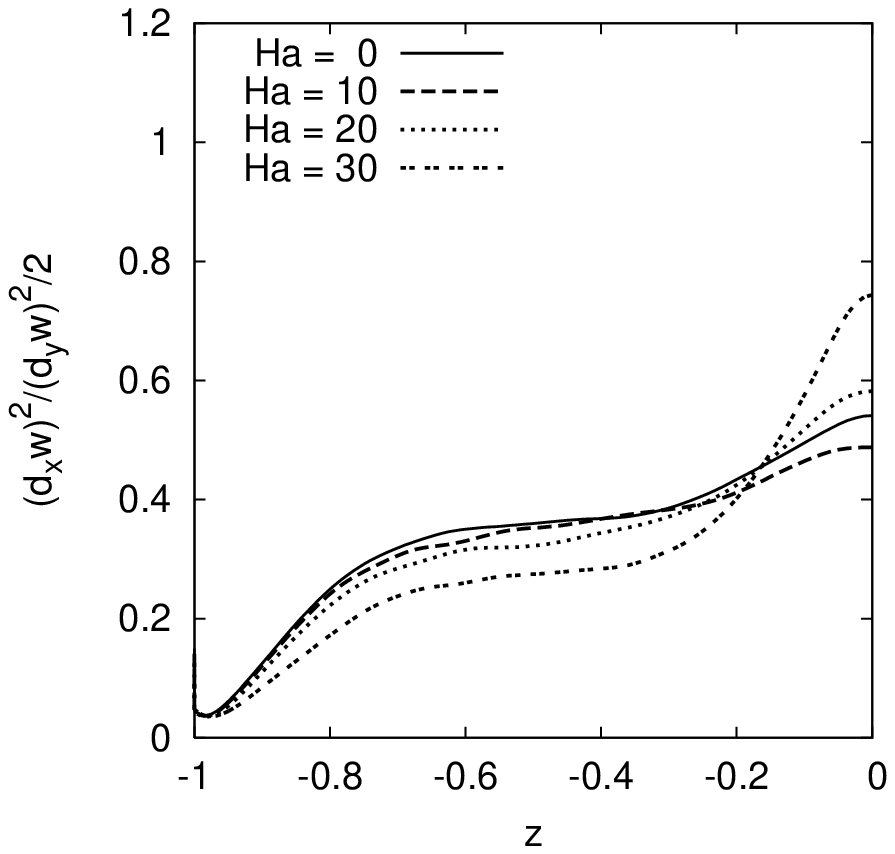}
}
}
\caption{Effect of magnetic field on flow
anisotropy. Time-averaged coefficients of the anisotropy of
gradients (\ref{eq:Gij}) are shown as computed in DNS  at
$\Rey=10000$.}
\label{fig8:re10_gij}
\end{figure}

\begin{figure}[t]
\centerline{
\includegraphics[width=0.48\textwidth]{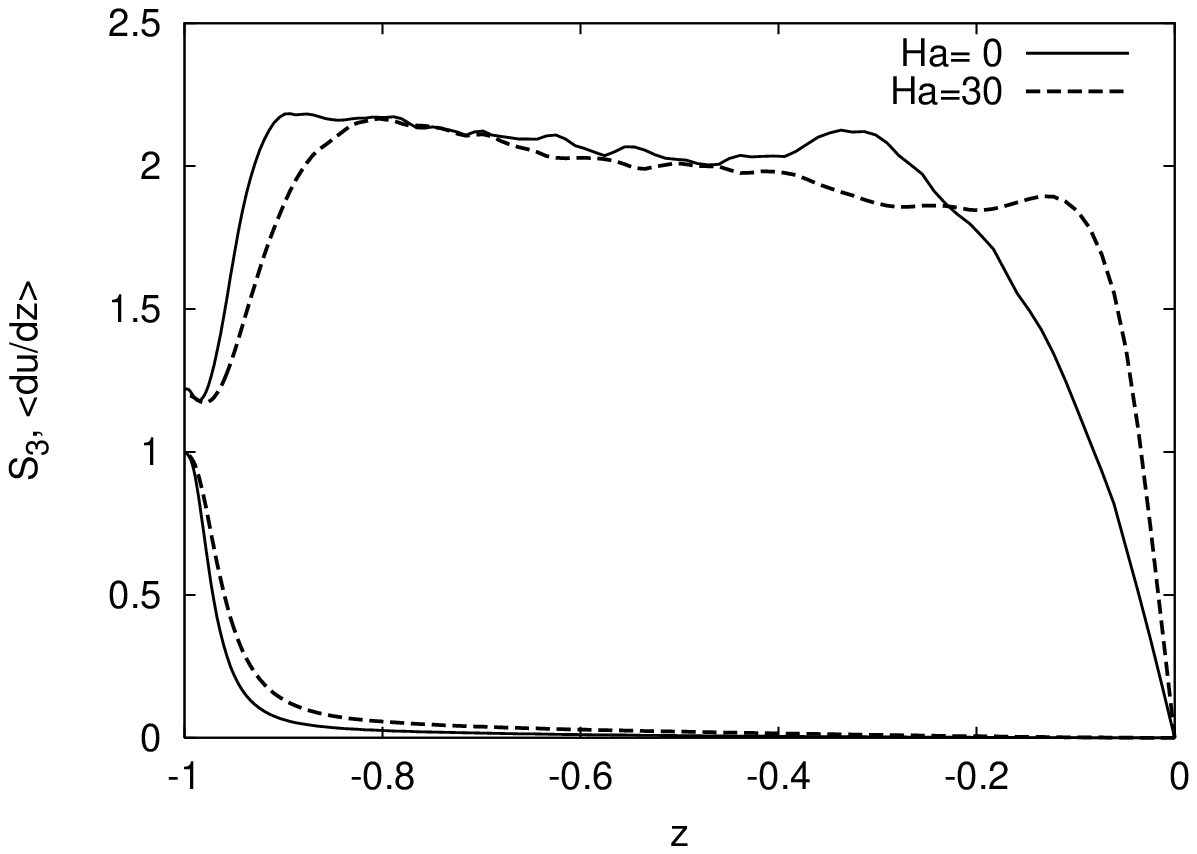}
\includegraphics[width=0.48\textwidth]{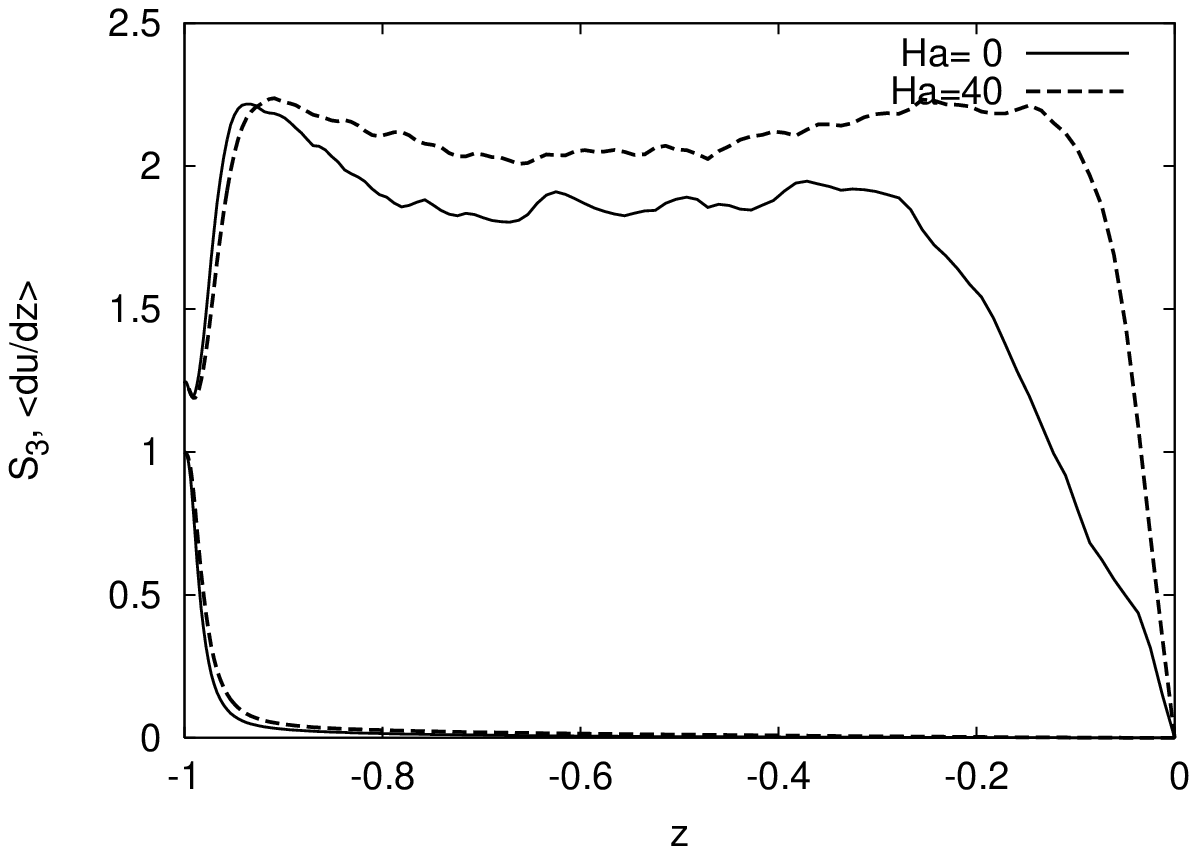}
}
\caption{Effect of magnetic field on flow anisotropy: wall-normal
distribution of skewness of the transverse derivative, $\partial u/
\partial z$, (upper curves) and wall-normal gradient of mean velocity
(lower ones) in physical units. Horizontally and time-averaged
quantities computed in DNS are shown for $Re=10000$ at $Ha = 0,30$
(left) and $Re=20000$ at $Ha = 0,40$ (right). For symmetry reasons,
only profiles for the lower half channel are shown.}
\label{fig8:re10_skew}
\end{figure}

Remarkably, we found that the transformation of the mean flow
profile is accompanied by the absence of the logarithmic layer
behavior. The reasons are not entirely clear to us.  In principle,
since the Reynolds balance equation for the streamwise momentum does
not include any additional terms associated with the magnetic field
and retains its classical hydrodynamic form, there is  no reason why
the arguments leading to the log-layer behavior should not be valid
in the MHD case. As a possible explanation, one may speculate that
the weakening of the turbulent stress renders the flow similar to
flows at lower Reynolds numbers, thus reducing the size or
completely eliminating the log-layer.  The curves in figures
\ref{fig4:re20_prof}(e,f) allow such an interpretation.  The
explanation presumes that the logarithmic law may recover in an MHD
channel flow at the same values of $Ha$ and  higher $\Rey$, a
possibility we have to leave for future investigations. As a final
note on this issue we remark that certain mixing-length models for
Hartmann flow have been made to work reasonably well with  heuristic
coefficients accounting for the damping of turbulent fluctuations by
the magnetic field. However, we found that the turbulent stress
ansatz by Lykoudis and Brouillette \cite{Lykoudis:1967} -- a fairly
successful model for turbulent Hartmann flow (see Ref.
\onlinecite{Boeck:2007}) -- fails to predict the shape of the mean
velocity profile for the spanwise field when it is applied to our
case.

We also analyzed the anisotropy of turbulent fluctuations and found
that, with exception of the central area of the channel, it is
dominated by the mean shear.  The effect of the magnetic field is
significantly less pronounced than observed in earlier studies of
homogeneous (periodic box)
turbulence.\cite{Schumann:1976,Zikanov:1998,Knaepen1:2004,Vorobev:2005}
The direct effect of the magnetic field, i.e. the suppression of
velocity gradients in the spanwise direction, was observed to a
certain degree. It was relatively large in the middle of the channel
and decreasing toward the walls.  Remarkably, we found a comparable
or even stronger effect of the magnetic field on the streamwise
gradients of the velocity.  The turbulent structures increase their
typical size in both horizontal directions.  As an explanation of
this phenomenon we suggest the suppression of fluctuations by the
magnetic field, which also leads to stabilization and growth in size
of the coherent structures.  Here, again, we can invoke an
(admittedly incomplete) analogy between the flow affected by the
magnetic field and the hydrodynamic flow at a lower Reynolds number.

\acknowledgments{We are grateful to Andr\'e Thess and Maurice Rossi
for interesting discussions and useful comments, and to the
organizers Bernard Knaepen, Daniele Carati and Stavros Kassinos of
the MHD Summer School 2007 at the Universit\'e Libre de Bruxelles,
where this work was started. TB, DK and OZ acknowledge financial
support from the Deutsche Forschungsgemeinschaft (Emmy--Noether
grant Bo 1668/2-2 and Gerhard-Mercator visiting professorship
program). OZ's work is supported by the grant DE FG02 03 ER46062
from the U.S. Department of Energy. Financial support for the
collaboration between the TU Ilmenau and the University of Michigan
- Dearborn was provided by the National Science Foundation (grant
OISE 0338713). Computer resources were provided by the computing
centers of TU Ilmenau and TU Dresden as well as by the
Forschungszentrum J\"ulich (NIC).}

\bibliography{les_pof_arxiv_Sept05.bib}

\begin{thebibliography}{49}
\expandafter\ifx\csname natexlab\endcsname\relax\def\natexlab#1{#1}\fi
\expandafter\ifx\csname bibnamefont\endcsname\relax
  \def\bibnamefont#1{#1}\fi
\expandafter\ifx\csname bibfnamefont\endcsname\relax
  \def\bibfnamefont#1{#1}\fi
\expandafter\ifx\csname citenamefont\endcsname\relax
  \def\citenamefont#1{#1}\fi
\expandafter\ifx\csname url\endcsname\relax
  \def\url#1{\texttt{#1}}\fi
\expandafter\ifx\csname urlprefix\endcsname\relax\def\urlprefix{URL }\fi
\providecommand{\bibinfo}[2]{#2}
\providecommand{\eprint}[2][]{\url{#2}}

\bibitem[{\citenamefont{Thomas and Zhang}(2001)}]{Thomas:Zhang:2001}
\bibinfo{author}{\bibfnamefont{B.~G.} \bibnamefont{Thomas}} \bibnamefont{and}
  \bibinfo{author}{\bibfnamefont{L.}~\bibnamefont{Zhang}},
  ``\bibinfo{title}{Mathematical modeling of fluid flow in continuous casting:
  a Review}'', \bibinfo{journal}{ISIJ Intern} \textbf{\bibinfo{volume}{41}},
  \bibinfo{pages}{1181} (\bibinfo{year}{2001}).

\bibitem[{\citenamefont{von Ammon et~al.}(2005)\citenamefont{von Ammon,
  Gelfgat, Gorbunov, Muhlbauer, Muiznieks, Makarov, Virbulis, and
  Muller}}]{Ammon:2005}
\bibinfo{author}{\bibfnamefont{W.}~\bibnamefont{von Ammon}},
  \bibinfo{author}{\bibfnamefont{Y.}~\bibnamefont{Gelfgat}},
  \bibinfo{author}{\bibfnamefont{L.}~\bibnamefont{Gorbunov}},
  \bibinfo{author}{\bibfnamefont{A.}~\bibnamefont{Muhlbauer}},
  \bibinfo{author}{\bibfnamefont{A.}~\bibnamefont{Muiznieks}},
  \bibinfo{author}{\bibfnamefont{Y.}~\bibnamefont{Makarov}},
  \bibinfo{author}{\bibfnamefont{J.}~\bibnamefont{Virbulis}}, \bibnamefont{and}
  \bibinfo{author}{\bibfnamefont{G.}~\bibnamefont{Muller}}, in
  \emph{\bibinfo{booktitle}{The $15^{th}$ Riga and $6^{th}$ {PAMIR} Conference
  on Fundamental and Applied {MHD} Modeling of {MHD} turbulence}}
  (\bibinfo{address}{Riga, Latvia}, \bibinfo{year}{2005}),
  vol.~\bibinfo{volume}{I}, pp. \bibinfo{pages}{41--54}.

\bibitem[{\citenamefont{Barleon et~al.}(2001)\citenamefont{Barleon, Burr, Mack,
  and Stieglitz}}]{Barleon:2001}
\bibinfo{author}{\bibfnamefont{L.}~\bibnamefont{Barleon}},
  \bibinfo{author}{\bibfnamefont{U.}~\bibnamefont{Burr}},
  \bibinfo{author}{\bibfnamefont{K.~J.} \bibnamefont{Mack}}, \bibnamefont{and}
  \bibinfo{author}{\bibfnamefont{R.}~\bibnamefont{Stieglitz}},
  ``\bibinfo{title}{Magnetohydrodynamic heat transfer research related to the
  design of fusion blankets}'', \bibinfo{journal}{Fusion Techn.}
  \textbf{\bibinfo{volume}{39(2)}}, \bibinfo{pages}{127}
  (\bibinfo{year}{2001}).

\bibitem[{\citenamefont{Roberts}(1967)}]{Roberts:1967}
\bibinfo{author}{\bibfnamefont{P.~H.} \bibnamefont{Roberts}},
  \emph{\bibinfo{title}{An introduction to Magnetohydrodynamics}}
  (\bibinfo{publisher}{Longmans, Green}, \bibinfo{address}{New York},
  \bibinfo{year}{1967}).

\bibitem[{\citenamefont{Davidson}(1997)}]{Davidson:1997}
\bibinfo{author}{\bibfnamefont{P.}~\bibnamefont{Davidson}},
  ``\bibinfo{title}{The role of angular momentum in the magnetic damping of
  turbulence}'', \bibinfo{journal}{J. Fluid Mech.}
  \textbf{\bibinfo{volume}{336}}, \bibinfo{pages}{123} (\bibinfo{year}{1997}).

\bibitem[{\citenamefont{Alemany et~al.}(1979)\citenamefont{Alemany, Moreau,
  Sulem, and Frisch}}]{Alemany:1979}
\bibinfo{author}{\bibfnamefont{A.}~\bibnamefont{Alemany}},
  \bibinfo{author}{\bibfnamefont{R.}~\bibnamefont{Moreau}},
  \bibinfo{author}{\bibfnamefont{P.~L.} \bibnamefont{Sulem}}, \bibnamefont{and}
  \bibinfo{author}{\bibfnamefont{U.}~\bibnamefont{Frisch}},
  ``\bibinfo{title}{Influence of an external magnetic field on homogeneous
  {MHD} turbulence}'', \bibinfo{journal}{J. de Mecanique}
  \textbf{\bibinfo{volume}{280}}, \bibinfo{pages}{18} (\bibinfo{year}{1979}).

\bibitem[{\citenamefont{Moffatt}(1967)}]{Moffatt:1967}
\bibinfo{author}{\bibfnamefont{H.~K.} \bibnamefont{Moffatt}},
  ``\bibinfo{title}{On the suppression of turbulence by a uniform magnetic
  field}'', \bibinfo{journal}{J. Fluid Mech.} \textbf{\bibinfo{volume}{28}},
  \bibinfo{pages}{571} (\bibinfo{year}{1967}).

\bibitem[{\citenamefont{Schumann}(1979)}]{Schumann:1976}
\bibinfo{author}{\bibfnamefont{U.}~\bibnamefont{Schumann}},
  ``\bibinfo{title}{Numerical simulation of the transition from three- to
  two-dimensional turbulence under a uniform magnetic field}'',
  \bibinfo{journal}{J. Fluid Mech.} \textbf{\bibinfo{volume}{31}},
  \bibinfo{pages}{74} (\bibinfo{year}{1979}).

\bibitem[{\citenamefont{Zikanov and Thess}(1998)}]{Zikanov:1998}
\bibinfo{author}{\bibfnamefont{O.}~\bibnamefont{Zikanov}} \bibnamefont{and}
  \bibinfo{author}{\bibfnamefont{A.}~\bibnamefont{Thess}},
  ``\bibinfo{title}{Direct numerical simulation of forced {MHD} turbulence at
  low magnetic {Reynolds} number}'', \bibinfo{journal}{J. Fluid Mech.}
  \textbf{\bibinfo{volume}{358}}, \bibinfo{pages}{299} (\bibinfo{year}{1998}).

\bibitem[{\citenamefont{Knaepen and Moin}(2004)}]{Knaepen1:2004}
\bibinfo{author}{\bibfnamefont{B.}~\bibnamefont{Knaepen}} \bibnamefont{and}
  \bibinfo{author}{\bibfnamefont{P.}~\bibnamefont{Moin}},
  ``\bibinfo{title}{Large-eddy simulation of conductive flows at low magnetic
  {Reynolds} number}'', \bibinfo{journal}{Phys. Fluids}
  \textbf{\bibinfo{volume}{16}}, \bibinfo{pages}{1255} (\bibinfo{year}{2004}).

\bibitem[{\citenamefont{Vorobev et~al.}(2005)\citenamefont{Vorobev, Zikanov,
  Davidson, and Knaepen}}]{Vorobev:2005}
\bibinfo{author}{\bibfnamefont{A.}~\bibnamefont{Vorobev}},
  \bibinfo{author}{\bibfnamefont{O.}~\bibnamefont{Zikanov}},
  \bibinfo{author}{\bibfnamefont{P.~A.} \bibnamefont{Davidson}},
  \bibnamefont{and} \bibinfo{author}{\bibfnamefont{B.}~\bibnamefont{Knaepen}},
  ``\bibinfo{title}{Anisotropy of magnetohydrodynamic turbulence at low
  magnetic Reynolds number}'', \bibinfo{journal}{Phys. Fluids}
  \textbf{\bibinfo{volume}{17}}, \bibinfo{pages}{125105}
  (\bibinfo{year}{2005}).

\bibitem[{\citenamefont{Vorobev and Zikanov}(2008)}]{Vorobev:2007}
\bibinfo{author}{\bibfnamefont{A.}~\bibnamefont{Vorobev}} \bibnamefont{and}
  \bibinfo{author}{\bibfnamefont{O.}~\bibnamefont{Zikanov}},
  ``\bibinfo{title}{Smagorinsky constant in {LES} modeling of anisotropic {MHD}
  turbulence}'', \bibinfo{journal}{Theor. Comp. Fluid Dyn.}
  \textbf{\bibinfo{volume}{22}}, \bibinfo{pages}{317} (\bibinfo{year}{2008}).

\bibitem[{\citenamefont{Hartmann and Lazarus}(1937)}]{Hartmann:1937}
\bibinfo{author}{\bibfnamefont{J.}~\bibnamefont{Hartmann}} \bibnamefont{and}
  \bibinfo{author}{\bibfnamefont{F.}~\bibnamefont{Lazarus}},
  ``\bibinfo{title}{Hg-Dynamics {II}: Experimental investigations on the flow
  of mercury in a homogeneous magnetic field}'', \bibinfo{journal}{K. Dan.
  Vidensk. Selsk. Mat. Fys. Medd.} \textbf{\bibinfo{volume}{15}},
  \bibinfo{pages}{1} (\bibinfo{year}{1937}).

\bibitem[{\citenamefont{Sommeria and Moreau}(1982)}]{Sommeria:Moreau:1982}
\bibinfo{author}{\bibfnamefont{J.}~\bibnamefont{Sommeria}} \bibnamefont{and}
  \bibinfo{author}{\bibfnamefont{R.}~\bibnamefont{Moreau}},
  ``\bibinfo{title}{Why, how and when {MHD}-turbulence becomes
  two-dimensional}'', \bibinfo{journal}{J. Fluid Mech.}
  \textbf{\bibinfo{volume}{118}}, \bibinfo{pages}{507} (\bibinfo{year}{1982}).

\bibitem[{\citenamefont{Reed and Lykoudis}(1978)}]{Reed:Lykoudis:1978}
\bibinfo{author}{\bibfnamefont{C.~B.} \bibnamefont{Reed}} \bibnamefont{and}
  \bibinfo{author}{\bibfnamefont{P.~S.} \bibnamefont{Lykoudis}},
  ``\bibinfo{title}{The effect of a transverse magnetic field on shear
  turbulence}'', \bibinfo{journal}{J. Fluid Mech.}
  \textbf{\bibinfo{volume}{89}}, \bibinfo{pages}{147} (\bibinfo{year}{1978}).

\bibitem[{\citenamefont{Krasnov et~al.}(2004)\citenamefont{Krasnov, Zienicke,
  Zikanov, Boeck, and Thess}}]{Krasnov:2004}
\bibinfo{author}{\bibfnamefont{D.~S.} \bibnamefont{Krasnov}},
  \bibinfo{author}{\bibfnamefont{E.}~\bibnamefont{Zienicke}},
  \bibinfo{author}{\bibfnamefont{O.}~\bibnamefont{Zikanov}},
  \bibinfo{author}{\bibfnamefont{T.}~\bibnamefont{Boeck}}, \bibnamefont{and}
  \bibinfo{author}{\bibfnamefont{A.}~\bibnamefont{Thess}},
  ``\bibinfo{title}{Numerical study of the instability of the {Hartmann}
  layer}'', \bibinfo{journal}{J. Fluid Mech.} \textbf{\bibinfo{volume}{504}},
  \bibinfo{pages}{183} (\bibinfo{year}{2004}).

\bibitem[{\citenamefont{Boeck et~al.}(2007)\citenamefont{Boeck, Krasnov, and
  Zienicke}}]{Boeck:2007}
\bibinfo{author}{\bibfnamefont{T.}~\bibnamefont{Boeck}},
  \bibinfo{author}{\bibfnamefont{D.}~\bibnamefont{Krasnov}}, \bibnamefont{and}
  \bibinfo{author}{\bibfnamefont{E.}~\bibnamefont{Zienicke}},
  ``\bibinfo{title}{Numerical study of turbulent magnetohydrodynamic channel
  flow}'', \bibinfo{journal}{J. Fluid Mech.} \textbf{\bibinfo{volume}{572}},
  \bibinfo{pages}{179} (\bibinfo{year}{2007}).

\bibitem[{\citenamefont{Kassinos et~al.}(2006)\citenamefont{Kassinos, Knaepen,
  and Wray}}]{Kassinos:Knaepen:Wray:2006}
\bibinfo{author}{\bibfnamefont{C.}~\bibnamefont{Kassinos}},
  \bibinfo{author}{\bibfnamefont{B.}~\bibnamefont{Knaepen}}, \bibnamefont{and}
  \bibinfo{author}{\bibfnamefont{A.}~\bibnamefont{Wray}}, ``\bibinfo{title}{The
  structure of {MHD} turbulence subjected to mean shear and frame rotation}'',
  \bibinfo{journal}{J. Turb.} \textbf{\bibinfo{volume}{7}}, \bibinfo{pages}{1}
  (\bibinfo{year}{2006}).

\bibitem[{\citenamefont{Germano et~al.}(1991)\citenamefont{Germano, Piomelli,
  Moin, and Cabot}}]{Germano:1991}
\bibinfo{author}{\bibfnamefont{M.}~\bibnamefont{Germano}},
  \bibinfo{author}{\bibfnamefont{U.}~\bibnamefont{Piomelli}},
  \bibinfo{author}{\bibfnamefont{P.}~\bibnamefont{Moin}}, \bibnamefont{and}
  \bibinfo{author}{\bibfnamefont{W.~H.} \bibnamefont{Cabot}},
  ``\bibinfo{title}{A dynamic subgrid-scale eddy viscosity model}'',
  \bibinfo{journal}{Phys. Fluids} \textbf{\bibinfo{volume}{3}},
  \bibinfo{pages}{1760} (\bibinfo{year}{1991}).

\bibitem[{\citenamefont{Lilly}(1992)}]{Lilly:1992}
\bibinfo{author}{\bibfnamefont{D.~K.} \bibnamefont{Lilly}}, ``\bibinfo{title}{A
  proposed modification to the {Germano} subgrid-scale closure model}'',
  \bibinfo{journal}{Phys. Fluids} \textbf{\bibinfo{volume}{4}},
  \bibinfo{pages}{633} (\bibinfo{year}{1992}).

\bibitem[{\citenamefont{Shimomura}(1991)}]{Shimomura:1991}
\bibinfo{author}{\bibfnamefont{Y.}~\bibnamefont{Shimomura}},
  ``\bibinfo{title}{Large eddy simulation of magnetohydrodynamic turbulent
  channel flows under a uniform magnetic field}'', \bibinfo{journal}{Phys.
  Fluids A} \textbf{\bibinfo{volume}{3}}, \bibinfo{pages}{3098}
  (\bibinfo{year}{1991}).

\bibitem[{\citenamefont{Kobayashi}(2006)}]{Kobayashi:2006}
\bibinfo{author}{\bibfnamefont{H.}~\bibnamefont{Kobayashi}},
  ``\bibinfo{title}{Large eddy simulation of magnetohydrodynamic turbulent
  channel flows with local subgrid-scale model based on coherent structures}'',
  \bibinfo{journal}{Phys. Fluids} \textbf{\bibinfo{volume}{18}},
  \bibinfo{pages}{045107} (\bibinfo{year}{2006}).

\bibitem[{\citenamefont{Sarris et~al.}(2007)\citenamefont{Sarris, Kassinos, and
  Carati}}]{Sarris:Kassinos:Carati:2007}
\bibinfo{author}{\bibfnamefont{I.~E.} \bibnamefont{Sarris}},
  \bibinfo{author}{\bibfnamefont{S.~C.} \bibnamefont{Kassinos}},
  \bibnamefont{and} \bibinfo{author}{\bibfnamefont{D.}~\bibnamefont{Carati}},
  ``\bibinfo{title}{Large-eddy simulations of the turbulent Hartmann flow close
  to the transitional regime}'', \bibinfo{journal}{Phys. Fluids}
  \textbf{\bibinfo{volume}{19}}, \bibinfo{pages}{085109}
  (\bibinfo{year}{2007}).

\bibitem[{\citenamefont{Lee and Choi}(2001)}]{Lee:Choi:2001}
\bibinfo{author}{\bibfnamefont{D.}~\bibnamefont{Lee}} \bibnamefont{and}
  \bibinfo{author}{\bibfnamefont{H.}~\bibnamefont{Choi}},
  ``\bibinfo{title}{Magnetohydrodynamic turbulent flow in a channel at low
  magnetic {Reynolds} number}'', \bibinfo{journal}{J. Fluid Mech.}
  \textbf{\bibinfo{volume}{429}}, \bibinfo{pages}{367} (\bibinfo{year}{2001}).

\bibitem[{\citenamefont{Votsish and
  Kolesnikov}(1977)}]{Votsish:Kolesnikov:1977}
\bibinfo{author}{\bibfnamefont{A.~D.} \bibnamefont{Votsish}} \bibnamefont{and}
  \bibinfo{author}{\bibfnamefont{Y.~B.} \bibnamefont{Kolesnikov}},
  ``\bibinfo{title}{An experimental investigation of two-dimensional turbulence
  characteristics in a plane channel with an azimuthal magnetic field}'',
  \bibinfo{journal}{Magnetohydrodynamics} \textbf{\bibinfo{volume}{13}},
  \bibinfo{pages}{27} (\bibinfo{year}{1977}).

\bibitem[{\citenamefont{Votyakov et~al.}(2007)\citenamefont{Votyakov,
  Kolesnikov, Andreev, Zienicke, and Thess}}]{Votyakov:2007}
\bibinfo{author}{\bibfnamefont{E.}~\bibnamefont{Votyakov}},
  \bibinfo{author}{\bibfnamefont{Y.}~\bibnamefont{Kolesnikov}},
  \bibinfo{author}{\bibfnamefont{O.}~\bibnamefont{Andreev}},
  \bibinfo{author}{\bibfnamefont{E.}~\bibnamefont{Zienicke}}, \bibnamefont{and}
  \bibinfo{author}{\bibfnamefont{A.}~\bibnamefont{Thess}},
  ``\bibinfo{title}{Structure of the wake of a magnetic obstacle}'',
  \bibinfo{journal}{Phys. Rev. Lett} \textbf{\bibinfo{volume}{98}},
  \bibinfo{pages}{144504} (\bibinfo{year}{2007}).

\bibitem[{\citenamefont{Piomelli}(1993)}]{Piomelli:1993}
\bibinfo{author}{\bibfnamefont{U.}~\bibnamefont{Piomelli}},
  ``\bibinfo{title}{High Reynolds number calculations using the dynamic
  subgrid-scale stress model}'', \bibinfo{journal}{Phys. Fluids}
  \textbf{\bibinfo{volume}{5}}, \bibinfo{pages}{1484} (\bibinfo{year}{1993}).

\bibitem[{\citenamefont{Krasnov et~al.}(2008)\citenamefont{Krasnov, Rossi,
  Zikanov, and Boeck}}]{Krasnov:2007}
\bibinfo{author}{\bibfnamefont{D.}~\bibnamefont{Krasnov}},
  \bibinfo{author}{\bibfnamefont{M.}~\bibnamefont{Rossi}},
  \bibinfo{author}{\bibfnamefont{O.}~\bibnamefont{Zikanov}}, \bibnamefont{and}
  \bibinfo{author}{\bibfnamefont{T.}~\bibnamefont{Boeck}},
  ``\bibinfo{title}{Optimal growth and transition to turbulence in channel flow
  with spanwise magnetic field}'', \bibinfo{journal}{J. Fluid Mech.}
  \textbf{\bibinfo{volume}{596}}, \bibinfo{pages}{73} (\bibinfo{year}{2008}).

\bibitem[{\citenamefont{Canuto et~al.}(1988)\citenamefont{Canuto, Hussaini,
  Quateroni, and Zang}}]{Canuto:1988}
\bibinfo{author}{\bibfnamefont{C.}~\bibnamefont{Canuto}},
  \bibinfo{author}{\bibfnamefont{M.~Y.} \bibnamefont{Hussaini}},
  \bibinfo{author}{\bibfnamefont{A.}~\bibnamefont{Quateroni}},
  \bibnamefont{and} \bibinfo{author}{\bibfnamefont{T.~A.} \bibnamefont{Zang}},
  \emph{\bibinfo{title}{Spectral Methods in Fluid Dynamics}}
  (\bibinfo{publisher}{Springer Verlag}, \bibinfo{year}{1988}).

\bibitem[{\citenamefont{Moin and Kim}(1982)}]{Moin:1982}
\bibinfo{author}{\bibfnamefont{P.}~\bibnamefont{Moin}} \bibnamefont{and}
  \bibinfo{author}{\bibfnamefont{J.}~\bibnamefont{Kim}},
  ``\bibinfo{title}{Numerical investigation of turbulent channel flow}'',
  \bibinfo{journal}{J. Fluid Mech.} \textbf{\bibinfo{volume}{118}},
  \bibinfo{pages}{341} (\bibinfo{year}{1982}).

\bibitem[{\citenamefont{van Driest}(1956)}]{vanDriest:1956}
\bibinfo{author}{\bibfnamefont{E.~R.} \bibnamefont{van Driest}},
  ``\bibinfo{title}{On turbulent flow near a wall}'', \bibinfo{journal}{J.
  Aeron. Sci.} \textbf{\bibinfo{volume}{23}}, \bibinfo{pages}{1007}
  (\bibinfo{year}{1956}).

\bibitem[{\citenamefont{Moser et~al.}(1999)\citenamefont{Moser, Kim, and
  Mansour}}]{Moser:1999}
\bibinfo{author}{\bibfnamefont{R.~D.} \bibnamefont{Moser}},
  \bibinfo{author}{\bibfnamefont{J.}~\bibnamefont{Kim}}, \bibnamefont{and}
  \bibinfo{author}{\bibfnamefont{N.~N.} \bibnamefont{Mansour}},
  ``\bibinfo{title}{Direct numerical simulation of turbulent channel flow up to
  ${Re_{\tau}}=590$}'', \bibinfo{journal}{Phys. Fluids}
  \textbf{\bibinfo{volume}{11}}, \bibinfo{pages}{943} (\bibinfo{year}{1999}).

\bibitem[{\citenamefont{Kim et~al.}(1987)\citenamefont{Kim, Moin, and
  Moser}}]{Kim:Moin:1987}
\bibinfo{author}{\bibfnamefont{J.}~\bibnamefont{Kim}},
  \bibinfo{author}{\bibfnamefont{P.}~\bibnamefont{Moin}}, \bibnamefont{and}
  \bibinfo{author}{\bibfnamefont{R.}~\bibnamefont{Moser}},
  ``\bibinfo{title}{Turbulence statistics in fully developed channel flow at
  low {Reynolds} number}'', \bibinfo{journal}{J. Fluid Mech.}
  \textbf{\bibinfo{volume}{177}}, \bibinfo{pages}{133} (\bibinfo{year}{1987}).

\bibitem[{\citenamefont{Piomelli et~al.}(1988)\citenamefont{Piomelli, Moin, and
  Ferziger}}]{Piomelli:1988}
\bibinfo{author}{\bibfnamefont{U.}~\bibnamefont{Piomelli}},
  \bibinfo{author}{\bibfnamefont{P.}~\bibnamefont{Moin}}, \bibnamefont{and}
  \bibinfo{author}{\bibfnamefont{J.~H.} \bibnamefont{Ferziger}},
  ``\bibinfo{title}{Model consistency in large eddy simulation of turbulent
  channel flows}'', \bibinfo{journal}{Phys. Fluids}
  \textbf{\bibinfo{volume}{31}}, \bibinfo{pages}{1884} (\bibinfo{year}{1988}).

\bibitem[{\citenamefont{Barenblatt et~al.}(1997)\citenamefont{Barenblatt,
  Chorin, and Prostokishin}}]{Barenblatt:1997}
\bibinfo{author}{\bibfnamefont{G.~I.} \bibnamefont{Barenblatt}},
  \bibinfo{author}{\bibfnamefont{A.}~\bibnamefont{Chorin}}, \bibnamefont{and}
  \bibinfo{author}{\bibfnamefont{V.~M.} \bibnamefont{Prostokishin}},
  ``\bibinfo{title}{Scaling laws for fully developed flow in pipes}'',
  \bibinfo{journal}{Appl. Mech. Rev.} \textbf{\bibinfo{volume}{50}},
  \bibinfo{pages}{413} (\bibinfo{year}{1997}).

\bibitem[{\citenamefont{Jeong and Hussein}(1995)}]{Jeong:Hussein:1995}
\bibinfo{author}{\bibfnamefont{J.}~\bibnamefont{Jeong}} \bibnamefont{and}
  \bibinfo{author}{\bibfnamefont{F.}~\bibnamefont{Hussein}},
  ``\bibinfo{title}{On the identification of a vortex}'', \bibinfo{journal}{J.
  Fluid Mech.} \textbf{\bibinfo{volume}{285}}, \bibinfo{pages}{69}
  (\bibinfo{year}{1995}).

\bibitem[{\citenamefont{Kassinos et~al.}(2001)\citenamefont{Kassinos, Reynolds,
  and Rogers}}]{Kassinos:2001}
\bibinfo{author}{\bibfnamefont{S.~C.} \bibnamefont{Kassinos}},
  \bibinfo{author}{\bibfnamefont{W.~C.} \bibnamefont{Reynolds}},
  \bibnamefont{and} \bibinfo{author}{\bibfnamefont{M.~M.}
  \bibnamefont{Rogers}}, ``\bibinfo{title}{One-point turbulence structure
  tensors}'', \bibinfo{journal}{J. Fluid Mech.} \textbf{\bibinfo{volume}{428}},
  \bibinfo{pages}{213} (\bibinfo{year}{2001}).

\bibitem[{\citenamefont{Warhaft}(2002)}]{Warhaft:2002}
\bibinfo{author}{\bibfnamefont{Z.}~\bibnamefont{Warhaft}},
  ``\bibinfo{title}{Turbulence in nature and the laboratory}'',
  \bibinfo{journal}{Proc. Nat. Acad. Sci.} \textbf{\bibinfo{volume}{99}},
  \bibinfo{pages}{2481} (\bibinfo{year}{2002}).

\bibitem[{\citenamefont{Gualtieri et~al.}(2002)\citenamefont{Gualtieri,
  Casciola, Benzi, Amati, and Piva}}]{Gualtieri:2002}
\bibinfo{author}{\bibfnamefont{P.}~\bibnamefont{Gualtieri}},
  \bibinfo{author}{\bibfnamefont{C.~M.} \bibnamefont{Casciola}},
  \bibinfo{author}{\bibfnamefont{R.}~\bibnamefont{Benzi}},
  \bibinfo{author}{\bibfnamefont{G.}~\bibnamefont{Amati}}, \bibnamefont{and}
  \bibinfo{author}{\bibfnamefont{R.}~\bibnamefont{Piva}},
  ``\bibinfo{title}{Scaling laws and intermittency in homogeneous shear
  flow}'', \bibinfo{journal}{Phys. Fluids} \textbf{\bibinfo{volume}{14}},
  \bibinfo{pages}{583} (\bibinfo{year}{2002}).

\bibitem[{\citenamefont{Schumacher et~al.}(2003)\citenamefont{Schumacher,
  Sreenivasan, and Yeung}}]{Schumacher:2003}
\bibinfo{author}{\bibfnamefont{J.}~\bibnamefont{Schumacher}},
  \bibinfo{author}{\bibfnamefont{K.~R.} \bibnamefont{Sreenivasan}},
  \bibnamefont{and} \bibinfo{author}{\bibfnamefont{P.~K.} \bibnamefont{Yeung}},
  ``\bibinfo{title}{Derivative moments in turbulent shear flows}'',
  \bibinfo{journal}{Phys. Fluids} \textbf{\bibinfo{volume}{15}},
  \bibinfo{pages}{84} (\bibinfo{year}{2003}).

\bibitem[{\citenamefont{Armenio and Sarkar}(2002)}]{Armenio:2002}
\bibinfo{author}{\bibfnamefont{V.}~\bibnamefont{Armenio}} \bibnamefont{and}
  \bibinfo{author}{\bibfnamefont{S.}~\bibnamefont{Sarkar}},
  ``\bibinfo{title}{An investigation of stably stratified turbulent channel
  flow using large-eddy simulation}'', \bibinfo{journal}{J. Fluid Mech.}
  \textbf{\bibinfo{volume}{459}}, \bibinfo{pages}{1} (\bibinfo{year}{2002}).

\bibitem[{\citenamefont{Toms}(1949)}]{Toms:1949}
\bibinfo{author}{\bibfnamefont{B.~A.} \bibnamefont{Toms}}, in
  \emph{\bibinfo{booktitle}{Proceedings of the International Congress on
  Rheology (Holland 1948)}} (\bibinfo{publisher}{North-Holland, Amsterdam},
  \bibinfo{year}{1949}), vol.~\bibinfo{volume}{2} of
  \emph{\bibinfo{series}{ERCOFTAC Series}}, pp. \bibinfo{pages}{135--141}.

\bibitem[{\citenamefont{Virk}(1975)}]{Virk:1975}
\bibinfo{author}{\bibfnamefont{P.~S.} \bibnamefont{Virk}},
  ``\bibinfo{title}{Drag reduction fundamentals}'', \bibinfo{journal}{AIChE J.}
  \textbf{\bibinfo{volume}{21}}, \bibinfo{pages}{625} (\bibinfo{year}{1975}).

\bibitem[{\citenamefont{Warholic et~al.}(1999)\citenamefont{Warholic, Massah,
  and Hanratty}}]{Warholic:1999}
\bibinfo{author}{\bibfnamefont{M.~D.} \bibnamefont{Warholic}},
  \bibinfo{author}{\bibfnamefont{H.}~\bibnamefont{Massah}}, \bibnamefont{and}
  \bibinfo{author}{\bibfnamefont{T.~J.} \bibnamefont{Hanratty}},
  ``\bibinfo{title}{Influence of drag-reducing polymers on turbulence: effects
  of {Reynolds} number, concentration and mixing}'', \bibinfo{journal}{Exp.
  Fluids} \textbf{\bibinfo{volume}{27}}, \bibinfo{pages}{461}
  (\bibinfo{year}{1999}).

\bibitem[{\citenamefont{Sureshkumar et~al.}(1997)\citenamefont{Sureshkumar,
  Beris, and Handler}}]{Sureshkumar:1997}
\bibinfo{author}{\bibfnamefont{R.}~\bibnamefont{Sureshkumar}},
  \bibinfo{author}{\bibfnamefont{A.~N.} \bibnamefont{Beris}}, \bibnamefont{and}
  \bibinfo{author}{\bibfnamefont{R.~A.} \bibnamefont{Handler}},
  ``\bibinfo{title}{Direct numerical simulation of the turbulent channel flow
  of a polymer solution}'', \bibinfo{journal}{Phys. Fluids}
  \textbf{\bibinfo{volume}{9}}, \bibinfo{pages}{743} (\bibinfo{year}{1997}).

\bibitem[{\citenamefont{Peters and Schumacher}(2007)}]{Peters:2007}
\bibinfo{author}{\bibfnamefont{T.}~\bibnamefont{Peters}} \bibnamefont{and}
  \bibinfo{author}{\bibfnamefont{J.}~\bibnamefont{Schumacher}},
  ``\bibinfo{title}{Two-way coupling of FENE dumbbells with a turbulent shear
  flow}'', \bibinfo{journal}{Phys. Fluids} \textbf{\bibinfo{volume}{19}},
  \bibinfo{pages}{065109} (\bibinfo{year}{2007}).

\bibitem[{\citenamefont{Dimitropoulos et~al.}(2005)\citenamefont{Dimitropoulos,
  Dubief, Shaqfeh, Moin, and Lele}}]{Dimitropoulos:2005}
\bibinfo{author}{\bibfnamefont{C.~D.} \bibnamefont{Dimitropoulos}},
  \bibinfo{author}{\bibfnamefont{Y.}~\bibnamefont{Dubief}},
  \bibinfo{author}{\bibfnamefont{E.~S.~G.} \bibnamefont{Shaqfeh}},
  \bibinfo{author}{\bibfnamefont{P.}~\bibnamefont{Moin}}, \bibnamefont{and}
  \bibinfo{author}{\bibfnamefont{S.~K.} \bibnamefont{Lele}},
  ``\bibinfo{title}{Direct numerical simulation of polymer-induced drag
  reduction in a turbulent shear flow}'', \bibinfo{journal}{Phys. Fluids}
  \textbf{\bibinfo{volume}{17}}, \bibinfo{pages}{011705}
  (\bibinfo{year}{2005}).

\bibitem[{\citenamefont{Benzi et~al.}(2006)\citenamefont{Benzi, De~Angelis,
  L`vov, Procaccia, and Tiberkevich}}]{Benzi:2006}
\bibinfo{author}{\bibfnamefont{R.}~\bibnamefont{Benzi}},
  \bibinfo{author}{\bibfnamefont{E.}~\bibnamefont{De~Angelis}},
  \bibinfo{author}{\bibfnamefont{V.~S.} \bibnamefont{L`vov}},
  \bibinfo{author}{\bibfnamefont{I.}~\bibnamefont{Procaccia}},
  \bibnamefont{and}
  \bibinfo{author}{\bibfnamefont{V.}~\bibnamefont{Tiberkevich}},
  ``\bibinfo{title}{Maximum drag reduction asymptotes and the cross-over to the
  {Newtonian} plug}'', \bibinfo{journal}{J. Fluid Mech.}
  \textbf{\bibinfo{volume}{551}}, \bibinfo{pages}{185} (\bibinfo{year}{2006}).

\bibitem[{\citenamefont{Lykoudis and Brouillette}(1967)}]{Lykoudis:1967}
\bibinfo{author}{\bibfnamefont{P.~S.} \bibnamefont{Lykoudis}} \bibnamefont{and}
  \bibinfo{author}{\bibfnamefont{E.~C.} \bibnamefont{Brouillette}},
  ``\bibinfo{title}{Magneto-Fluid-Mechanic Channel Flow. II. Theory}'',
  \bibinfo{journal}{Phys. Fluids} \textbf{\bibinfo{volume}{10}},
  \bibinfo{pages}{1002} (\bibinfo{year}{1967}).

\end{thebibliography}
\newpage


\begin{table}[t]
\begin{center}
\begin{ruledtabular}
\scriptsize{
\begin{tabular}{@{}ccccccc@{}}
{$Re$} & {$Re_q$} & Ha & N & {Simulation} &
{$N_x\times N_y\times N_z$} & {$L_x\times L_y\times L_z$}\\[6pt]
\hline
$10000$ & $6.67\times 10^3$ &0, 10, 20, 30 &0, 0.01, 0.04, 0.09 &
DNS & $256\times 256\times 256$ & $2\pi\times \pi \times 2 $\\
$ $ & $ $ & $ $ & $ $ & LES & $64\times 64\times 64$ & $ $\\
$ $ & $ $ & $ $ & $ $ & UDNS & $64\times 64\times 64$ & $ $\\
\hline
$20000$ & $1.33\times 10^4$ &0, 20, 30, 40 &0, 0.02, 0.045, 0.08 &
DNS & $512\times 512\times 256$ & $2\pi\times \pi \times 2$\\
$ $ & $ $ & $ $ & $ $ & LES & $128\times 128\times 128$, $64\times 64\times 64$ & $ $\\
$ $ & $ $ & $ $ & $ $ & UDNS & $128\times 128\times 128$ & $ $
\end{tabular}
} 
\end{ruledtabular}
\end{center}
\caption{Parameters of numerical experiments.} \label{table:param1}
\end{table}

\begin{table}[t]
\begin{center}
\begin{ruledtabular}
\scriptsize{
\begin{tabular}{@{}cccccc@{}}
{Run} & {$Ha$} & {$U_{cl}$} & {$Re_\tau$} & {$C_f$} & {$\epsilon$}\\[6pt]

\hline \multicolumn{6}{c}{$Re=10000$} \\ \hline

 DNS &  $0$ & $1.1469$ & $381.34$ & $6.543\times 10^{-3}$ & $82.8802$ \\
 DSM &  $0$ & $1.1464$ & $376.23$ & $6.370\times 10^{-3}$ & $60.8874$ \\
  SM &  $0$ & $1.1472$ & $373.76$ & $6.286\times 10^{-3}$ & $59.3454$ \\
UDNS &  $0$ & $1.1728$ & $405.57$ & $7.402\times 10^{-3}$ & $103.2484$ \\
\hline
 DNS & $10$ & $1.1627$ & $373.14$ & $6.265\times 10^{-3}$ & $77.4391$ \\
 DSM & $10$ & $1.1600$ & $369.02$ & $6.128\times 10^{-3}$ & $57.3222$ \\
  SM & $10$ & $1.1613$ & $364.97$ & $5.994\times 10^{-3}$ & $55.1329$ \\
UDNS & $10$ & $1.1962$ & $394.72$ & $7.011\times 10^{-3}$ & $95.9467$ \\
\hline
 DNS & $20$ & $1.1990$ & $350.21$ & $5.519\times 10^{-3}$ & $60.9412$ \\
 DSM & $20$ & $1.2031$ & $344.04$ & $5.326\times 10^{-3}$ & $47.1586$ \\
  SM & $20$ & $1.2011$ & $338.12$ & $5.144\times 10^{-3}$ & $43.2758$ \\
UDNS & $20$ & $1.2359$ & $371.84$ & $6.222\times 10^{-3}$ & $78.5375$ \\
\hline
 DNS & $30$ & $1.2542$ & $306.69$ & $4.232\times 10^{-3}$ & $38.8687$ \\
 DSM & $30$ & $1.2695$ & $302.27$ & $4.112\times 10^{-3}$ & $32.7482$ \\
  SM & $30$ & $1.2676$ & $288.85$ & $3.755\times 10^{-3}$ & $25.9266$ \\
UDNS & $30$ & $1.2887$ & $331.38$ & $4.941\times 10^{-3}$ & $54.2468$ \\

\hline \multicolumn{6}{c}{$Re=20000$} \\ \hline

 DNS &  $0$ & $1.1314$ & $702.00$ & $5.544\times 10^{-3}$ & $153.777$ \\
 DSM &  $0$ & $1.1341$ & $696.55$ & $5.458\times 10^{-3}$ & $116.885$ \\
  SM &  $0$ & $1.1388$ & $704.66$ & $5.586\times 10^{-3}$ & $123.347$ \\
UDNS &  $0$ & $1.1422$ & $753.29$ & $6.384\times 10^{-3}$ & $192.581$ \\
\hline
 DNS & $20$ & $1.1668$ & $669.54$ & $5.043\times 10^{-3}$ & $134.258$ \\
 DSM & $20$ & $1.1629$ & $669.25$ & $5.038\times 10^{-3}$ & $103.991$ \\
  SM & $20$ & $1.1669$ & $675.88$ & $5.139\times 10^{-3}$ & $108.525$ \\
DSM64 & $20$ & $1.1692$ & $634.78$ & $4.533\times 10^{-3}$ & $72.6128$ \\
 SM64 & $20$ & $1.1685$ & $604.75$ & $4.115\times 10^{-3}$ & $58.1954$ \\
UDNS & $20$ & $1.1853$ & $713.11$ & $5.721\times 10^{-3}$ & $165.251$ \\
\hline
 DNS & $30$ & $1.2037$ & $633.67$ & $4.517\times 10^{-3}$ & $112.074$ \\
 DSM & $30$ & $1.1979$ & $632.80$ & $4.505\times 10^{-3}$ & $89.6866$ \\
  SM & $30$ & $1.1990$ & $639.19$ & $4.596\times 10^{-3}$ & $91.8488$ \\
UDNS & $30$ & $1.2167$ & $675.13$ & $5.127\times 10^{-3}$ & $138.615$ \\
\hline
 DNS & $40$ & $1.2392$ & $593.79$ & $3.966\times 10^{-3}$ & $89.7640$ \\
 DSM & $40$ & $1.2412$ & $589.60$ & $3.911\times 10^{-3}$ & $74.5571$ \\
  SM & $40$ & $1.2414$ & $588.59$ & $3.897\times 10^{-3}$ & $72.0002$ \\
DSM64 & $40$ & $1.2484$ & $559.55$ & $3.522\times 10^{-3}$ & $52.5176$ \\
 SM64 & $40$ & $1.2398$ & $523.84$ & $3.087\times 10^{-3}$ & $38.4884$ \\
UDNS & $40$ & $1.2583$ & $625.55$ & $4.402\times 10^{-3}$ & $109.628$
\end{tabular}
} 
\end{ruledtabular}
\end{center}
\caption{Time-averaged integral parameters of the flow.
Listed are the Reynolds and Hartmann numbers (\ref{reynolds})
and (\ref{hartmann}), the channel centerline velocity $U_{cl}$ (relative to
the mean velocity $U_q$),
the friction Reynolds number (\ref{retau}), the friction
coefficient (\ref{cfri}), and the total resolved dissipation
rate (\ref{dissrate}). DSM and SM stand for dynamic and classical
Smagorinsky models conducted at higher resolution: $64^3$ at
$\Rey=10000$ and $128^3$ at $\Rey=20000$. DSM64 and SM64 are for
the results obtained with the resolution $64^3$ at $\Rey=20000$
and UDNS is for unresolved DNS (Table \ref{table:param1}).}
\label{table:param2}
\end{table}

\end{document}